\documentclass[a4paper,11pt]{article}
\pdfoutput=1 
 \usepackage{jheppub} 
\usepackage[T1]{fontenc} 
\usepackage{mathtools}
\usepackage{subcaption}
\usepackage{slashed}
\usepackage{cancel}
\usepackage{bm}
\usepackage{upgreek}
\usepackage[normalem]{ulem}
\makeatletter
\newsavebox{\@brx}
\newcommand{\llangle}[1][]{\savebox{\@brx}{\(\m@th{#1\langle}\)}
  \mathopen{\copy\@brx\kern-0.5\wd\@brx\usebox{\@brx}}}
\newcommand{\rrangle}[1][]{\savebox{\@brx}{\(\m@th{#1\rangle}\)}
  \mathclose{\copy\@brx\kern-0.5\wd\@brx\usebox{\@brx}}}
\makeatother

\newcommand{\sgn}[1]{{\text{sgn}(#1)}}

\title{Where Non-Invertible Symmetries End:\\
Twist Defects for Electromagnetic Duality}

\makeatletter
\gdef\@fpheader{MIT-CTP/5930, YITP-SB-2025-17}
\makeatother

\author[1]{Shu-Heng Shao}
\author[2,3]{and Siwei Zhong}

\affiliation[1]{Center for Theoretical Physics - a Leinweber Institute, Massachusetts Institute of Technology, \\
Cambridge, MA 02139, USA}
\affiliation[2]{Simons Center for Geometry and Physics, Stony Brook University, \\
Stony Brook, NY 11794, USA}
\affiliation[3]{C. N. Yang Institute for Theoretical Physics, Stony Brook University, \\Stony Brook, NY 11794, USA}

\emailAdd{shuheng@mit.edu}
\emailAdd{siwei.zhong@stonybrook.edu}

\abstract{We study novel conformal twist defects in 4d Maxwell theory, around which electric and magnetic fields are exchanged. These are codimension-2 defects living at the end of topological defects for certain non-invertible global symmetries.  
We determine the operator spectrum of the twist defect by solving classical electromagnetic wave equations subject to a twisted boundary condition. 
Using techniques from defect CFT, we show that correlation functions of these defect operators factorize into two sectors: a universal generalized free-field sector, and a chiral current sector analogous to edge modes in Chern-Simons theory.
In a similar setup, we also revisit the twist fields attached to non-invertible line defects in the 2d compact boson CFT. We discuss a defect 't Hooft anomaly involving a chiral $O(2)$ symmetry, highlighting its dynamical implications.}

\begin{document} 
\maketitle
\flushbottom

\section{Introduction and summary}

\subsection{Global symmetries and twist defects}

Conformal field theories (CFTs) with an ordinary internal global symmetry $G$ have many interesting observables beyond those of local operators. 
Of particular importance are \textit{twist defects}, also commonly known as \textit{monodromy defects}, associated with a global symmetry element $g\in G$. Twist defects are codimension-2, extended defects in spacetime. 
The defining property of a twist defect is that any local (gauge-invariant) operator $\Phi$ undergoes a $g$-symmetry transformation when going around the defect:
\begin{equation}
\Phi(\theta + 2\pi) = \Phi^g(\theta),
\end{equation}
where $\theta$ denotes the angular coordinate around the twist defect and we have suppressed the remaining spacetime coordinates. 
$\Phi^g$ is the image of $\Phi$ under the global symmetry transformation $g\in G$. 

Importantly, twist defects are not genuine codimension-2 defects; rather, they are attached to a codimension-1 defect  $\mathcal{D}_g$ associated with the global symmetry element $g\in G$. 
In other words, twist defects live at the end (or boundary) of the symmetry defect. 
Let us compare these two kinds of defects:
\begin{itemize}
\item The symmetry defect $\mathcal{D}_g$ is \textit{topological}, in the sense that it commutes with the stress-energy tensor. 
Consequently, its locus may be infinitesimally deformed without affecting correlation functions. 
In contrast, a twist defect is non-topological as long as its associated symmetry acts faithfully on local operators.\footnote{In topological quantum field theories, such as Chern-Simons theories, with a unique local operator, 0-form global symmetries act nontrivially on extended operators but trivially on local operators. In these theories twist defects are topological \cite{Barkeshli:2014cna}. 
See, for example, \cite{Brennan:2025acl, Seiberg:2025bqy} and references therein, for other examples.} Throughout this work, we will assume that twist defects preserve a proper subgroup of the conformal symmetry, and hence are  \textit{conformal defects}.
\item The symmetry defect $\mathcal{D}_g$ is a \textit{genuine} codimension-1 defect, in the sense that any correlation function depends only on the codimension-1 manifold on which it is supported. In contrast, twist defects are not genuine codimension-2 defects: their correlation functions additionally require specifying the locus of the codimension-1 symmetry defect to which they are attached.
\end{itemize}
We refer the reader to \cite{Kapustin:2014gua, Gaiotto:2014kfa} for more general discussions on topological defects and genuine extended defects.

Below we review some well-known examples of twist defects in diverse spacetime dimensions.

In two spacetime dimensions, the symmetry defect $\mathcal{D}_g$ is a line in spacetime, and twist defects are point-like operators living at the end of $\mathcal{D}_g$. 
In this context, twist defects are also known as disorder operators \cite{Kadanoff:1970kz}, twist fields \cite{Dixon:1986qv,Knizhnik:1987xp}, or twisted sector operators.\footnote{In CFT, twisted sector operators or twist fields usually refer to genuine local operators in the CFT obtained from gauging (also known as orbifolding) a finite global symmetry of another CFT. Here, as a slight abuse of terminology, we use these terms to refer to the non-genuine, point-like operators attached to a topological line in the CFT before gauging. See also \cite{Doyon:2025xvo} for a recent review of twist fields.} 
As an example, consider the 2d Ising CFT, which has a $G=\mathbb{Z}_2$ global symmetry. 
The associated twist fields include the disorder operator, commonly denoted as $\mu$, whose conformal weights are $(h,\bar h)= (1/16,1/16)$, as well as the free fermion fields $\psi, \bar\psi$, whose conformal weights are $(1/2,0)$ and $(0,1/2)$, respectively (see, e.g., \cite{Ginsparg:1988ui}). 
These are not genuine local operators; rather, they are attached to the $\mathbb{Z}_2$ topological line defect, which creates a branch cut for the $\mathbb{Z}_2$-odd, genuine local operators, such as the spin field/order operator $\sigma$. 
Its correlation functions with other local operators are not single-valued functions of their positions, but can have branch cuts. 
In general fermionic CFTs, the Ramond–Ramond sector operators play an analogous role: they are twist fields associated with the fermion parity symmetry generated by $(-1)^F$.

In higher spacetime dimensions, a well-known example is the twist/monodromy line defect in the 3d Ising CFT. 
This twist defect is associated with the $G=\mathbb{Z}_2$ global symmetry of the Ising CFT, and is attached to the topological $\mathbb{Z}_2$ surface defect in spacetime.\footnote{This twist line defect is not to be confused with the pinning line defect in the 3d Ising CFT. In spacetime, the former is not a genuine line defect, while the latter is. If we gauge the $\mathbb{Z}_2$ global symmetry of the Ising CFT, the twist defect becomes a genuine line defect in the Ising$/\mathbb{Z}_2$ CFT. 
This genuine line defect in Ising$/\mathbb{Z}_2$ is charged under a $\mathbb{Z}_2$ 1-form global symmetry. 
It follows that it is ``unbreakable" in the sense that it does not admit an endpoint (see, e.g., \cite{Rudelius:2020orz,Harlow:2025cqc} for a general argument). 
By contrast, the pinning line defect in the original Ising CFT is breakable, and its endpoint spectrum has been studied recently in \cite{Lanzetta:2025xfw}.} 
This line defect has been studied by Monte Carlo in \cite{Billo:2013jda} and by the conformal bootstrap in \cite{Gaiotto:2013nva}.  
Twist defects are also essential for computing the R\'enyi and entanglement entropies when using the replica trick \cite{Calabrese:2004eu}. 
See, for example,  \cite{Hung:2014npa,Bianchi:2015liz,Cordova:2017mhb,Soderberg:2017oaa,Giombi:2021uae,Wang:2021lmb,Bianchi:2021snj,Wang:2021yaf,Gimenez-Grau:2021wiv,Dowker:2022mac,Dowker:2022mex,Gimenez-Grau:2022czc,SoderbergRousu:2023pbe,Barkeshli:2025cjs} for a highly incomplete selection of references on twist defects.

\subsection{Non-invertible symmetry and electromagnetic duality}

So far, we have assumed $G$ to be an ordinary invertible global symmetry. 
What if $G$ is a generalized global symmetry?

In recent years, it has become increasingly clear that symmetries in quantum field theory and lattice systems need not be invertible.   
In Euclidean correlation functions, 
these (0-form) non-invertible symmetries are realized by 
codimension-1 topological defects whose fusion rules are not group-like. 
In Lorentzian signature, they correspond to conserved operators acting on the Hilbert space; however, unlike ordinary global symmetries, which are implemented by (anti-)unitary operators with inverses, these operators do not have an inverse. 
Non-invertible symmetries exist ubiquitously in familiar quantum systems and have far-reaching dynamical consequences. See \cite{McGreevy:2022oyu,Cordova:2022ruw,Schafer-Nameki:2023jdn,Brennan:2023mmt,Shao:2023gho,Costa:2024wks,PT} for recent reviews.

The simplest example of non-invertible symmetries in four spacetime dimensions can be found in free Maxwell $U(1)$ gauge theory without matter fields at special values of the coupling constant $e$. 
Although the spectrum of local operators is independent of $e$,  partition functions and extended operators depend nontrivially on $e$ (see, e.g., \cite{Witten:1995gf}).
This non-invertible global symmetry is closely related to electromagnetic duality, or S-duality, as we explain below. 

Generally, duality and symmetry are distinct notions. 
Duality states that two seemingly different descriptions (e.g., Lagrangians) define the same quantum field theory. 
A classic example is electromagnetic duality: 4d Maxwell theory at coupling $e$ is equivalent to the theory at $\tilde e = 2\pi /e$.  
Under such a duality, the abstract operator content is preserved, but the labeling changes -- for instance, the 
electric and magnetic fields $\vec{E}$ and $\vec{B}$ are exchanged. 
Other well-known examples of exact dualities include T-duality in 2d \cite{Kikkawa:1984cp, Sakai:1985cs} and the Montonen–Olive duality \cite{Montonen:1977sn,Goddard:1976qe} in 4d $\mathcal{N}=4$ super Yang-Mills theory. 

On the other hand, global symmetries are intrinsic to the quantum system. They are realized by conserved operators or equivalently by topological defects, and act by mapping one operator to another within the same description. Duality can sometimes lead to a global symmetry. 
For instance, at the self-dual coupling where $e^2=\tilde e^2=2\pi$, electromagnetic duality maps the Lagrangian back to itself, and the duality transformation becomes a $\mathbb{Z}_4$ global symmetry.\footnote{Note that the electromagnetic transformation \eqref{intromonodromy} is of order 4. It squares to charge conjugation $C$.}

Recently, it has been shown \cite{Choi:2021kmx,Choi:2022zal,Niro:2022ctq,Cordova:2023ent} that electromagnetic duality gives rise to new global symmetries not only at the self-dual point, but more generally whenever the fine-structure constant takes a rational value,
\begin{equation}
{e^2\over 2\pi } \in \mathbb{Q}^+ \,.
\end{equation}
Although electromagnetic duality does not leave the Maxwell Lagrangian invariant at these values of $e$, the dual Lagrangian can be brought back to the original coupling by gauging an appropriate 1-form global symmetry. 
Because this procedure involves gauging, the resulting symmetry is not an ordinary global symmetry but a non-invertible one. 
We refer to this new symmetry in Maxwell theory as the \textit{non-invertible duality symmetry}.

Concretely, the duality symmetry acts invertibly on the field strength, implementing the familiar transformation $\vec{E} \to \vec{B}$ and $\vec{B} \to -\vec{E}$. 
However, its action on extended operators is more subtle. 
A genuine Wilson line operator is mapped not to another genuine line operator, but rather to a non-genuine loop attached to a surface carrying a fractional charge \cite{Choi:2021kmx}. 
In the special case where $2\pi /e^2=N$ with some positive integer $N$, 
the associated conserved operator $\mathcal{D}_N$ obeys \cite{Choi:2021kmx,Kaidi:2021xfk}
\begin{align}\label{introfusion}
    \mathcal{D}_N\times \mathcal{D}_N^\dagger=\mathcal{D}_N^\dagger\times \mathcal{D}_N
    ={1\over N}\sum_{\mathcal{M}_2}\exp{(i\oint_{\mathcal{M}_2} \star F)}~,
\end{align}
where the sum is over all nontrivial $\mathbb{Z}_N$-valued 2-cycles ${\cal M}_2$ in 3d space and $F$ is the field strength two-form. 
If the space has nontrivial 2-cycles (e.g., $S^2\times S^1$), then from \eqref{introfusion} it follows that $\mathcal{D}_N$ annihilates states whose electric flux is nonzero modulo $2\pi N$. 
Therefore, $\mathcal{D}_N$ is a non-invertible conserved operator.
For $N=2$, this topological defect/operator has been realized in Euclidean lattice models in \cite{Koide:2021zxj} and in  Hamiltonian lattice models in \cite{Gorantla:2024ocs}.

\begin{figure}[thb]
\centering
\includegraphics[width=.95\textwidth]{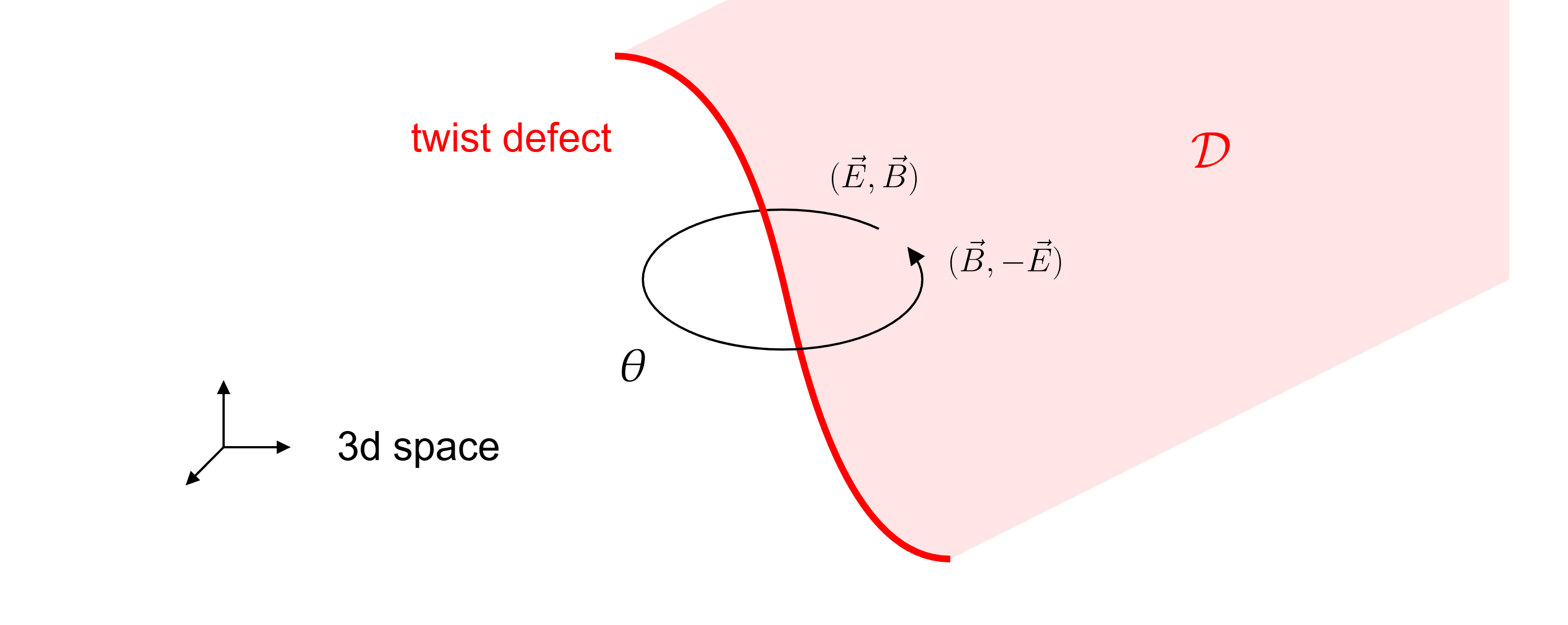}
  \caption{Twist defect for electromagnetic duality. In 3d space, the twist defect (in red) is a string attached to a topological membrane $\cal D$ (in pink). } \label{fig:intro}
\end{figure}

\subsection{Twist defects for non-invertible duality symmetries}

In this paper, we initiate the study of twist defects for non-invertible global symmetries. 
In section \ref{sec_4d Maxwell theory}, 
we consider a novel twist defect associated with the non-invertible duality symmetry in 4d Maxwell theory. 
At a fixed time, this twist defect is a string, attached to a membrane $\mathcal{D}$ in 3d space. 
See figure \ref{fig:intro}. 
As we go around the string and cross the membrane, the electromagnetic fields experience a duality transformation:
\begin{equation}\label{intromonodromy}
\vec E(\theta+2\pi)= \vec B(\theta) \,,\quad \vec B (\theta+2\pi)= -\vec E(\theta)\,,
\end{equation}
where $\theta$ denotes the angular coordinate around the string, and we have suppressed the remaining three spacetime coordinates. 
In spacetime, $\mathcal{D}$ is a 3d topological defect associated with the non-invertible duality symmetry, ending along the 2d twist defect, parametrized by the complex coordinates $(z,\bar z)$.\footnote{Our setting in figure \ref{fig:intro} is different from those considered in \cite{Ganor:2008hd,Ganor:2010md,Ganor:2012mu,Ganor:2014pha,Hsieh:2019iba,Kaidi:2022uux}, where an electromagnetic duality transformation is performed when going around a nontrivial path in spacetime. In those cases, the non-invertible duality defect $\mathcal{D}$ wraps a closed 3-manifold with no boundary. For instance, $\mathcal{D}$ may wrap a $T^3$ with coordinates $(t,y,z)$ in the $T^4$ spacetime, and $\vec E(t,x+2\pi L,y,z ) =\vec B(t,x,y,z), \,\vec B(t,x+2\pi L,y,z ) =-\vec E(t,x,y,z)$.}

While Maxwell theory is conformally invariant, the twist defect dynamics can depend on the intrinsic scale associated with the degrees of freedom on the defect. 
In general, the twist defect theory can admit different effective descriptions at different wavelengths, whose parameters are subject to the defect renormalization group (RG) flow. 
We will be interested in the universal defect dynamics in the long-wavelength limit, which is typically described by a defect CFT (DCFT) under mild assumptions. 
More specifically, we are interested in the spectrum of defect conformal primary operators ${\cal O}(z,\bar z)$, which are point operators confined on the 2d twist defect. 

We solve a universal sector of this DCFT using two complementary approaches:
\begin{enumerate}
\item In section \ref{sec_defect Hilbert space}, we solve the classical electromagnetic wave equations on $S^3\times \mathbb{R}$ in the presence of a twist defect. The frequency spectrum of these electromagnetic waves is presented in \eqref{eq_eigenfrequencies}. 

\item In section \ref{sec_Generalized free field sector}, we analyze the two-point function $\langle F\mathcal{O}\rangle$ between the bulk field strength $F$ and a defect primary operator $\cal O$. Using conformal symmetry, unitarity, and the monodromy condition \eqref{intromonodromy}, we determine the scaling dimensions and other quantum numbers of those defect primaries with nonzero $\langle F\mathcal{O}\rangle$, generalizing the argument in  \cite{Herzog:2022jqv}. The results are presented in \eqref{eq_defect vector primary}.  
\end{enumerate}
By the state/operator correspondence, the energies of the states on $S^3$ in the presence of a twist defect (computed in approach 1) are identified as the scaling dimensions of the defect primaries (computed in approach 2). 
Indeed, we find perfect agreement between the two calculations. 
These defect primaries have the following conformal weights:
\begin{equation}
    \begin{aligned}
\mathcal{V}_{z}^{s}~:&~~~(h,\bar{h})=\left(1+\frac{|s|}{2},\frac{|s|}{2}\right)~,~~~\text{where}~|s|\in \mathbb{N}+\frac{1}{4}~,\\
\mathcal{V}_{\bar{z}}^{s}~:&~~~(h,\bar{h})=\left(\frac{|s|}{2},1+\frac{|s|}{2}\right)~,~~~\text{where}~|s|\in \mathbb{N}+\frac{3}{4}~,
    \end{aligned}
\end{equation}
and $s\in \mathbb{Z}\pm\frac14$ is the spin associated with the spatial rotation transverse to the twist defect. 
Note that the operator spectrum is chiral in the sense that it is not invariant under exchanging $h$ and $\bar{h}$. 
Furthermore, by considering the three-point function $\langle F{\cal OO}\rangle$ of a bulk field strength and two defect primaries, we show that this universal sector forms a generalized free field theory of $\mathcal{V}_{z}^{s}, \mathcal{V}_{\bar z}^{s}$, i.e., their correlation functions can be computed using Wick's theorem.

While the generalized free field sector is independent of the rational value of $e^2/2\pi$, the DCFT contains an additional sector that does depend on the fine-structure constant. We refer to this as the chiral current sector, described by a chiral compact boson whose radius is fixed by the value of the fine-structure constant (see section \ref{sec_Chiral current sector}). 
This sector arises in a similar way to the standard chiral edge modes for the 3d Chern-Simons theory.

\subsection{Twist fields in two spacetime dimensions}

In two spacetime dimensions, the simplest example of non-invertible symmetries is the Kramers-Wannier duality symmetry of the Ising CFT \cite{Grimm:1992ni,Oshikawa:1996dj,Petkova:2000ip,Frohlich:2004ef,Aasen:2016dop,Chang:2018iay}. 
The associated twist-field spectrum consists of four Virasoro primaries, of conformal weights $(1/16,0)$, $(1/16,1/2)$, $ (0,1/16)$, $(1/2,1/16)$ \cite{Petkova:2000ip,Ho:2014vla,Hauru:2015abi,Aasen:2016dop,Chang:2018iay,Lin:2023uvm,Seiberg:2023cdc,Seiberg:2024gek}. 

Section \ref{sec:2d} discusses an analogous non-invertible global symmetry associated with T-duality in the 2d compact boson CFT at rational radius square $R^2\in \mathbb{Q}^+$ \cite{Fuchs:2007tx,Kapustin:2009av,Thorngren:2021yso,Choi:2021kmx,Niro:2022ctq,Pace:2024oys}. 
We point out that the defect action commonly used in the literature to describe this symmetry is imprecise and leads to an incorrect fusion rule. 
To address this, we derive a more precise defect action by adding a boundary correction term (see \eqref{eq_duality defect action 1} and \eqref{eq_duality defect action 2}). 
We then use this new defect action to compute the spectrum of twist fields for this non-invertible symmetry, and find agreement with the literature. We discuss an anomaly involving a chiral $O(2)$ symmetry and the non-invertible symmetry associated with T-duality. The anomaly has nontrivial dynamical consequences even when the non-invertible symmetry is broken.  
We also discuss conformal line defects closely related to this non-invertible symmetry in the compact boson CFT with generic radius $R$.

\subsection{Outline}

The sections and appendices of this paper are organized as follows. In section \ref{sec_4d Maxwell theory}, we investigate the twist defect associated with electromagnetic duality in 4d Maxwell theory. Section \ref{sec:2d} studies the twist fields associated with T-duality in the 2d compact boson CFT. 

In Appendix \ref{sec_DB cohomology}, we review differential cohomology, which plays a central role in formulating the precise actions of duality defects. Appendix \ref{app:CS} focuses on the $(2+1)$d Chern–Simons action, presented in the framework of differential cohomology. In Appendix \ref{app:mode}, we analyze the mode expansion in both two and four dimensions, and carry out a Euclidean path integral to evaluate the partition functions in the presence of a duality defect. Appendix \ref{sec_hypergeometric solutions} provides details on the differential equations and their solutions that arise in investigating the twist defects in 4d Maxwell theory. Finally, Appendix \ref{sec_dcft tensors} examines the tensor structures that appear in the DCFT.

\section{4d Maxwell theory}
\label{sec_4d Maxwell theory}

We now discuss non-invertible global symmetries associated with electromagnetic duality and their twist defects in 4d Maxwell theory. 
The duality defects are supported on a 3d world-volume, ending on a 2d surface where the twist defects are defined.

This section is organized as follows: In section \ref{sec_review of the Maxwell theory}, we review the global symmetries of 4d Maxwell theory. 
We introduce the duality defect action and discuss its associated fusion rules in section \ref{sec_Maxwell defect action and fusion rules}. 
Sections \ref{sec_defect Hilbert space}, \ref{sec_Generalized free field sector}, and \ref{sec_Chiral current sector} study the conformal twist defect of the non-invertible duality symmetry. 
We first solve the spectrum of photon excitations on a 3-sphere in section \ref{sec_defect Hilbert space}, where the spatial slice of the twist defect extends along a circular direction. 
These states are mapped to the DCFT operators via the state/operator correspondence. 
We prove in section \ref{sec_Generalized free field sector} that these DCFT operators form a generalized free field sector. 
Section \ref{sec_Chiral current sector} discusses a chiral current sector of the DCFT. 
Finally, we comment on generalizations to general rational values of the fine-structure constant in section \ref{sec_4d general duality defects}.

\subsection{Review of the Maxwell theory}
\label{sec_review of the Maxwell theory}

\subsubsection{Invertible global symmetries}

We first review 4d Maxwell theory and its invertible global symmetries. 
The Maxwell action for a $U(1)$ abelian gauge field $A$ on a closed 4d Euclidean manifold $\mathcal{M}_4$ is 
\begin{equation}
\label{eq_review maxwell action}
    S[A]=\frac{1}{2e^2}\int_{\mathcal{M}_{4}}F\wedge \star F~,
\end{equation}
where $e\in \mathbb{R}^+$ denotes the gauge coupling, and $F=dA$ is the field strength.\footnote{More generally, one can include a theta-angle term in the Maxwell action \eqref{eq_review maxwell action}. 
Even though this topological term doesn't affect the equations of motion, it changes the global properties of the theory and leads to new non-invertible symmetries \cite{Choi:2022zal,Choi:2022rfe, Niro:2022ctq}. 
However, we will not study this generalization here.  } 
Here $\star$ is the Hodge dual of a differential form. 
We normalize the gauge field so that the magnetic flux through any 2-cycle is quantized, i.e., $\oint_{{\cal M}_2} dA \in 2\pi\mathbb{Z}$.

The action \eqref{eq_review maxwell action} has an ordinary $\mathbb{Z}_2$ symmetry, which acts on the gauge field by charge conjugation $C:~ A\to -A$. Furthermore, it has a continuous 1-form global symmetry $U(1)_\text{e}^{(1)}\times U(1)_\text{m}^{(1)}$ generated by the electric current $J_\text{e}$ and the magnetic current $J_\text{m}$ \cite{Gaiotto:2014kfa}. Explicitly, the 1-form symmetry currents $J_{\text{e}/\text{m}}$ and their defects $\eta_{\text{e}/\text{m}}$ are defined as follows:
\begin{equation}
\begin{aligned}
    \label{eq_maxwell U1xU1}&U(1)_\text{e}^{(1)}~:~~~J_\text{e}=\frac{i}{ e^2}\star F~, &&\eta_\text{e}(\alpha)=\exp{(i\alpha \int_{\mathcal{M}_2}J_\text{e})}~;\\
    &U(1)_\text{m}^{(1)}~:~~~J_\text{m}=\frac{1}{2\pi}F~,&&\eta_\text{m}(\alpha)=\exp{(i\alpha \int_{\mathcal{M}_2}J_\text{m})}~,
\end{aligned}
\end{equation}
where $\mathcal{M}_2 \subset \mathcal{M}_4$ is a closed 2d surface in spacetime\footnote{Throughout this paper, we adopt the convention that the current $J$ associated with a continuous $p$-form global symmetry  in $d$-dimensional spacetime is a differential $(d-p-1)$-form \cite{Gaiotto:2014kfa}. Current conservation in this convention takes the form of $dJ=0$.}. 
They are genuine 2d surface defects in spacetime, and are special examples of the Gukov-Witten defects \cite{Gukov:2006jk,Gukov:2008sn}. 
These defects are topological, which follow from the equation of motion  $dJ_\text{e}=0$ and the Bianchi identity $dJ_\text{m}=0$. 
(They are not to be confused with the twist defects discussed below, which are non-genuine 2d conformal surface defects.)

These two 1-form global symmetries have a mixed 't Hooft anomaly, which we review below. Their currents couple to two 2-form background gauge fields, which we denote as $B_\text{e}$ and $B_\text{m}$, respectively. 
The  Maxwell action coupled  to  $B_\text{e}$ and $B_\text{m}$ reads
\begin{equation}
\label{eq_covariant maxwell action}
    S[A;B_\text{e},B_\text{m}]=\frac{1}{2e^2}\int_{\mathcal{M}_{4}}(F-B_\text{e})\wedge \star (F-B_\text{e})+\frac{i}{2\pi}\int_{\mathcal{M}_{4}}B_\text{m}\wedge (F-B_\text{e})~.
\end{equation}
The background  gauge transformations for $U(1)^{(1)}_\text{e}\times U(1)^{(1)}_\text{m}$ are:
\begin{equation}
\label{eq_1form gauge transformation}
    A\to A+a_\text{e}~,~~~B_\text{e}\to B_\text{e}+da_\text{e}~,~~~B_\text{m}\to B_\text{m}+da_\text{m}~,
\end{equation}
where $a_\text{e}$ and $a_\text{m}$  are two 1-form gauge parameters. 
Under these gauge transformations, the Maxwell action \eqref{eq_covariant maxwell action} is not invariant and transforms as
\begin{equation}
\label{eq_1form thooft anomaly}
    S[A;B_\text{e},B_\text{m}] \to  S[A;B_\text{e},B_\text{m}]-\frac{i}{2\pi}\int_{\mathcal{M}_4} a_\text{m}\wedge dB_\text{e}~.
\end{equation}
The variation term in \eqref{eq_1form thooft anomaly} signals the mixed 't Hooft anomaly between $U(1)_\text{e}^{(1)}$ and $U(1)_\text{m}^{(1)}$, which can be canceled by a 5d anomaly inflow action.

The Maxwell theory can alternatively be formulated in terms of a dual gauge field $\tilde{A}$ with the field strength $\tilde{F}=d\tilde{A}$ and gauge coupling $\tilde{e}=2\pi/e$. The dual action takes the form 
\begin{equation}
\label{eq_review maxwell action dual}
    S[\tilde{A}]=\frac{e^2}{8\pi^2}\int_{\mathcal{M}_{4}}\tilde{F}\wedge \star \tilde{F}~. 
\end{equation}
Electromagnetic duality (also known as S-duality) states that the theory \eqref{eq_review maxwell action} is equivalent to \eqref{eq_review maxwell action dual} with the following identification between 1-form symmetry currents:
\begin{equation}
\label{eq_S-dual U1 group}
\begin{aligned}
    &\tilde{U}(1)_\text{e}^{(1)}~:~~~\tilde{J}_\text{e}=\frac{ie^2}{ 4\pi^2}\star \tilde{F}=J_\text{m}~,\\
    &\tilde{U}(1)_\text{m}^{(1)}~:~~~\tilde{J}_\text{m}=\frac{1}{2\pi}\tilde{F}=-J_\text{e}~.
\end{aligned}
\end{equation}
The dual action \eqref{eq_review maxwell action dual} can also be covariantly coupled to the background gauge fields $B_\text{e}$ and $B_\text{m}$. The gauge transformation \eqref{eq_1form gauge transformation} for the dual field is $\tilde{A}\to \tilde{A}+a_\text{m}$, and it is straightforward to reproduce the mixed 't Hooft anomaly \eqref{eq_1form thooft anomaly}.

The 1-form symmetries $U(1)_\text{e}^{(1)}$ and $U(1)_\text{m}^{(1)}$ act the Wilson loop $W_{\text{e}}$ and the 't Hooft loop $W_{\text{m}}$, which are defined as
\begin{equation}
    W_\text{e}=\exp(i\oint_{\mathcal{M}_1} A) ~,~~~W_\text{m}=\exp(i\oint_{\mathcal{M}_1} \tilde{A}) ~,
\end{equation}
where $\mathcal{M}_1 \subset \mathcal{M}_4$ is a closed loop in spacetime. 
Physically, $W_\text{e}$ can be interpreted as the worldline of an infinitely heavy, electrically charged particle, whereas $W_\text{m}$ represents that of an infinitely heavy magnetic monopole. 
The insertion of the Wilson loop $W_{\text{e}}$ modifies the equation of motion in \eqref{eq_review maxwell action}, and its linking with $\eta_\text{e}(\alpha)$ leads to a $U(1)$ phase $e^{i\alpha}$. 
Thus, the Wilson loop $W_{\text{e}}$ carries $+1$ charge under $U(1)_\text{e}^{(1)}$ while it is neutral under $U(1)_\text{m}^{(1)}$.  
Dually, the 't Hooft loop $W_{\text{m}}$ carries $+1$ charge under $U(1)_\text{m}^{(1)}$ and is neutral under $U(1)_\text{e}^{(1)}$. 
S-duality \eqref{eq_S-dual U1 group} identifies the Wilson loop $\tilde{W}_{\text{e}}$ and the 't Hooft loop $\tilde{W}_{\text{m}}$ in the dual description with the loops in the original description as follows: 
\begin{equation}
\label{eq_S-daul loop operators}
(\tilde{W}_\text{e},\tilde{W}_\text{m})=(W_\text{m},W_\text{e}^\dagger)~.
\end{equation}

When $e^2=\tilde{e}^2=2\pi$, S-duality becomes an automorphism of the Maxwell theory, and it is implemented by an invertible, 0-form, $\mathbb{Z}_4$  symmetry element. More generally, the Maxwell theory \eqref{eq_review maxwell action} with gauge coupling $e^2\in 2\pi\mathbb{Q}$ has a non-invertible 0-form symmetry associated with  S-duality, which we review next.

\subsubsection{Non-invertible duality defect}
\label{sec_Maxwell defect action and fusion rules}

To define the non-invertible duality defect, we assume the Euclidean spacetime to be $T^3\times [0, 2\pi]$, with coordinates $x_i\sim x_i+2\pi$ for $i=1,2,3$, and $0\leq x_4\leq 2\pi$. As in \eqref{eq_review maxwell action}, the Maxwell action reads
\begin{equation}
\label{eq_maxwell bulk action}
    S_\text{bulk}[A]=\frac{1}{2e^2}\int_{T^3\times [0, 2\pi]}F\wedge \star F~.
\end{equation}
We impose the Neumann boundary condition $F_{14}=F_{24}=F_{34}=0$ on the two boundaries at $x_4=0$ and $x_4=2\pi$. 
We denote the 1-form gauge fields on the boundaries as:
\begin{equation}
\label{eq_boundary gauge fields}
    A^+(x_i)\equiv A(x_i,x_4=0)~, ~\text{and}~A^-(x_i)\equiv A(x_i,x_4=2\pi)~.
\end{equation}
The duality defect, denoted by $\mathcal{D}_N$, is defined by a gluing action that couples the gauge fields on the two boundaries \cite{Gaiotto:2008ak,Kapustin:2009av,Choi:2021kmx, Niro:2022ctq, Cordova:2023ent,Kim:2025zdy}:
\begin{equation}
\label{eq_maxwell duality defect action}
\mathcal{D}_N~:~~~S_\text{defect}[A^-,A^+]=\frac{iN}{2\pi}\int_{T^3} A^-\wedge dA^+~~~~~(\text{Euclidean}).
\end{equation}
The duality defect wraps around the 3-manifold $T^3$, which has no boundary. (In section \ref{sec_defect Hilbert space}, we will generalize the discussion to the case where the 3-manifold has a boundary where the twist defect lives.)
Similar to the standard argument in Chern-Simons gauge theories, gauge invariance requires $N\in \mathbb{Z}$. For simplicity, we focus on the $N>0$ case for now, and comment on the $N<0$ case later. 

It is well-known that Chern-Simons-type actions, such as \eqref{eq_maxwell duality defect action}, are not precise globally. The naive action exhibits an unphysical dependence on the choice of the trivialization for the gauge fields. 
This issue has a standard resolution using differential cohomology, which we review in appendix \ref{sec_DB cohomology}. 
The precise defect action is presented in appendix \ref{sec_4d torus partition functions}. 
For the purpose of this work, however, the distinction between the naive defect action \eqref{eq_maxwell duality defect action} and its refined version \eqref{eq_maxwell precise defect} does not play a direct role in the analysis. We will therefore use the simplified form \eqref{eq_maxwell duality defect action} for convenience.

It follows from the equation of motion that the stress-energy tensor is conserved across the duality defect \eqref{eq_maxwell duality defect action} if 
\begin{equation}
\label{eq_defect topo cond}
    e^2=\frac{2\pi }{N}~.
\end{equation}
From now on until section \ref{sec_4d general duality defects}, we will always assume this condition so that $\mathcal{D}_N$ is a topological defect. 
(The more general $e^2/2\pi \in \mathbb{Q}^+$ case will be discussed in section \ref{sec_4d general duality defects}.)

The variation of the bulk action \eqref{eq_maxwell bulk action} is:
\begin{align}\label{dSbulk}
\delta S_\text{bulk} =  {N\over 2\pi} \int_{T^3\times [0,2\pi]} \delta A\wedge d\star F
-{N\over 2\pi} \int_{T^3} \delta A^+ \wedge \star F^+
+{N\over 2\pi} \int_{T^3} \delta A^{-} \wedge \star F^-\,.
\end{align}
The first term vanishes when the equations of motion are imposed, while the second and the third terms have to be canceled against the variation of the defect action \eqref{eq_maxwell bulk action}:
\begin{align}
\delta S_\text{defect}= {iN\over 2\pi} \int_{T^3} (\delta A^{-} \wedge dA^+ +A^- \wedge d\delta A^+)
\end{align}
We therefore find,
\begin{equation}
\label{eq_4d modified Neumann}
    F^+=i\star F^-~.
\end{equation}
This gluing condition across $\mathcal{D}_N$ implements the electromagnetic duality (or S-duality). This is the reason why $\mathcal{D}_N$ is referred to as the duality defect.

Equation \eqref{eq_4d modified Neumann} implies the following fusion rule:
\begin{equation}
\label{eq_maxwell fusion rule 1}
\begin{aligned}
    \eta_\text{e}(\alpha)\times \mathcal{D}_N=&\mathcal{D}_N\times \eta_\text{m}(N\alpha)~,\\
     \mathcal{D}_N\times \eta_\text{e}(\alpha)=&\eta_\text{m}(-N\alpha)\times\mathcal{D}_N ~.
\end{aligned}
\end{equation}
It follows that the duality defect can absorb the electric 1-form symmetry defects associated with the subgroup $\mathbb{Z}^{(1)}_N\subset U(1)^{(1)}_\text{e}$:
\begin{equation}
\label{eq_maxwell fusion corollary}
    \eta_\text{e}(\alpha)\times \mathcal{D}_N=\mathcal{D}_N\times \eta_\text{e}(\alpha)=\mathcal{D}_N~,~~~\text{if } \alpha\in \frac{2\pi \mathbb{Z}}{N}~.
\end{equation}
This implies that $\mathcal{D}_N$ is non-invertible when $N>1$.\footnote{To see this, suppose $\mathcal{D}_N$ were invertible, then we multiply $\mathcal{D}_N^{-1}$ on both sides and find $\eta_\text{e}(2\pi/N)=1$, which is a contradiction if $N>1$.} 
At $N=1$, \eqref{eq_defect topo cond} reduces to the self-dual radius, and the duality defect $\mathcal{D}_{N=1}$ is invertible. 

Next, we discuss the dual of the topological defect, denoted as $\bar{\cal D}_N$. 
For this purpose, it will be convenient to do a Wick rotation and go to Lorentzian signature, where the defect action becomes
\begin{equation}
\label{eq_LorentzianD}
\mathcal{D}_N~:~~~S_\text{defect}[A^-,A^+]=-\frac{N}{2\pi}\int_{x=0} A^-\wedge dA^+~~~~~(\text{Lorentzian}).
\end{equation}
Here we have chosen the defect to be localized at $x=0$, and extends in the $t,y,z$ directions. 
In any relativistic quantum field theory in Lorentzian signature, there is a canonical antiunitary symmetry $\Theta$ that reflects one spatial coordinate (say, $x$) and reverses the time coordinate. 
This operator is commonly referred to as the CRT operator, which becomes a $\pi$ rotation in spacetime in Euclidean signature. 
(See \cite{Witten:2015aba,Hason:2020yqf,Harlow:2023hjb,Goodhew:2024eup,Witten:2025ayw,Seiberg:2025zqx}, for examples, for some recent discussions of the CRT symmetry and its relation to CPT.) 
In the Maxwell theory, $\Theta$ acts on the gauge fields as 
\begin{align}
&\Theta A_t (t,x,y,z) \Theta^{-1} = -A_t(-t,-x,y,z)\,,\quad 
&&\Theta A_x (t,x,y,z) \Theta^{-1} = -A_x(-t,-x,y,z)\,,\quad \notag\\
&\Theta A_y (t,x,y,z) \Theta^{-1} = A_y(-t,-x,y,z)\,,\quad
&&\Theta A_z (t,x,y,z) \Theta^{-1} = A_z(-t,-x,y,z)\,.
\end{align}
The dual of a defect is defined by its image under $\Theta$ in Lorentzian signature, with the reflection direction $x$ chosen to align with the normal direction of the codimension-1 defect. 
The dual of an invertible defect is its own inverse. For example, $\overline{\eta_{\text{m}}(\alpha)}=\eta_{\text{m}}(-\alpha)$. 
By applying the $\Theta$ transformation on \eqref{eq_LorentzianD}, we find\footnote{Note that since $\Theta$ reflects in the $x$ direction, it exchanges $A^+$ with $A^-$.}
\begin{equation}
\label{eq_4d orientation reverse}
    \bar{\mathcal{D}}_N=\mathcal{D}_{-N}.
\end{equation}
Note that $\mathcal{D}_{-N}=C\times \mathcal{D}_{N}=\mathcal{D}_{N}\times C$, where $C$ is the charge conjugation defect. 

If we instead choose $\mathcal{D}_N$ to extend in all three spatial coordinates and localized in time, it is an operator acting on the Hilbert space. 
Then $\bar {\mathcal{D}}_N$ becomes the adjoint of the operator $\mathcal{D}_N$, i.e., $\bar {\mathcal{D}}_N=\mathcal{D}_N^\dagger$ as operators.

For duality defect $\mathcal{D}_N$ placed on a general 3-manifold $M_3$, it obeys the following fusion rule in Lorentzian signature \cite{Choi:2021kmx,Kaidi:2021xfk}:
\begin{align}
\label{eq_maxwell fusion rule 2}
    \mathcal{D}_N\times \bar{\mathcal{D}}_N=\bar{\mathcal{D}}_N\times \mathcal{D}_N
    &={1\over N }\sum_{\mathcal{M}_2\in H_2(\mathcal{M}_3,\mathbb{Z}_N)}\eta_\text{e} ({2\pi\over N})\notag\\
    &={1\over N}\sum_{\mathcal{M}_2\in H_2(\mathcal{M}_3,\mathbb{Z}_N)}\exp{(i\oint_{\mathcal{M}_2} \star F)}~,
\end{align}
where the sum is over all 2-cycles on $\mathcal{M}_3$. The RHS of \eqref{eq_maxwell fusion rule 2} is known as the condensation defect \cite{Gaiotto:2019xmp, Roumpedakis:2022aik,Choi:2022zal}.

\subsection{Twist defects for non-invertible duality symmetries}
\label{sec_defect Hilbert space}

Previously, the duality defect $\mathcal{D}_N$ was assumed to be supported on a closed 3-manifold without boundaries in spacetime. 
We now assume $\mathcal{D}_N$ is supported on a 3-manifold with a boundary, which defines a 2d twist defect. 
The latter is not a genuine 2d surface defect; rather, it is attached to the 3d topological defect $\mathcal{D}_N$. 
See figure \ref{fig:intro} for this configuration in space at a fixed time. 
As before, we impose the condition $e^2 = \tfrac{2\pi}{N}$.

\begin{figure}[thb]
\centering
\includegraphics[width=.95\textwidth]{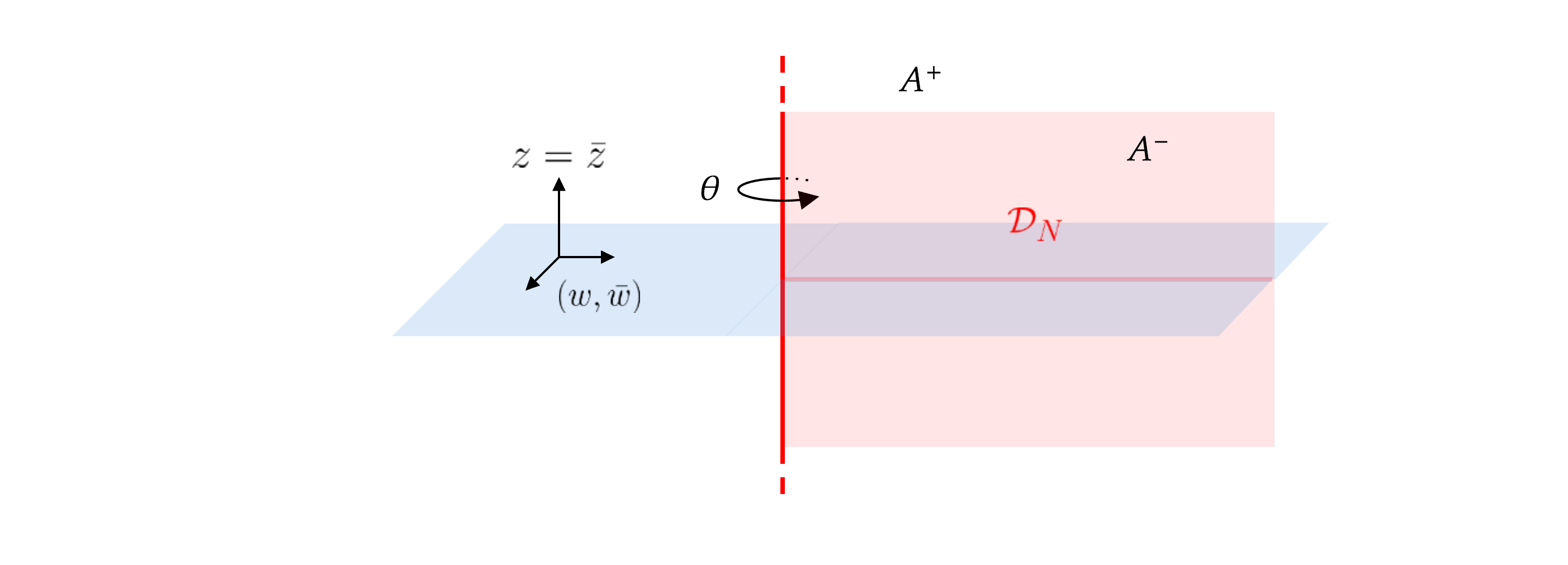}
  \caption{\label{pic_complex coordinate} The configuration for the conformal twist defect and the non-invertible duality defect in $\mathbb{R}^4$.  
  The vertical red line is the 2d twist defect, which is localized at $w=\bar{w}=0$ and extends along the $(z,\bar{z})$ plane. The twist defect is attached to the 3d non-invertible topological defect $\mathcal{D}_N$ shown as the half-infinite red plane. This figure only displays the $\mathbb{R}^3$ slice defined by $z=\bar{z}$.} 
\end{figure}

Let the Euclidean spacetime be $\mathbb{R}^4$ with complex coordinates $z,w\in \mathbb{C}$ and metric
\begin{equation}
\label{eq_Euclidean metric}
    ds^2=2dzd\bar{z}+2dwd\bar{w}~.
\end{equation}
We place the topological duality defect $\mathcal{D}_N$, defined in \eqref{eq_maxwell duality defect action}, so that it covers the complex $z$-plane and extends along the positive real axis $w>0$ of the complex $w$-plane, terminating at the origin $w=\bar{w}=0$.
Thus, the twist defect is localized at $w=\bar{w}=0$ and extends across the entire $z$-plane. 
See figure \ref{pic_complex coordinate}.
More concretely, the action for the duality defect is
\begin{equation}
\label{eq_maxwell twist defect attach}
\mathcal{D}_N~:~~~S_\text{defect}[A^-,A^+]=\frac{iN}{2\pi}\int_{w\in \mathbb{R}^+} A^-\wedge dA^+~.
\end{equation}
where $A_\pm$ are the gauge fields on the two sides of $\mathcal{D}_N$:
\begin{equation}
\label{eq_gauge fields across defect}
A^\pm(z,\bar{z},|w|)=\lim_{\epsilon\to0^{\pm}}A(z,\bar{z},w=|w|+ i\epsilon,\bar{w}=|w|- i\epsilon)~.
\end{equation}

\begin{figure}[thb]
\centering
\includegraphics[width=.95\textwidth]{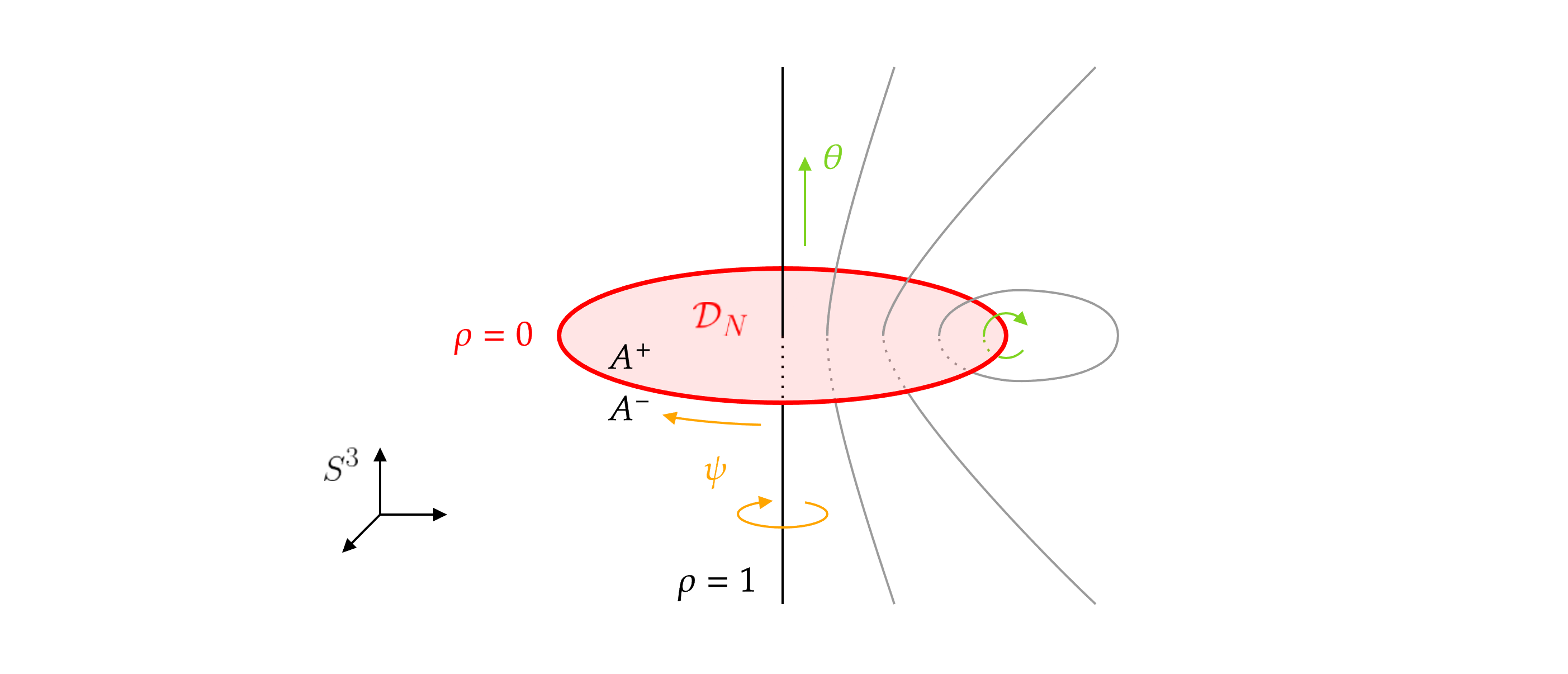}
  \caption{\label{pic_S3 coordinate} 
  An $S^3$ slice of the defect configuration in figure \ref{pic_complex coordinate}. The topological defect $\mathcal{D}_N$ is supported on the red disk, whose boundary $\rho=0$ (the red circle) is the twist defect.} 
\end{figure}

We will assume that this 2d twist defect theory is scale invariant and use the state/operator correspondence to study its operator spectrum. 
Define the coordinates $(\tau,\rho,\psi,\theta)$ as
\begin{equation}
    z=\sqrt{1-\rho} \, e^{\tau+i\psi}~,~~~
    w=\sqrt{\rho} \, e^{\tau+i\theta}~, 
\end{equation}
where $\tau \in \mathbb{R}$, $0\leq \rho\leq 1$, $\psi\sim \psi+2\pi$, and $\theta\sim \theta+2\pi$. 
The flat space metric \eqref{eq_Euclidean metric} written in terms of these new coordinates takes the form:
\begin{equation}
    ds^2=2e^{2\tau}\left[d\tau^2+(1-\rho)d\psi^2+\frac{d\rho^2}{4\rho (1-\rho)}+\rho d\theta^2\right]~.
\end{equation}
The hypersurface $\tau=$ const defines an $S^3$ slice parametrized by the Hopf coordinates $(\rho,\psi,\theta)$. 
The twist defect intersects with these slices at $\rho=0$ and extends along the $\psi$-direction, corresponding to a great circle of $S^3$.\footnote{A great $k$-sphere of $S^n$ is defined by the intersection of $S^n$ with a $\mathbb{R}^{k+1}$ hyperplane passing through its origin in $\mathbb{R}^{n+1}$.} See figure~\ref{pic_S3 coordinate}.

Next, we perform a Weyl transformation to map $\mathbb{R}^4$ to $\mathbb{R} \times S^3$.
We further perform a Wick rotation $\tau=it$ to Lorentzian signature. 
The final metric is
\begin{equation}
\label{eq_Lorenztian metric}
    ds^2=dt^2-\left[(1-\rho)d\psi^2+\frac{d\rho^2}{4\rho (1-\rho)}+\rho d\theta^2
    \right]~,
\end{equation}

 State/operator correspondence maps the dilation $\tau\to\tau+$ const of $S^3$ to the time evolution of the states in the Hilbert space with the twist defect. 
We will refer to this Hilbert space on $S^3$ as the twist defect Hilbert space. 
In the following,  we solve for the frequencies $\omega$ for the electromagnetic waves on $S^3$ with a twist defect, which are identified as the scaling dimensions of the twist defect operators $\Delta$ by the state/operator correspondence. These are point operators localized on the twist defect, and their spectrum is insensitive to the spacetime topology. This is analogous to the fact that the spectrum of local operators in an ordinary CFT does not depend on the topology of the underlying manifold.

In Lorentzian signature, the monodromy condition \eqref{eq_4d modified Neumann}  becomes 
\begin{equation}\label{eq_selfdual}
F_{\mu\nu}(t,\psi,\rho,\theta+2\pi)= \frac12 \epsilon_{\mu\nu\rho\sigma} F^{\rho\sigma}(t,\psi,\rho,\theta)\,.
\end{equation}
Note that $\theta\to \theta+2\pi$ is an order 4 action:
\begin{equation}
\label{eq_Lorentzian twist condition}
F \longrightarrow \star F\longrightarrow -F\longrightarrow -\star F \longrightarrow F~.
\end{equation}
This instanton-like condition imposes strong constraints on the electromagnetic wave profiles. 
In particular, it implies that a closed form $dF=0$ automatically satisfies the equation of motion $d\star F=0$. We perform a Fourier mode expansion of the field strength as follows:
\begin{equation}
    F_{\mu \nu }(t,\psi,\rho ,\theta)=\sum_{s\in \mathbb{Z}\pm \frac{1}{4}}\sum_{n\in \mathbb{Z}}\int \frac{d\omega}{(2\pi)^3}e^{i (-\omega t+n\psi+s \theta)}f_{\mu\nu}(\rho;\omega,n,s)~,
\end{equation}
where the electric components $(f_{t\psi},f_{t\rho},f_{t\theta})$ and the magnetic comonents $(f_{\rho\theta},f_{\theta \psi},f_{\psi \rho})$ are functions of $\rho$ as well as the  wavenumbers $(\omega,n,s)$. 
Importantly,  \eqref{eq_Lorentzian twist condition} imposes an interesting  spin selection rule 
\begin{align}\label{spinselection}
s\in \mathbb{Z}\pm \frac{1}{4}\,.
\end{align}

From \eqref{eq_selfdual}, the magnetic components 
$(f_{\rho\theta}, f_{\theta\psi}, f_{\psi\rho})$ 
are solved in terms of the electric components
$(f_{t\psi}, f_{t\rho}, f_{t\theta})$:
\begin{equation}
    \left. (f_{\rho\theta}, f_{\theta\psi}, f_{\psi\rho})\right|_{s\in \mathbb{Z}\pm \frac{1}{4}}=\mp i\left(\frac{f_{t\psi}}{2(1-\rho)},\rho(1-\rho)f_{t\rho},\frac{f_{t\theta}}{2\rho}\right)~.
\end{equation}
On the other hand, from the Bianchi identity $dF=0$ we can algebraically solve $f_{t\rho}$ in terms of $(f_{t\psi},f_{t\theta})$:
\begin{equation}
    \left. f_{t\rho}\right|_{s\in \mathbb{Z}\pm \frac{1}{4}}=\pm \frac{i(n f_{t\theta}-sf_{t\psi})}{2\omega \rho(1-\rho)}~.
\end{equation}
The Bianchi identity further yields the following ordinary differential equations:
\begin{equation}
\label{eq_maxwell ODE}
    2\omega \rho(1-\rho)\partial_\rho \left.\begin{pmatrix}
 f_{t\psi}\\
 f_{t\theta}
\end{pmatrix}\right|_{s\in \mathbb{Z}\pm \frac{1}{4}}=\pm\begin{pmatrix}
ns & (1-\rho)\omega^2-n^2\\
s^2-\omega^2 \rho & -ns \\
\end{pmatrix}\begin{pmatrix}
 f_{t\psi}\\
 f_{t\theta}
\end{pmatrix}~.
\end{equation}

Equation \eqref{eq_maxwell ODE} admits two linearly independent hypergeometric solutions, as we detailed in Appendix \ref{sec_hypergeometric solutions}. Near the twist defect at $\rho\to 0$, these two solutions exhibit different scaling behaviors $(f_{t\psi},f_{t\theta})\sim \rho^{s/2}$ and $(f_{t\psi},f_{t\theta})\sim \rho^{-s/2}$. We note that it is common for solutions to the wave equations to diverge close to a (non-topological) defect. For example, in free scalar and free fermion theories, such modes are created by the alternative quantization scheme \cite{Bianchi:2021snj, Giombi:2021uae, Barkeshli:2025cjs}. However, the divergent solution of \eqref{eq_maxwell ODE} yields a non-normalizable profile of $f_{t\rho}$ for all values of $s$. As we will discuss in Section \ref{sec_Generalized free field sector}, these solutions correspond to DCFT operators that violate the unitarity bound. In what follows, we will focus on the convergent solution, such that $(f_{t\psi},f_{t\theta}) \sim \rho^{|s|/2}$.

On the other hand, $\rho=1$ defines a great circle of $S^3$ that lies far from the conformal twist defect. We therefore require the wave profile to be finite as $\rho\to1$. 
See \eqref{appendix_eq_maxwell ODE solutions expand 2} for the detailed discussion. 
This requirement enforces the frequency $\omega$ of the wave profile to take discrete values labeled by $m\in \mathbb{N}$, which physically counts the number of wave nodes along the $\rho$-direction. 
The positive allowed values of $\omega$ are\footnote{More precisely, these frequencies are multiplied by the inverse of the $S^3$ radius, which was set to be 1 in \eqref{eq_Lorenztian metric}.} 
\begin{equation}
\label{eq_eigenfrequencies}
    \omega=\begin{dcases}
        2+|s|+|n|+2m~,& \text{if } |s|\in \mathbb{N}+\frac{1}{4} ,~ n\geq 0\text{ or }|s|\in \mathbb{N}+\frac{3}{4} ,~ n\leq 0~,\\
        |s|+|n|+2m~,& \text{if } |s|\in \mathbb{N}+\frac{1}{4} ,~ n< 0\text{ or }|s|\in \mathbb{N}+\frac{3}{4} ,~ n>0~.\\
    \end{dcases}
\end{equation}
Upon quantization, they correspond to the energy spectrum of the single photon states in the presence of the twist defect. 
We assume that these states are determined by the linear equations \eqref{eq_maxwell ODE}, independent of the possible interaction terms on the defect.
We will justify this assumption in section \ref{sec_Generalized free field sector} by showing that these states correspond to a universal generalized free field sector of the DCFT, determined entirely by conformal symmetry and unitarity.

\begin{figure}[thb]
\centering
\subfloat[$s=\frac{1}{4},n=-1,\omega=\frac{5}{4}$]{\includegraphics[width=.3\textwidth]{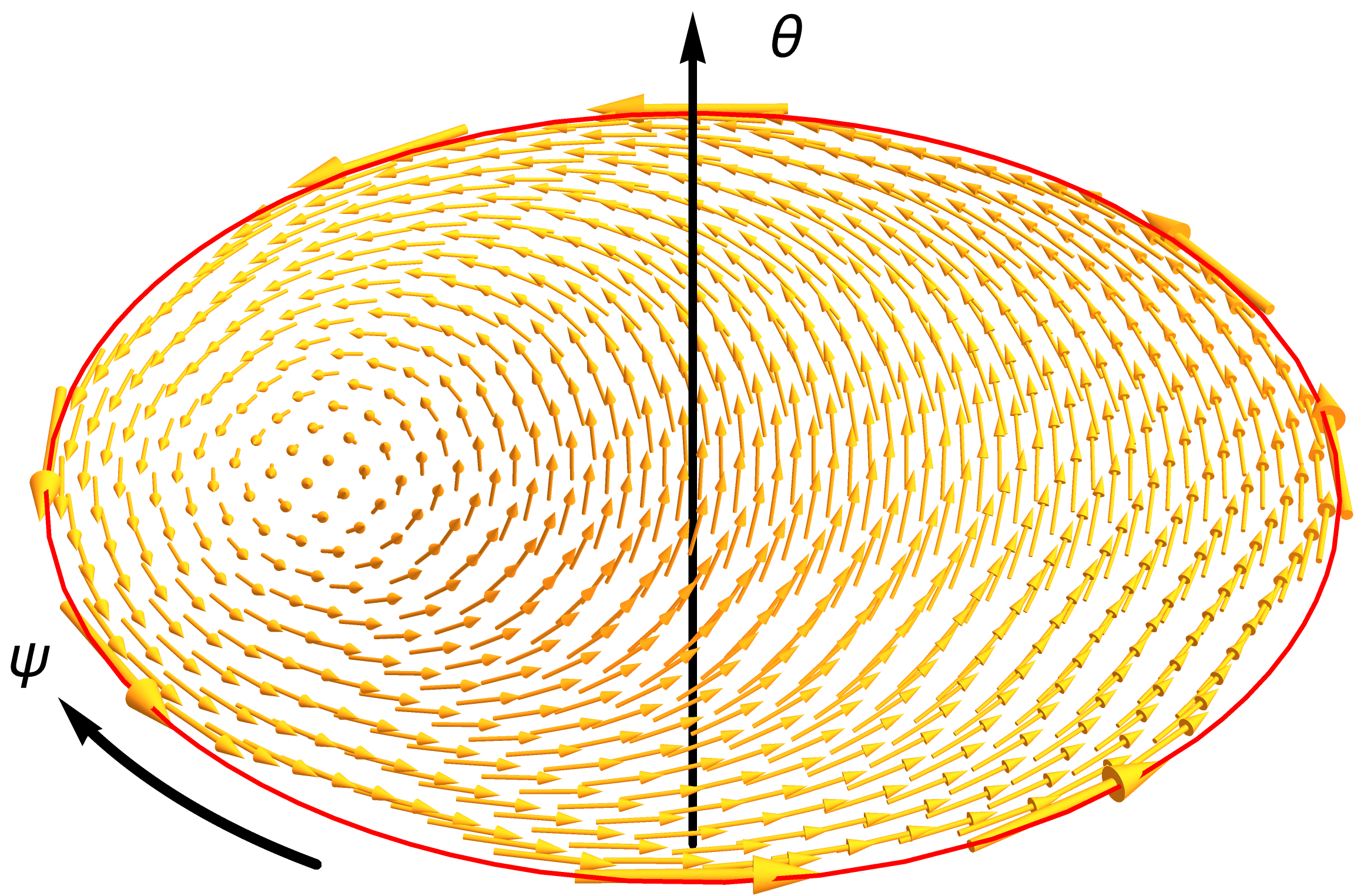}}\hspace{.04 \textwidth }
\subfloat[$s=\frac{1}{4},n=0,\omega=\frac{9}{4}$]{\includegraphics[width=.3\textwidth]{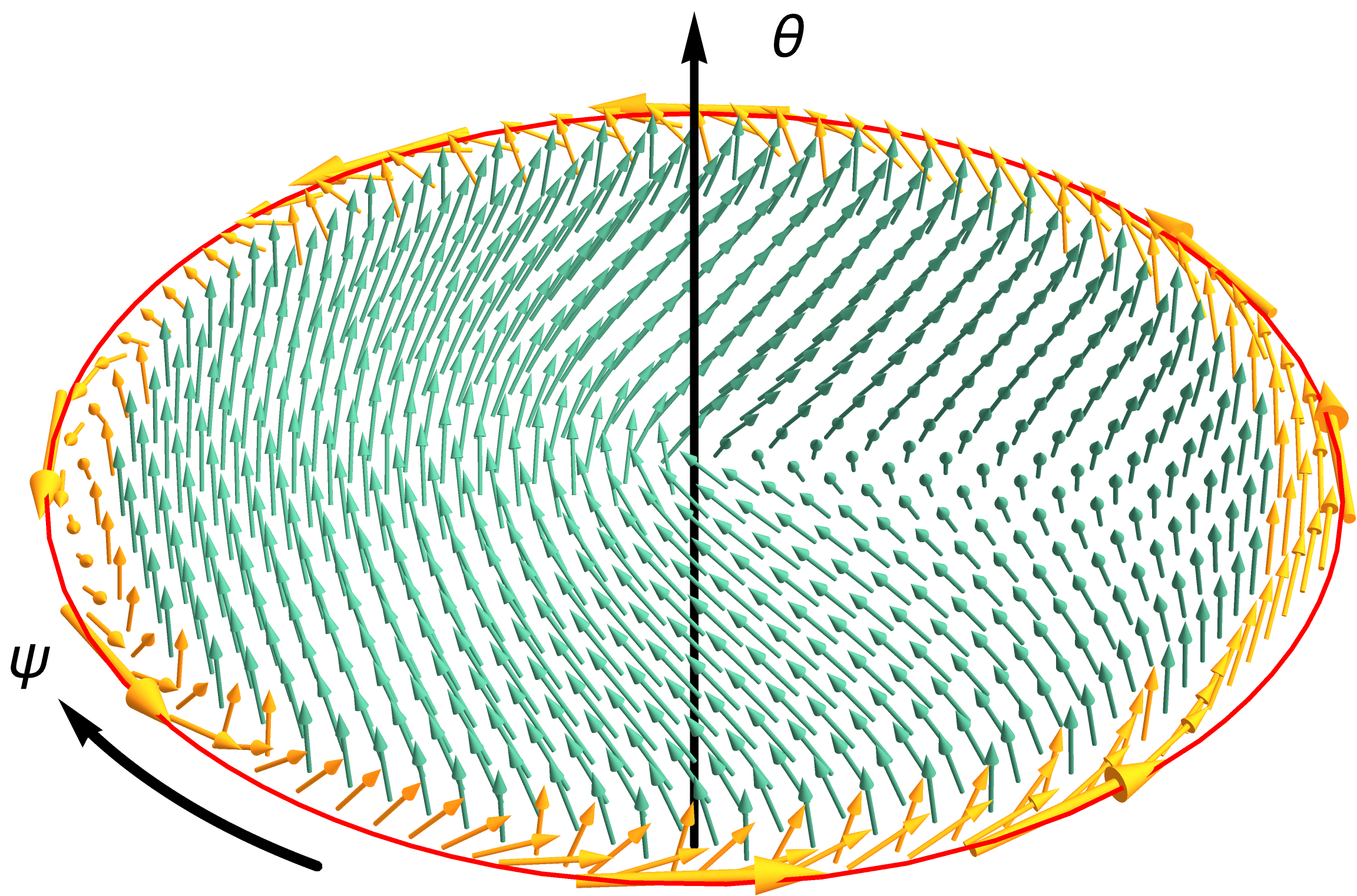}}\hspace{.04 \textwidth }
\subfloat[$s=\frac{1}{4},n=1,\omega=\frac{13}{4}$]{\includegraphics[width=.3\textwidth]{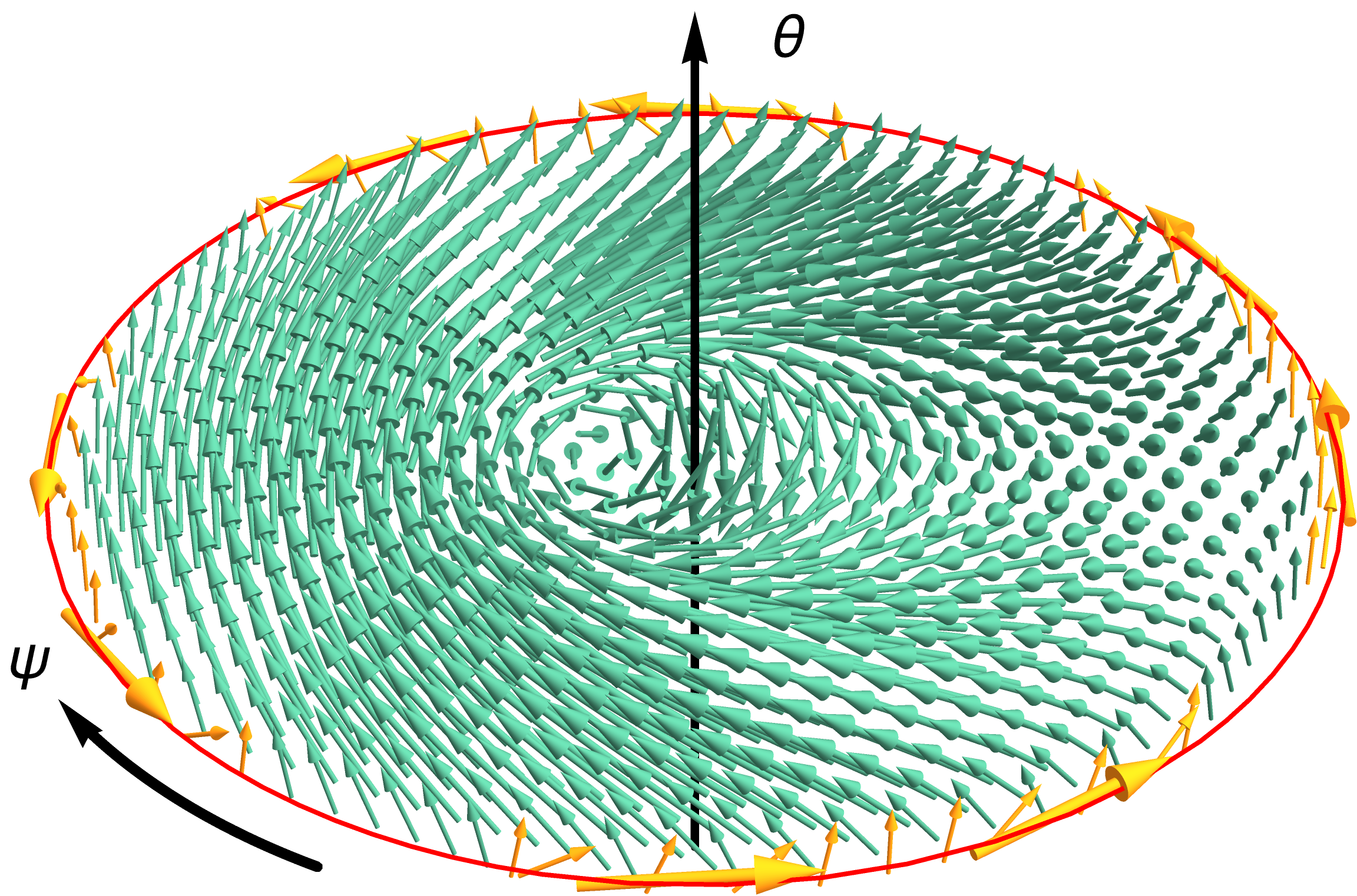}}\\
\subfloat[$s=\frac{3}{4},n=-1,\omega=\frac{15}{4}$]{\includegraphics[width=.3\textwidth]{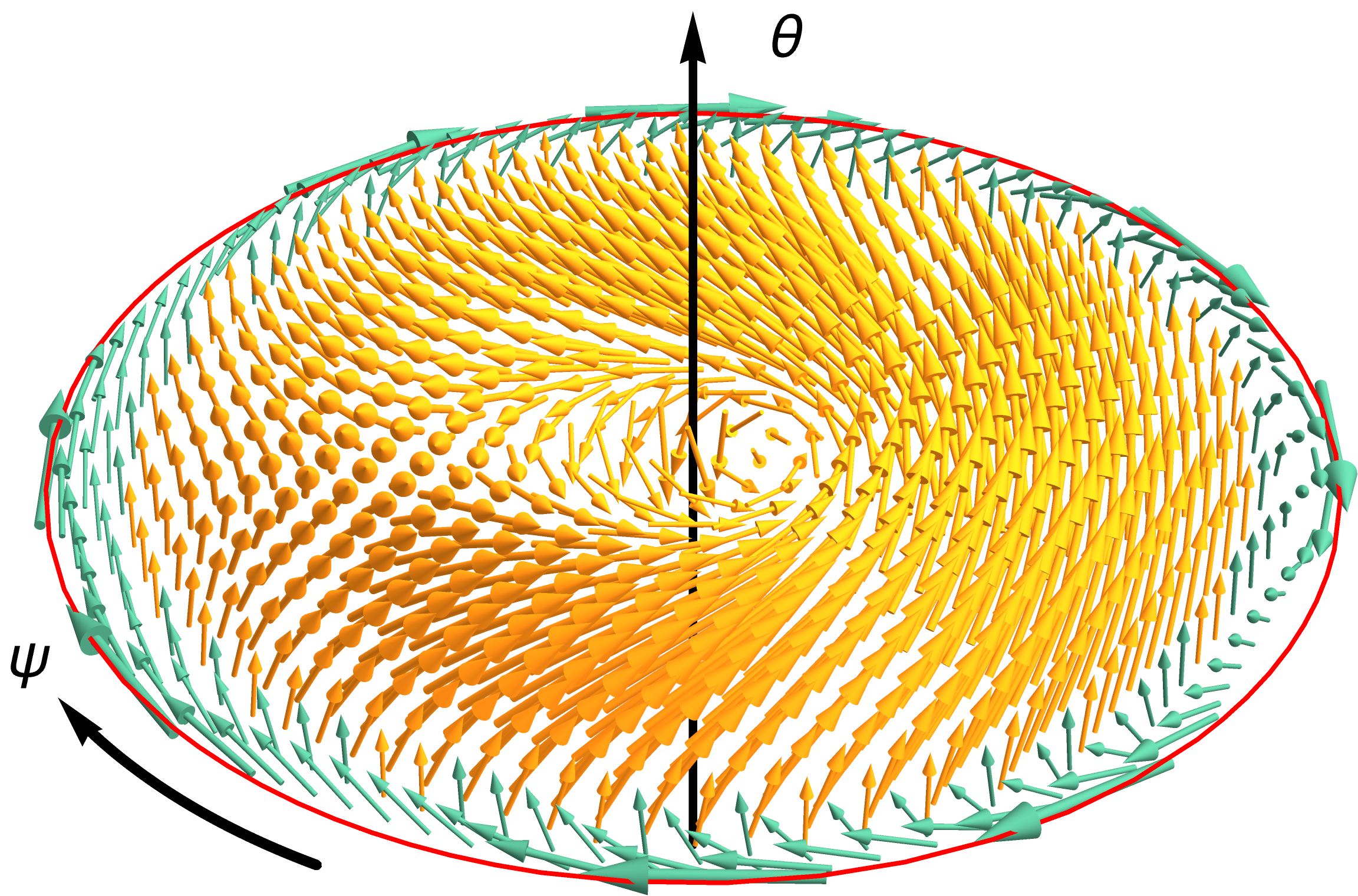}}\hspace{.04 \textwidth }
\subfloat[$s=\frac{3}{4},n=0,\omega=\frac{11}{4}$]{\includegraphics[width=.3\textwidth]{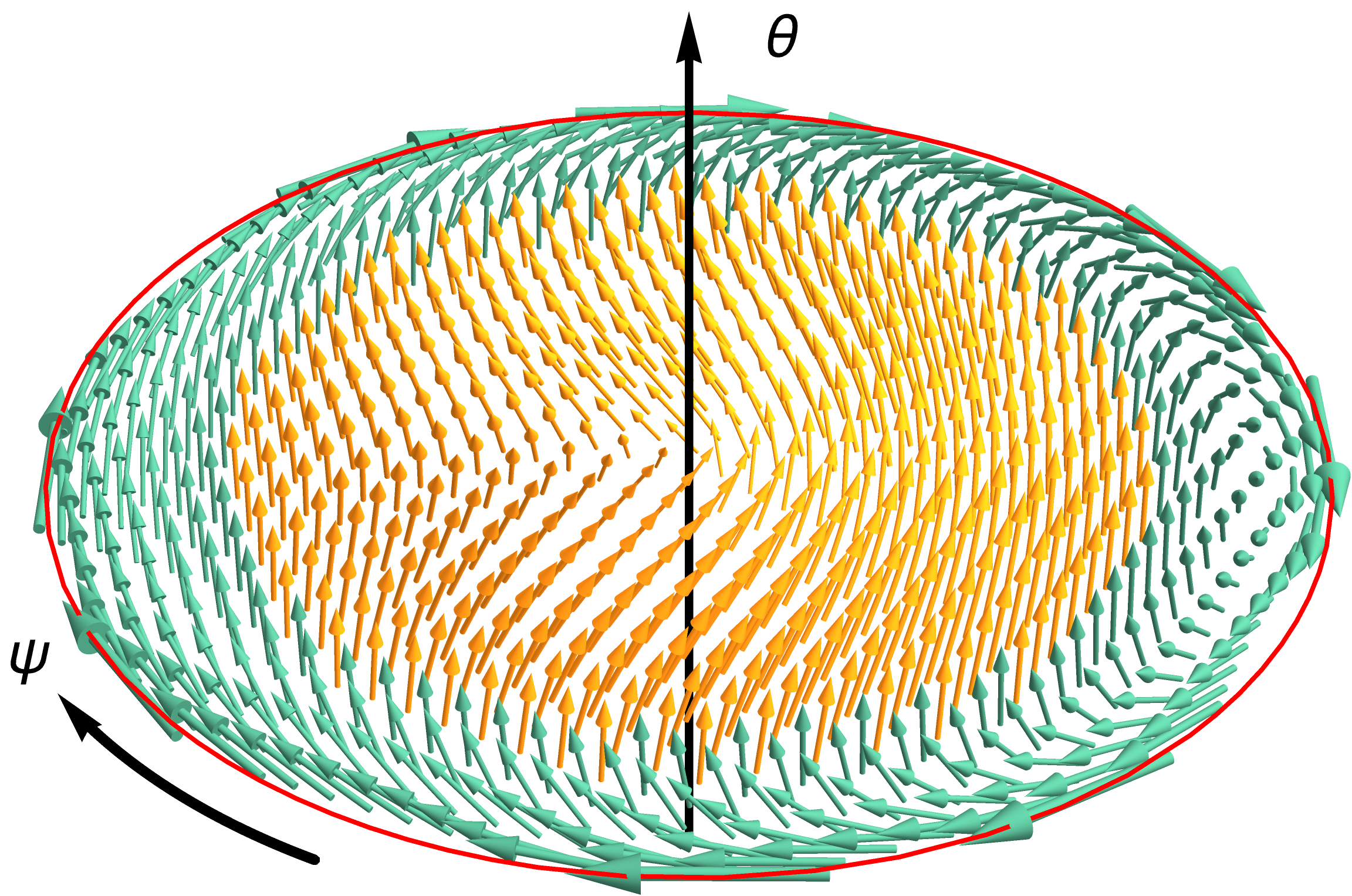}}\hspace{.04 \textwidth }
\subfloat[$s=\frac{3}{4},n=1,\omega=\frac{7}{4}$]{\includegraphics[width=.3\textwidth]{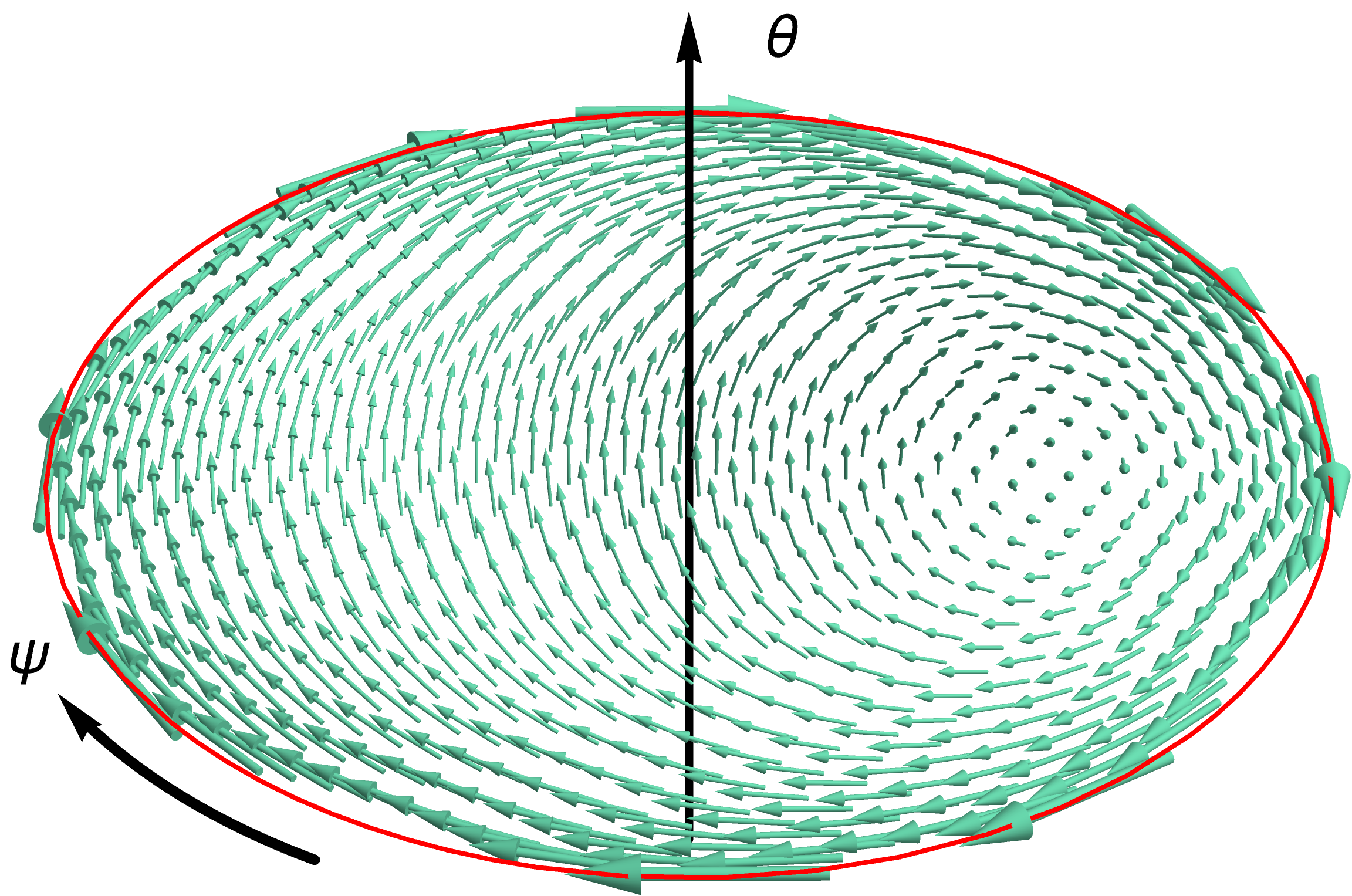}}
\caption{
\label{pic_eigenmode profiles}
Poynting vector $\vec S$ for the electromagnetic waves in the presence of a conformal twist defect.  
We color those Poynting vectors with positive energy flux density in the $\psi$-direction $S_\psi>0$ in green, and those with negative $S_\psi<0$ in yellow.
Given an $|s|\in \mathbb{N}+\frac{1}{4}$ or $|s|\in \mathbb{N}+\frac{3}{4}$, the energy flux density near the twist defect flows in a fixed direction along the twist defect, shown as the red circle, independent of $n$ and $\omega$.
(Here we only display the Poynting vector on a $\theta=$ const slice.)
}
\end{figure}

From the spectrum, we see that the conformal twist defect is chiral, i.e., there is no spatial reflection symmetry in the $\psi$-direction. 
Indeed, the dispersion relation \eqref{eq_eigenfrequencies} is asymmetric between the left-moving modes ($n<0$) and the right-moving modes ($n>0$) in the $\psi$-direction along the twist defect. 
 Given a fixed $|n|$, the lowest right-moving excitation moves faster than the lowest left-moving one for spin $|s|\in \mathbb{N}+\frac{1}{4}$, while the opposite holds for $|s|\in \mathbb{N}+\frac{3}{4}$. 
 In figure \ref{pic_eigenmode profiles}, we visualize these solutions by plotting their Poynting vectors.

Note that for any fixed $n,m$, there are two possible branches of $s$, i.e., $s\in \mathbb{Z}+\frac 14$ and $s\in \mathbb{Z}-\frac 14$. 
This agrees with the counting of propagating degrees of freedom in 4d free Maxwell theory, where one finds two independent polarizations in flat spacetime.
 
The photon states contribute to the total Casimir momentum along the $\psi$-direction is\footnote{\label{ft_Casimir}The Casimir momentum along the $\theta$-direction is identically zero, and the Casimir energy on $S^3$ reads
\begin{equation*}
    E_{0,\text{free}}= \lim_{\alpha\to 1}\frac{1}{2}\sum_{s\in \mathbb{Z}\pm \frac{1}{4}} \sum_{n\in \mathbb{Z}}\sum_{m\in \mathbb{N}} \omega^{\alpha}=-\frac{157}{15360}~.
\end{equation*}
The subscript ``free'' will become clear in the next subsection, where we show that this sector of defect operators forms a generalized free field theory. For a $(1+1)$d CFT on a spatial circle, recall that the Casimir energy and momentum are determined by the left and right central charges $c$ and $\bar{c}$. Explicitly, in units of the circle radius, they are $E_0=-(c+\bar{c})/24$ and $P_0=(c-\bar{c})/24$.}
\begin{equation}
\label{eq_Casimir momentum}
    P_{0,\text{free}}=\lim_{\alpha \to0}\frac{1}{2}\sum_{s\in \mathbb{Z}\pm \frac{1}{4}} \sum_{n\in \mathbb{Z}}\sum_{m\in \mathbb{N}} n \omega^{\alpha}=\frac{11}{192}~,
\end{equation}
where we have used the $\zeta$-function regularization. 
The non-zero Casimir momentum is another indication that the conformal twist defect is chiral.

Under our working assumption, we have solved a subsector of the twist defect operator spectrum \eqref{eq_eigenfrequencies} via the state/operator correspondence. The next step is to assess the validity of this assumption. At first glance, one might expect that generic interactions localized on the twist defect could modify the single-photon spectrum \eqref{eq_eigenfrequencies}.  However, in section \ref{sec_Generalized free field sector}, we demonstrate that there exists a universal generalized free field sector whose spectrum is precisely given by \eqref{eq_eigenfrequencies}. 
Of course, the complete spectrum can contain additional operators. The most straightforward example arises from stacking a decoupled two-dimensional CFT on the twist defect at $w=\bar{w}=0$. There can also be gauge field modes whose profiles are localized on the twist defect. We analyze these modes, along with their global symmetry properties, in section \ref{sec_Chiral current sector}.

\subsection{Generalized free field sector}
\label{sec_Generalized free field sector}

In this section, we follow techniques from DCFT \cite{Herzog:2020bqw, Herzog:2022jqv}  to show that the modes in \eqref{eq_eigenfrequencies} correspond to a universal, generalized free field sector of the twist defect.

We assume the twist defect preserves a $so(3,1)\times so(2)$ subalgebra of the 4d Euclidean conformal algebra $so(5,1)$.
Here $so(3,1)$ is the residual conformal algebra along the $(z,\bar{z})$ plane of the twist defect, and $so(2)$ is the transverse rotation symmetry that shifts the angle $\theta$ around the twist defect. 

One might be concerned that the presence of the non-invertible defect $\mathcal{D}_N$ breaks the transverse rotation symmetry $so(2)$. 
However, because $\mathcal{D}_N$ is topological, the $so(2)$ is not broken; rather, the monodromy condition \eqref{eq_4d modified Neumann} under $\theta\to \theta+2\pi$
 \begin{equation}
 \label{eq_Euclidean twist condition}
 F \longrightarrow -i\star F\longrightarrow -F\longrightarrow i\star F \longrightarrow F~,
\end{equation}
implies that the $\theta\sim \theta+2\pi$ periodicity is lifted to $\theta\sim \theta+8\pi$ in the presence of $\mathcal{D}_N$.

We now discuss the DCFT data on this twist defect. 
Our discussion below is a straightforward generalization of \cite{Herzog:2022jqv}, but we further impose a stronger monodromy condition \eqref{eq_Euclidean twist condition}. 

Let $\mathcal{O}(z,\bar z)$ be a defect primary operator localized at a point $(z,\bar z)$ on the twist defect (which itself is located at $w=\bar w=0$).
It is labeled by the irreducible representations of $so(3,1)\times so(2)$. 
We denote the conformal weights with respect to the Euclidean 2d conformal algebra $so(3,1)$ by $(h,\bar{h})$. 
Moreover,  $\Delta=h+\bar{h}$ is the scaling dimension and $\ell=h-\bar{h}$ is the spin parallel to the twist defect.  
We denote the transverse rotational spin under $so(2)$ by $s$.  
Furthermore, we use the shorthand notation $\textbf{x}_\mu=(z,\bar{z},w,\bar{w})$ for the Euclidean coordinates \eqref{eq_Euclidean metric}.
\begin{figure}[thb]
\centering
\includegraphics[width=1\textwidth]{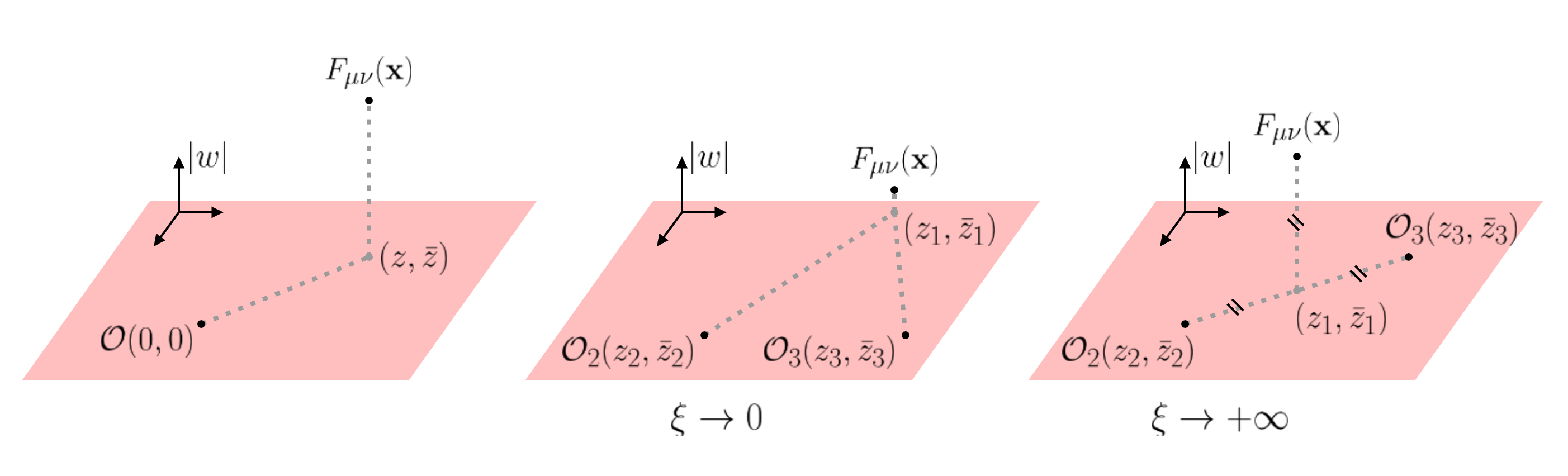}
  \caption{\label{pic_DCFT points}DCFT two-point and three-point functions. The red plane represents the twist defect at $w=\bar{w}=0$. Left: the bulk-defect two-point function $\langle F \mathcal{O}\rangle$. Middle: the bulk-defect-defect three-point function $\langle F \mathcal{O}_2\mathcal{O}_3\rangle$ in the limit $\xi\to 0$, where the bulk point $\textbf{x}$ approaches the defect. Right: the three-point function $\langle F \mathcal{O}_2\mathcal{O}_3\rangle$ in the limit $\xi \to +\infty$, where the projection of the bulk point onto the defect plane coincides with the midpoint between $(z_2,\bar{z}_2)$ and $(z_3,\bar{z}_3)$. } 
\end{figure}

\subsubsection{Two-point function}
 
Consider the two-point function $\langle F \mathcal{O}\rangle$ between the bulk field strength operator $F_{\mu\nu}(\textbf{x}) $ and a defect primary $\mathcal{O}(z=0,\bar{z}=0)$. See figure \ref{pic_DCFT points}. 
For this two-point function to be nontrivial, the transverse spin $s$ of $\mathcal{O}$ has to satisfy the spin selection rule \eqref{spinselection}, i.e., 
\begin{align}\label{spinselection2}
\langle F\mathcal{O}\rangle \neq 0 \Rightarrow s\in \mathbb{Z}\pm \frac 14\,.
\end{align}
The functional form of $\langle F \mathcal{O}\rangle$ is strongly constrained by the   $so(3,1)\times so(2)$ symmetry. These constraints can be systematically solved as we review in Appendix \ref{sec_dcft tensors}. 
In particular, covariant tensor structures in the two-point function between a bulk primary and a defect primary are spanned by polynomials of the following basis tensors: 
\begin{equation}
\label{eq_tensor building blocks}
\begin{aligned}
    X_\mu=&\frac{|w|}{|z|^2+|w|^2}\left(\bar{z},z,\frac{\bar{w}}{2}(1-|z/w|^2),\frac{w}{2}(1-|z/w|^2)\right)~,\\
    Y_\mu=&\frac{1}{2}\left(0,0,\frac{\bar{w}}{|w|},-\frac{w}{|w|}\right)~,\\
   I_{\mu z}=&-\frac{1}{2(|z|^2+|w|^2)}\left(\bar{z}^2,-|w|^2,\bar{z} \bar{w},\bar{z}w\right)~,\\
   I_{\mu \bar{z}}=&-\frac{1}{2(|z|^2+|w|^2)}\left(-|w|^2,z^2,z\bar{w},zw\right) ~,
\end{aligned}
\end{equation}
We have $X^\mu X_\mu=\frac{1}{2}$, $Y^\mu Y_\mu=-\frac{1}{2}$, and $I^\mu_{\phantom{\mu} z}I_{\mu \bar{z}}=\frac{1}{4}$, while the index contractions between other pairs of tensors are zero. Since one cannot build a tensor out of \eqref{eq_tensor building blocks} with a pair of anti-symmetrized indices $[\mu\nu]$ and parallel spin $|\ell|\geq 2$,  the two-point function $\langle F \mathcal{O}\rangle$ vanishes unless $\mathcal{O}$ has $|\ell|<2$. 
We therefore learned that the defect primary $\mathcal{O}$ is either a scalar ($\ell=0$) or a vector ($\ell=\pm1$) to admit a non-zero two-point function $\langle F \mathcal{O}\rangle$. It is further shown in \cite{Herzog:2022jqv} that a scalar primary $\mathcal{O}$ yields $\langle F \mathcal{O}\rangle=0$ unless $s=0$ and $\Delta(\mathcal{O})=2$, which contradicts the spin selection rule $s\in \mathbb{Z}\pm \frac{1}{4}$.  
We are therefore left with the vector primaries.

Let $\mathcal{V}_{z}^{s}(z,\bar z)$ be the vector primary of parallel spin $\ell=1$ and transverse spin $s$ on the twist defect. 
There are two independent tensor structures in $ \langle F_{\mu\nu}\mathcal{V}^{s}_z\rangle$, namely $X_{[\mu}I_{\nu] z}$ and $Y_{[\mu}I_{\nu] z}$. Note that the linear combination $X_{[\mu}I_{\nu] z}+Y_{[\mu}I_{\nu] z}$ is self-dual under the Hodge star operation, while $X_{[\mu}I_{\nu] z}-Y_{[\mu}I_{\nu] z}$ is anti-self-dual. The residual conformal group $so(3,1)\times so(2)$ and the monodromy condition \eqref{eq_Euclidean twist condition} fix the functional form of the two-point function $ \langle F_{\mu\nu}\mathcal{V}^{s}_z\rangle$ up to an overall constant: 
\begin{equation}
\label{eq_bulktodefect 2pt 1}
     \left. \langle F_{\mu\nu}(\textbf{x}) \mathcal{V}_{z}^{s}(0) \rangle\right|_{s\in \mathbb{Z}\pm \frac{1}{4}}=\frac{C_{F \mathcal{V}_{z}^{s}}}{|w|^2}\left(\frac{|w|}{|w|^2+|z|^2}\right)^{\Delta\left(\mathcal{V}_{z}^{s}\right)}\left(X_{[\mu}I_{\nu] z} \mp Y_{[\mu}I_{\nu] z}\right)\left(\frac{w}{|w|}\right)^{s}~,
\end{equation}
where $C_{F\mathcal{V}_{z}^{s}}\in \mathbb{C}$ denotes the bulk-to-defect OPE coefficient. 
A similar analysis can be applied to the defect vector primary $\mathcal{V}_{\bar{z}}^{s}$ of parallel spin $\ell=-1$, and we find 
\begin{equation}
\label{eq_bulktodefect 2pt 2}
\left. \langle F_{\mu\nu}(\textbf{x}) \mathcal{V}_{\bar{z}}^{s}(0) \rangle\right|_{s\in \mathbb{Z}\pm \frac{1}{4}}=\frac{C_{F\mathcal{V}_{\bar{z}}^{s}}}{|w|^2}\left(\frac{|w|}{|w|^2+|z|^2}\right)^{\Delta\left(\mathcal{V}_{\bar{z}}^{s}\right)}\left(X_{[\mu}I_{\nu] \bar{z}} \pm Y_{[\mu}I_{\nu] \bar{z}}\right)\left(\frac{w}{|w|}\right)^{s}~.
\end{equation}

Imposing the Bianchi identity $dF=0$ on \eqref{eq_bulktodefect 2pt 1} and \eqref{eq_bulktodefect 2pt 2}, we find that non-zero OPE coefficients are only compatible with the following scaling dimensions 
\begin{align}
\Delta(\mathcal{V}_{z}^{s})|_{s\in \mathbb{Z}\pm \frac{1}{4}}=1\pm s \,,~~~~~
\Delta(\mathcal{V}_{\bar{z}}^{s})|_{s\in \mathbb{Z}\pm \frac{1}{4}}=1\mp s\,.
\end{align}
Note that the unitarity bound for a defect vector is $\Delta \geq 1$, which is saturated by the conserved spin-1 currents. We therefore conclude that the defect primaries that appear in the OPE of the bulk field strength $F$ with the twist defect are:
\begin{equation}
\label{eq_defect vector primary}
    \begin{aligned}
\mathcal{V}_{z}^{s}~:&~~~(h,\bar{h})=\left(1+\frac{|s|}{2},\frac{|s|}{2}\right)~,~~~\text{where}~|s|\in \mathbb{N}+\frac{1}{4}~,\\
\mathcal{V}_{\bar{z}}^{s}~:&~~~(h,\bar{h})=\left(\frac{|s|}{2},1+\frac{|s|}{2}\right)~,~~~\text{where}~|s|\in \mathbb{N}+\frac{3}{4}~.
    \end{aligned}
\end{equation}
We emphasize that scaling dimensions in \eqref{eq_defect vector primary} are completely determined by conformal symmetry and unitarity. 
They are universal and insensitive to the interactions localized on the twist defect. 

By the state/operator correspondence, the defect vector primaries $\mathcal{V}_{z}^{s}$, $\mathcal{V}_{\bar{z}}^{s}$, and their descendants are mapped to the single-photon states in the twist defect Hilbert space on $S^3$.  
Specifically, we identify the frequency $\omega$ as the scaling dimension $\Delta$, and the wavenumber $n$ along the $\psi$-direction as the spin $\ell$ along the defect:
\begin{equation}
\omega=\Delta =h+\bar h\,,\quad 
n=-\ell = \bar h-h.
\end{equation}
The first branch of the modes in \eqref{eq_eigenfrequencies} is mapped to the operators $\partial_{\bar{z}}^{|n|+1}(\partial_z\partial_{\bar{z}})^m\mathcal{V}_{z}^s$ and $\partial_{z}^{|n|+1}(\partial_z\partial_{\bar{z}})^m\mathcal{V}_{\bar{z}}^s$, while the second branch is mapped to $\partial_{z}^{|n|-1}(\partial_z\partial_{\bar{z}})^m\mathcal{V}_{z}^s$ and $\partial_{\bar{z}}^{|n|-1}(\partial_z\partial_{\bar{z}})^m\mathcal{V}_{\bar{z}}^s$.
We thus find exact agreement between the spectrum of defect primaries obtained here and the spectrum of states in the twisted Hilbert space on $S^3$ computed in section~\ref{sec_defect Hilbert space}.

\subsubsection{Three-point function}

Next, we consider the three-point function $\langle F\mathcal{O}_2 \mathcal{O}_3\rangle$ between a bulk field strength operator $F$ at $\textbf{x}_1=(z_1,\bar{z}_1,w,\bar{w})$ and two defect primaries $\mathcal{O}_2(z_2,\bar{z}_2)$ and $\mathcal{O}_3(z_3,\bar{z}_3)$. See figure \ref{pic_DCFT points}. Let $(\Delta_i,\ell_i,s_i)$ be the quantum numbers of the defect primaries $\mathcal{O}_i$ with $i=2,3$. 
It follows from the spin selection rule that $\langle F\mathcal{O}_2 \mathcal{O}_{3}\rangle= 0$ unless $s=s_2+s_3\in \mathbb{Z}\pm \frac{1}{4}$.\footnote{Note that $s_2$ and $s_3$ themselves need not satisfy the spin selection rule \eqref{spinselection}. In other words, the two-point functions $\langle F \mathcal{O}_i\rangle$ can be zero.} 
Moreover, the functional form of $\langle F\mathcal{O}_2 \mathcal{O}_3\rangle$ is fixed by the residual conformal group $so(3,1)\times so(2)$ up to functions of the cross ratio $\xi$, defined as:
\begin{equation}
\label{eq_crossration}
    \xi=\frac{|z_{23}|^2|w|^2}{(|w|^2+z_{12}\bar{z}_{13})(|w|^2+\bar{z}_{12}z_{13})}>0~,
\end{equation}
where $z_{ab}=z_a-z_b$. Tensor structures in this three-point function can be enumerated in a way similar to \eqref{eq_bulktodefect 2pt 1} and \eqref{eq_bulktodefect 2pt 2}. 
In \eqref{eq_app_rank2 antisymmetric tensor def}, we present a complete basis for rank-2 anti-symmetric tensors $\mathcal{X}_{a,\mu \nu}^{\pm}$ with $a=1,2,3$. Among the 6 covariant tensors,  3 tensors $\mathcal{X}_{a,\mu \nu}^{+}$ are self-dual under the Hodge star operation, and the other 3 tensors $\mathcal{X}_{a,\mu \nu}^{-}$ are anti-self-dual. 
Thus, a general three-point function $\langle F\mathcal{O}_2 \mathcal{O}_3\rangle$ takes the form
 \begin{equation}
 \label{eq_3pt function}
\mathtoolsset{multlined-width=0.9\displaywidth}
\begin{multlined}
\left. \langle F_{\mu \nu}(\textbf{x}_1)\mathcal{O}_2(z_2,\bar{z}_2) \mathcal{O}_{3}(z_3,\bar{z}_3)\rangle\right|_{s\in \mathbb{Z}\pm \frac{1}{4}}=\frac{(w/|w|)^{s}|w|^{\Delta_2+\Delta_3-2}}{(|z_{12}|^2+|w|^2)^{\Delta_2}(|z_{13}|^2+|w|^2)^{\Delta_3}}\hfill\\
\times \left(\frac{|w|^2+\bar{z}_{12}z_{13}}{z_{23}|w|}\right)^{\ell_2}\left(\frac{|w|^2+\bar{z}_{13}z_{12}}{z_{23}|w|}\right)^{\ell_3}\left(\sum_{a=1}^3f^\mp_a(\xi) \mathcal{X}_{a,\mu\nu}^{\mp}\right)~,
\end{multlined}
\end{equation}
where $f^\pm_{a}$ are real-valued functions of the cross ratio \eqref{eq_crossration}.

Imposing the Bianchi identity $dF=0$ on \eqref{eq_3pt function}, we obtain an algebraic relation among the functions $f^\pm_{a}$: 
\begin{equation}
\label{eq_three point linear}
   \left(
   \frac{s}{4}\pm (h_2-h_3)
   \right)
   f_1^{\pm}+sf_2^{\pm}+(\ell_2-\ell_3)f_3^\pm=0~,\text{ for }s\in \mathbb{Z}\mp \frac{1}{4}~.
\end{equation}
This relation allows us to express $f_2^{\pm}$ in terms of $f_1^{\pm}$ and $f_3^{\pm}$. Furthermore, the Bianchi identity yields an ordinary differential equation:
\begin{equation}
\label{eq_three point closure condition}
\mathtoolsset{multlined-width=0.9\displaywidth}
\begin{multlined}
    s\xi(1+\xi)\partial_\xi f_1^\pm=2\left(s^2+\xi(\ell_2-\ell_3)^2\right)f_3^\pm\hfill\\
    \hfill+\left[\left(\frac{s}{2}(2+\ell_2+\ell_3)\pm(h_2-h_3)(\ell_2-\ell_3)\right)\xi+s(1\pm \frac{s}{2}-\bar{h}_2-\bar{h}_3)\right]f^\pm_1~,\\
    s\xi(1+\xi)\partial_\xi f_3^\pm=\frac{\xi}{8}(s^2-4(h_2-h_3)^2)f_1^\pm\hfill\\
    \hfill+\left[\left(\frac{s}{2}(2+\ell_2+\ell_3)\mp(h_2-h_3)(\ell_2-\ell_3)\right)\xi+s(1\mp\frac{s}{2}-\bar{h}_2-\bar{h}_3)\right]f^\pm_3~.
\end{multlined}
\end{equation}
The two hypergeometric solutions to \eqref{eq_three point closure condition} are discussed in appendix \ref{sec_hypergeometric solutions}. These solutions are subject to constraints from unitarity (at $\xi \to 0$) as well as convergence (at $\xi \to \infty$). We find that $\langle F\mathcal{O}_2 \mathcal{O}_{3}\rangle=0$ unless the following conditions are satisfied:
\begin{equation}
\label{eq_double twist condition}
\begin{aligned}
\text{if}~|s|\in \mathbb{N}+\frac{1}{4}~,~~~ \text{either}~(h_2,\bar{h}_2)-(h_3,\bar{h}_3) \in & (1+\frac{|s|}{2}+\mathbb{N},\frac{|s|}{2}+\mathbb{N})\\
\text{or}~(h_3,\bar{h}_3)-(h_2,\bar{h}_2) \in &(1+\frac{|s|}{2}+\mathbb{N},\frac{|s|}{2}+\mathbb{N})~;\\
\text{if}~|s|\in \mathbb{N}+\frac{3}{4}~,~~~ \text{either}~(h_2,\bar{h}_2)-(h_3,\bar{h}_3) \in & (\frac{|s|}{2}+\mathbb{N},1+\frac{|s|}{2}+\mathbb{N})\\
\text{or}~(h_3,\bar{h}_3)-(h_2,\bar{h}_2) \in &(\frac{|s|}{2}+\mathbb{N},1+\frac{|s|}{2}+\mathbb{N})~.\\
\end{aligned}
\end{equation}
Crucially, we see that the differences between the allowed values of $(h_2,\bar h_2)$ and $(h_3,\bar h_3)$ correspond to the conformal weights of $\mathcal{V}_{z}^s$ and  $\mathcal{V}_{\bar z}^s$, or their descendants.

This result constrains the OPE between the defect primaries. 
To see this, we start with a nonzero three-point function $\langle F\mathcal{O}_2\mathcal{O}_3\rangle$, and bring the bulk field strength operator $F$ close to the twist defect. 
This induces the defect vector primaries $\mathcal{V}_{z}^s$ and  $\mathcal{V}_{\bar z}^s$ as well as their descendants.  
Then the selection rule in \eqref{eq_double twist condition}  implies that the operator spectrum in the OPE between $\mathcal{V}_{z}^s$, $\mathcal{V}_{\bar{z}}^s$, and a generic defect primary $\mathcal{O}$ is of the double-twist type, with no anomalous dimension. 
It follows that  correlation functions of $\mathcal{V}_{z}^s$, $\mathcal{V}_{\bar{z}}^s$ are computed using the Wick contraction \cite{Lauria:2020emq, Herzog:2022jqv}. 
We therefore conclude that $\mathcal{V}_{z}^s$,  $\mathcal{V}_{\bar{z}}^s$, and their descendants form a universal sector of generalized free fields on the twist defect.

Finally, we note that there are two defect operators with transverse spin $s=\pm1$ and scaling dimension $\Delta=3$:
\begin{equation}
    D^{+ 1}=\mathcal{V}_z^{+ \frac{1}{4}}\mathcal{V}_{\bar{z}}^{+ \frac{3}{4}}~,~~~D^{- 1}=\mathcal{V}_z^{- \frac{1}{4}}\mathcal{V}_{\bar{z}}^{- \frac{3}{4}}~.
\end{equation}
They are identified as the defect displacement operator \cite{McAvity:1993ue,McAvity:1995zd,Liendo:2012hy,Gliozzi:2015qsa,Billo:2016cpy}, whose existence marks the mobility of the twist defect. 

\subsection{Chiral current sector and anyonic branes}
\label{sec_Chiral current sector}

We have determined a universal sector of the DCFT using just the monodromy condition \eqref{eq_Euclidean twist condition} and conformal symmetry. 
Importantly, our analysis thus far does not depend on the specific action that defines the twist defect. In general, the DCFT also contains additional operator sectors that are not visible in the OPEs between the bulk local operators (e.g., the field strength $F$) and the twist defect. 
As a trivial example, we can stack an arbitrary, decoupled 2d CFT on the twist defect. 
These other sectors depend on the detailed definition of the twist defect itself, beyond what is fixed by the monodromy condition. 
Such freedom is common in the study of boundary conditions in quantum systems, for instance, in the context of edge modes in 3d Chern–Simons theories \cite{Witten:1988hf, Elitzur:1989nr, Wen:1992uk, Wen:1992vi}. 

In this section, we argue that there has to be another sector, which we call the chiral current sector, enforced by anomaly inflow and the 1-form global symmetry in the bulk. 
This is intuitively clear, since the action \eqref{eq_maxwell twist defect attach} for the 3d topological duality defect $\mathcal{D}_N$ takes the form of a Chern-Simons action. 
Our discussion below follows closely the standard treatment of conformal boundary conditions of chiral 3d Chern-Simons theory in \cite{Witten:1988hf,Elitzur:1989nr}. 

Consider again the defect configuration in figure \ref{pic_complex coordinate}. The total action, $S_\text{bulk}+S_\text{defect}$, is given by the sum of the bulk Maxwell term \eqref{eq_maxwell bulk action} and the topological defect term \eqref{eq_maxwell twist defect attach}. Since the topological defect $\mathcal{D}_N$ is supported on a three-dimensional manifold with a boundary, variation of its action yields an extra boundary contribution:
\begin{align}
\delta S_\text{defect}= {iN\over 2\pi} \int_{w\in \mathbb{R}^+} (\delta A^{-} \wedge dA^+ + \delta A^+ \wedge dA^- ) 
+{i N\over 2\pi} \int_{w=\bar w=0} A\wedge \delta A \,.
\end{align}
With the gluing condition \eqref{eq_4d modified Neumann}, the first two terms are canceled by the variation of the bulk action in \eqref{dSbulk}. 
Following \cite{Elitzur:1989nr}, we pick a complex structure along the twist defect at $w=\bar w=0$ and introduce a local counterterm:
\begin{equation}
\label{eq_counter term}
    S_{\text{counter}}[A]=-\frac{N}{2\pi}\int_{w=\bar{w}=0} A_z A_{\bar{z}}dzd\bar{z}~. 
\end{equation}
The variation of the total action is then
\begin{equation}
\label{eq_twist defect current}
\delta(S_\text{bulk}+S_\text{defect}+S_\text{counter})=-\frac{N}{\pi}\int_{w=\bar{w}=0} A_{\bar{z}}\delta A_{z} dzd\bar{z}~.
\end{equation}
To ensure a well-defined variational principle, we impose a Dirichlet condition on the holomorphic component of the gauge field $A_{z}$ at $w=\bar{w}=0$:
\begin{equation}
\label{eq_chiral defect condition}
    \left. A_{z}\right|_{w=\bar{w}=0}=0~.
\end{equation}
This is precisely the usual conformal boundary condition for the chiral WZW model in the chiral Chern-Simons theory \cite{Witten:1988hf, Elitzur:1989nr}. 
The standard analysis implies that there is a chiral compact boson $\phi$ living along the twist defect, which, on-shell, is related to the gauge field as $A_{\bar z} |_{w=\bar w=0}= \partial_{\bar z}\phi$. 
From \eqref{eq_twist defect current}, we define a current operator as:
\begin{equation}\label{chiralcurrent}
    J_{\bar{z}}=-\frac{i N}{\pi} A_{\bar{z}} = - {iN\over \pi} \partial_{\bar z}\phi~. 
\end{equation}
It has conformal weights $(h,\bar{h})=(0,1)$ and zero transverse spin $s=0$. 
From the spin selection rule in \eqref{spinselection2}, its two-point function with the bulk field strength vanishes, i.e., $\langle F J_{\bar{z}}\rangle=0$. 
We have also shown in \eqref{eq_double twist condition} that fusion of defect operators $\mathcal{V}_z^s$, $\mathcal{V}_{\bar{z}}^s$, and their descendants cannot produce current operators with either $h=0$ or $\bar{h}=0$. 
We therefore conclude that correlation functions associated with the chiral current sector and the generalized free field sector factorize. 
 Physically, this can be intuitively understood by the fact that the centrifugal force repels photons away from the chiral boson localized on the twist defect.

To analyze the chiral current $J_{\bar{z}}$, it is convenient to perform a conformal map from $\mathbb{R}^4$ to $S^3\times \mathbb{R}$ and go to Lorentzian signature as in section \ref{sec_defect Hilbert space}. 
In this setup, the twist defect is supported on a Lorentzian cylinder parametrized by $(t,\psi)$, as shown in figure \ref{pic_S3 coordinate}. 
The chiral boson action is \footnote{For manifestly Lorentz invariant formulations of the chiral boson action, see, e.g. \cite{Floreanini:1987as,Sen:2019qit,Mkrtchyan:2019opf}.}
\begin{equation}
\label{eq_chiral compact boson action}
    S_{\text{chiral}}[\phi]=-\frac{N}{2\pi}\int_{\rho=0}dtd\psi[(\partial_\psi\phi)^2+\partial_t \phi\partial_\psi \phi]~.
\end{equation}
Here $\phi$ is identified as $\phi+2\pi$. 
This theory has a chiral $U(1)_{2N}$ global symmetry that shifts the chiral boson, i.e., $\phi\to \phi+\lambda$. 
(The subscript $2N$ will become clear later.)
The $U(1)_{2N}$ symmetry is generated by the chiral current in \eqref{chiralcurrent}, which in these coordinates takes the form $J=\frac{N}{\pi}\partial_\psi\phi(dt-d\psi)$.  

\begin{figure}[thb]
\centering
\includegraphics[width=1\textwidth]{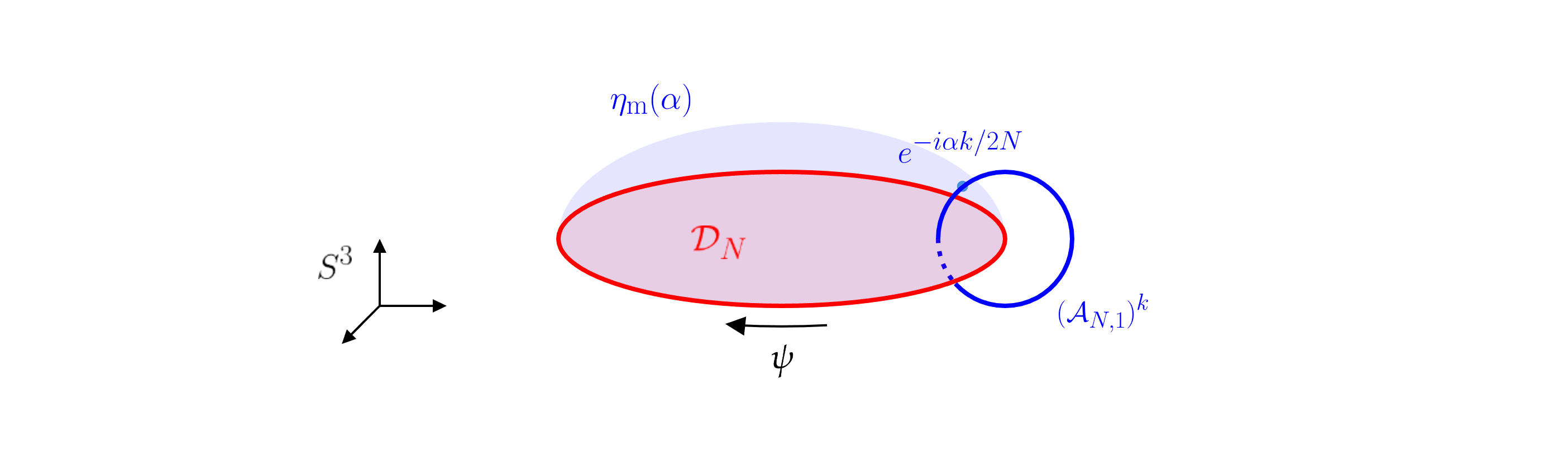}
  \caption{\label{pic_linking} The insertion of the anyonic brane $(\mathcal{A}_{N,1})^k$ changes the $U(1)_{2N}$ charge, analogous to the insertion of an anyon line in the context of 3d Chern-Simons theory. 
  As in figure \ref{pic_S3 coordinate}, the red circle denotes the twist defect at $\rho=0$, and the red disk denotes the topological duality defect $\mathcal{D}_N$. 
  The anyonic brane is shown as a blue circle, which links with the twist defect (red circle). 
  The magnetic 1-form symmetry operator $\eta_m(\alpha)$ is shown as the blue surface.  This figure only shows the spatial $S^3$, and every defect except for $\eta_\text{m}(\alpha)$ is extended in the time direction as well.} 
\end{figure}

Next, we discuss the charge quantization for this chiral $U(1)_{2N}$ symmetry. 
The total $U(1)_{2N}$ charge on the spatial circle of the twist defect is
\begin{equation}
\label{eq_chiral current charge}
    Q_{\text{R}}=\int_{S^1}d\psi J_t 
    = {2N\over 2\pi} \int_{S^1} d\psi \partial_\psi \phi={2N\over 2\pi}\int_{S^1}d\psi A_\psi~.
\end{equation}
The charge $Q_{\text{R}}$ can be understood in two complementary ways: In the second equation of \eqref{eq_chiral current charge}, we interpret it as the winding number of $\phi$ multiplied by $2N$. In the third equation of \eqref{eq_chiral current charge}, we interpret it as the gauge field holonomy, which equals the magnetic 1-form symmetry defect ending on the twist defect. See also figure \ref{pic_linking}. 

Note that a fractional winding number of $\phi$ would result in $Q_\text{R} \notin 2N\mathbb{Z}$. 
This signals a discontinuity in the chiral boson field and can be interpreted as the insertion of a 1-form symmetry defect, which we now explain. 

Consider the following discrete $\mathbb{Z}_{2N}^{(1)}$ subgroup of the $U(1)^{(1)}_\text{e} \times U(1)^{(1)}_\text{m}$ 1-form global symmetry \eqref{eq_maxwell U1xU1} generated by the operator $\mathcal{A}_{N,1}$ supported on a 2d  surface $\mathcal{M}_2$: 
\begin{equation}\label{anyonic}
\mathcal{A}_{N,1}=\eta_{\text{e}}(\frac{\pi}{N})\eta_{\text{m}}(\pi)=\exp\left(\frac{i}{2}\int_{\mathcal{M}_2}(F-\star F)\right)~.
\end{equation}
It is a certain linear combination of the electric and magnetic fluxes and is of order $2N$, i.e., $(\mathcal{A}_{N,1})^{2N}=1$. 
It follows from the fusion rule \eqref{eq_maxwell fusion rule 1} that 
$\mathcal{A}_{N,1}$ commutes with $\mathcal{D}_{N}$:
\begin{equation}
\label{eq_maxwell commutation relation 1}
    \mathcal{A}_{N,1}\times \mathcal{D}_N=\mathcal{D}_N\times \mathcal{A}_{N,1}~.
\end{equation}
In other words, $\mathbb{Z}_{2N}^{(1)}$ is the 1-form symmetry subgroup preserved by the duality symmetry defect $\mathcal{D}_{N}$. 
We can therefore consider the defect configuration in figure \ref{pic_linking}, where $\mathcal{D}_N$ intersects topologically with the $k$-th power of $\mathcal{A}_{N,1}$ at a point. (Here $k=1,2,\cdots, 2N$.) 
Furthermore, $\mathcal{A}_{N,1}$ links with the twist defect in this $S^3$ spatial slice. The configuration of these defects is summarized in Table \ref{tab_geometry}.
\begin{table}[h!]
\centering
\begin{tabular}{ c| c c c c }
  & $t$ & $\psi$ & $\rho$ & $\theta$ \\ 
 \hline
 duality defect $\mathcal{D}_N$ &  $\times$  & $\times$ & $\times$ & \\  
 twsit defect & $\times$ & $\times$ & & \\
 anyonic brane $\mathcal{A}_{N,1}$ & $\times$& & & $\times$
\end{tabular}
\caption{\label{tab_geometry} We use $\times$ to denote the directions occupied by the various defects in figure \ref{pic_linking}.}
\end{table}

Let us consider the magnetic 1-form symmetry operator $\eta_\text{m}(\alpha)$, defined on an open surface that terminates on the twist defect as in figure \ref{pic_linking}. 
The mixed 't Hooft anomaly between $\mathcal{A}_{N,1}$ and $\eta_\text{m}(\alpha)$ (which follows from \eqref{eq_1form thooft anomaly}) modifies the latter operator from its original form in \eqref{eq_maxwell U1xU1} to:\footnote{Strictly speaking, the symmetry operators in the presence of the $\mathcal{A}_{N,1}$ defect on an open surface with boundaries should be denoted by a different symbol. To simplify the notation, we will not make this distinction for  $\eta_\text{m}(\alpha)$ in what follows. The configuration here is similar to  \eqref{eq_defect gauge coupling} and figure \ref{pic_stateoperator and junction} in the context of the 2d compact boson CFT. \label{fn:modified}}
\begin{equation}
    \eta_\text{m}(\alpha)=\exp{\left(\frac{i\alpha}{2\pi}\int F-\frac{i\alpha k}{2N}\right)}=\exp\left(\frac{i\alpha}{2N}(Q_\text{R}-k)\right)~.
\end{equation}
Since the $\eta_\text{m}(2\pi)$ surface can be annihilated with $\mathcal{D}_N$, we demand $\eta_\text{m}(2\pi)=1$, leading to the charge quantization condition $Q_\text{R}\in 2N\mathbb{Z}+k$. 
The insertions of $(\mathcal{A}_{N,1})^k$ correspond to the $2N$ primaries of the compact chiral boson \eqref{eq_chiral compact boson action}, whose conformal weights are
\begin{equation}
\label{eq_chiral primaries}
    (h,\bar{h})=(0,\frac{Q_\text{R}^2}{4N})~,~~~\text{where}~ Q_\text{R}\in 2N\mathbb{Z}+k~.
\end{equation}
In the context of 3d chiral Chern-Simons theories, these primaries of the chiral boson correspond to the insertions of the bulk anyon lines in a similar way. 
For this reason, we will refer to $\mathcal{A}_{N,1}$ as an \textit{anyonic brane}. 

 We comment on the contribution to the Casimir momentum along the twist defect from this chiral current sector. 
Since the compact boson \eqref{eq_chiral compact boson action} has chiral central charges $c=0$ and $\bar{c}=1$, they contribute to the Casimir momentum by $P_{0,\text{chiral}}={c-\bar c\over24} = -{1\over 24}$. 
Combining with the contribution from the generalized free field sector in \eqref{eq_Casimir momentum}, we find
\begin{equation}
\label{eq_final Cas momentum}
P_0=P_{0,\text{free}}+P_{0,\text{chiral}}=\frac{11}{192}+\frac{c-\bar{c}}{24}=\frac{1}{64}~.
\end{equation}

Finally, we note that the DCFT is stable against perturbations from the generalized free field sector and the chiral current sector. The lowest-lying primary operator that preserves both the defect Lorentz symmetry and the transverse rotation symmetry is $J_{\bar{z}}J_{\bar{z}}\mathcal{V}^{+1/4}_z\mathcal{V}^{-1/4}_z$, which has scaling dimension $\Delta=4.5>2$ and is therefore irrelevant. Even if we relax constraints from the transverse rotation symmetry, the lowest-lying primaries $J_{\bar{z}}\mathcal{V}^{+1/4}_z$ and $J_{\bar{z}}\mathcal{V}^{-1/4}_z$ have $\Delta=2.25>2$, and the DCFT remains stable against anitropic perturbations.

\subsection{Non-invertible symmetries at rational fine-structure constants}\label{sec_4d general duality defects}

So far, we have assumed that the electric coupling constant takes some special quantized values $e^2= {2\pi/N}$ as in \eqref{eq_defect topo cond} for some positive integer $N$. The infrared fine-structure constant of our real world is approximately $e^2 \sim {4\pi \over 137.04}$, which clearly does not fit the quantized form of~\eqref{eq_defect topo cond}. 
In this section, we show that the duality defect \eqref{eq_maxwell duality defect action} and its associated twist defect can in fact be generalized to any rational $e^2 \in 2\pi \mathbb{Q}^+$. 
By choosing an appropriate sequence of rational numbers, one can thus approximate the observed fine-structure constant to arbitrary accuracy.

For any pair of coprime positive integers $N_\text{e}$, $N_\text{m}$, we define a defect $\mathcal{D}_{N_\text{e}/N_\text{m}}$  as follows \cite{Niro:2022ctq,Cordova:2023ent}: 
\begin{equation}
\label{eq_maxwell generalization}
\mathtoolsset{multlined-width=0.9\displaywidth}
\begin{multlined}
\mathcal{D}_{N_\text{e}/N_\text{m}}~:~~~S_{\text{defect}}[A^-,A^+;a_1,a_2]=\frac{iN_\text{e}}{2\pi }\int_{w\in \mathbb{R}^+} A^-\wedge da_1\\
+\frac{i }{2\pi}\int_{w\in \mathbb{R}^+} A^+ \wedge da_2-\frac{iN_\text{m}}{2\pi}\int_{w\in \mathbb{R}^+} a_1\wedge da_2~,
\end{multlined}
\end{equation}
where $a_1$, $a_2$ are auxliary $U(1)$ gauge fields localized on the 3d duality defect. 
See figure \ref{pic_complex coordinate} for the defect configuration, and  \eqref{eq_gauge fields across defect} for the definition of $A^\pm$. The path integral of $a_1$ and $a_2$ yields an effective defect action that reads
\begin{equation}
\mathcal{D}_{N_\text{e}/N_\text{m}}~:~~~S_{\text{defect}}[A^-,A^+]\sim`` \frac{iN_\text{e}}{2\pi N_\text{m} }\int_{w\in \mathbb{R}^+} A^-\wedge dA^+~."
\end{equation}
However, this is not gauge-invariant (hence the quotation marks), and only serves as an intuitive way to understand the more precise, gauge-invariant action in \eqref{eq_maxwell generalization}. 
(See \cite{Choi:2022jqy} for similar discussions.)
Conservation and continuity of the stress-energy tensor require
\begin{equation}
\label{eq_general defect topo cond}
    e^2=\frac{2\pi N_\text{m}}{N_\text{e}}~.
\end{equation}
We will assume \eqref{eq_general defect topo cond} from now on, such that the defect $\mathcal{D}_{N_\text{e}/N_\text{m}}$ is topological. The special case in \eqref{eq_defect topo cond} is reproduced with $N_\text{e}=N$ and $N_\text{m}=1$.

The fusion rules between $\mathcal{D}_{N_\text{e}/N_\text{m}}$ and the 1-form symmetry operators $\eta_\text{e}$, $\eta_\text{m}$ are \cite{Niro:2022ctq, Cordova:2023ent}: 
\begin{equation}
\label{eq_maxwell fusion rule 3}
\begin{aligned}
    \eta_\text{e}(N_\text{m}\alpha)\times \mathcal{D}_{N_\text{e}/N_\text{m}}=&\mathcal{D}_{N_\text{e}/N_\text{m}}\times \eta_\text{m}(N_\text{e}\alpha)~,\\
     \mathcal{D}_{N_\text{e}/N_\text{m}}\times \eta_\text{e}(N_\text{m} \alpha)=&\eta_\text{m}(-N_\text{e}\alpha)\times\mathcal{D}_{N_\text{e}/N_\text{m}} ~.
\end{aligned}
\end{equation}
Furthermore, 
\begin{align}
\mathcal{D}_{N_\text{e}/N_\text{m}} &=\eta_\text{e}({2\pi\over N_\text{e}})
\times\mathcal{D}_{N_\text{e}/N_\text{m}} = \mathcal{D}_{N_\text{e}/N_\text{m}}  \times\eta_\text{e}({2\pi\over N_\text{e}})\\
&
=\eta_\text{m}({2\pi\over N_\text{m}})
\times
\mathcal{D}_{N_\text{e}/N_\text{m}} = \mathcal{D}_{N_\text{e}/N_\text{m}}  \times\eta_\text{m}({2\pi\over N_\text{m}}).
\end{align}
The duality defect \eqref{eq_maxwell generalization} preserves the discrete 1-form $\mathbb{Z}_{2N_\text{e}N_\text{m}}^{(1)}$ subgroup generated by the operator \footnote{It should be noted that such 1-form symmetry subgroups are not unique when $N_\text{m}>1$. For instance, ${\mathcal{A}}'_{N_\text{e},N_\text{m}}=\eta_{\text{e}}(\frac{\pi}{N_\text{e}})\eta_{\text{m}}(-\frac{\pi}{N_\text{m}})$ generates another $\mathbb{Z}_{2N_\text{e}N_\text{m}}^{(1)}$ subgroup that is preserved by the duality defect.}
\begin{equation}
\label{eq_discrete 1form subgroup}
\mathcal{A}_{N_\text{e},N_\text{m}}=\eta_{\text{e}}(\frac{\pi}{N_\text{e}})\eta_{\text{m}}(\frac{\pi}{N_\text{m}})=\exp{\left(\frac{i}{2N_\text{m}}\int_{\mathcal{M}_2}(F-\star F)\right)}~,
\end{equation}
such that $(\mathcal{A}_{N_\text{e},N_\text{m}})^{2N_\text{e}N_\text{m}}=1$. Indeed, it follows from  \eqref{eq_maxwell fusion rule 3} that $\mathcal{A}_{N_\text{e},N_\text{m}}$ commutes with the duality defect:
\begin{equation}
\mathcal{A}_{N_\text{e},N_\text{m}}\times \mathcal{D}_{N_\text{e}/N_\text{m}}=\mathcal{D}_{N_\text{e}/N_\text{m}}\times \mathcal{A}_{N_\text{e},N_\text{m}}~.
\end{equation}
 
The discussions in section \ref{sec_defect Hilbert space} and section \ref{sec_Generalized free field sector} are insensitive to $N_\text{e},N_\text{m}$ and apply equally to the twist defect of $\mathcal{D}_{N_\text{e}/N_\text{m}}$. 
In particular, there is a universal generalized free field sector \eqref{eq_defect vector primary} on the twist defect for $\mathcal{D}_{N_\text{e}/N_\text{m}}$ for any $N_\text{e}$ and $N_\text{m}$. 
Note that the operator spectrum in \eqref{eq_defect vector primary} is independent of the gauge coupling value.

On the other hand, the chiral current sector of the twist defect depends on $N_\text{e},N_\text{m}$. 
Similar to the standard analysis of the edge modes in frational quantum Hall systems \cite{Wen:1992uk, Wen:1992vi, Levin:2013gaa}, one finds that the chiral current sector has a right-moving chiral global symmetry  $U(1)_{2N_\text{e}N_\text{m}}$, which can be effectively realized by a chiral compact boson as in \eqref{eq_chiral compact boson action} with $N$ replaced by $N_\text{e}N_\text{m}$. 
By inserting the anyonic brane $(\mathcal{A}_{N_\text{e},N_\text{m}})^k$ as in figure \ref{pic_linking}, we find the following conformal primaries with $U(1)_{2N_\text{e}N_\text{m}}$ symmetry charge $Q_\text{R}$: 
\begin{equation}
(h,\bar{h})=(0, \frac{Q_\text{R}^2}{4N_\text{e}N_\text{m}})~,~~~\text{where}~ Q_\text{R}\in 2N_\text{e}N_\text{m}\mathbb{Z}+k~,
\end{equation}
with $k=1,2,\cdots, 2N_\text{e}N_\text{m}$. 

\section{2d free compact boson}\label{sec:2d}

In the 2d free compact boson theory, the non-invertible symmetry associated with T-duality is implemented by a topological, 1d defect line in spacetime. 
Its endpoint is the twist defect, which is a 0d point in spacetime. 
The state/operator correspondence maps the operator content of the twist defect to the spectrum in the Hilbert space on a circle twisted by the topological defect line. 
We will refer to these non-local, point operators that are attached to a topological line as twist fields, which are also commonly known as disorder operators or twisted sector operators.

This section is organized as follows: In section \ref{sec_review}, we review the compact boson CFT. 
Section \ref{sec_duality defect action revisted} presents a puzzle in the naive expression for the duality defect action that appeared in the literature. By deriving a more precise defect action, we resolve the puzzle and determine the fusion rules of the duality defect unambiguously. We proceed to analyze the spectrum of twist fields for the non-invertible duality defects in section \ref{sec_twisted sector operators}. 
Finally, we explore generalizations of the duality defect and their twist fields in section \ref{sec_2d general duality defects}.

\subsection{Review of the free compact boson}\label{sec_review}

We start with a review of the compact boson theory.  
We refer readers to \cite{Ginsparg:1988ui, Thorngren:2021yso, Cheng:2022sgb, Pace:2024oys} for more detailed discussions about $c=1$ CFTs and their symmetries. The action of the compact boson field $\phi \sim \phi+2\pi$ on a  $2$d Euclidean  manifold $\mathcal{M}_2$ is 
\begin{equation}
\label{eq_compact boson action}
S[\phi]=\frac{R^2}{4\pi}\int_{\mathcal{M}_2} d\phi\wedge\star d\phi~,
\end{equation}
where $R\in \mathbb{R}^+$ is an exactly marginal parameter known as the radius. 
The free compact boson theory \eqref{eq_compact boson action} has a continuous global symmetry $U(1)_\text{m}\times U(1)_\text{w}$, known as the momentum and winding symmetries in the string theory literature. 
Their corresponding Noether currents $J_{\text{m}}, J_{\text{w}}$ and the unitary $U(1)$ symmetry operators/defects $\eta_\text{m,w}(\alpha)$ are given by
\begin{equation}
\label{eq_compact boson U1xU1 def}
    \begin{aligned}
        &U(1)_{\text{m}}:~~~ J_\text{m}=\frac{i R^2}{2\pi }\star d\phi~, &\eta_{\text{m}}(\alpha)= \exp(i \alpha\int_{\mathcal{M}_1} J_{\text{m}})~;\\
         &U(1)_{\text{w}}:~~~ J_\text{w}=\frac{1}{2\pi }d\phi~, &\eta_{\text{w}}(\alpha)= \exp(i \alpha\int_{\mathcal{M}_1} J_{\text{w}})~,
    \end{aligned}
\end{equation}
where $\mathcal{M}_1$ is a one-dimensional closed loop in spacetime. 
Both line defects $\eta_\text{m}$ and $\eta_\text{w}$ are topological. 
For the momentum symmetry $U(1)_\text{m}$, it follows from equation of motion $dJ_\text{m}=0$, while for the winding symmetry $U(1)_\text{w}$, it follows from  $dJ_\text{w}= {1\over 2\pi }d^2\phi=0$. 

As operators, both $\eta_\text{m}$ and $\eta_\text{w}$ are conserved. 
The momentum symmetry operator $\eta_\text{m}(\alpha)$ acts on the compact boson field as $\phi\to \phi+\alpha$. 
The winding symmetry operator $\eta_\text{w}(\alpha)$ acts on local fields in a more subtle way, as insertions of $\exp{(i\phi)}$ do not modify $dJ_\text{w}=0$. T-duality \cite{Kikkawa:1984cp, Sakai:1985cs} states that there exists an alternative description of the compact boson theory \eqref{eq_compact boson action}, with the dual action:
\begin{equation}
\label{eq_compact boson action dual}
    S[\tilde{\phi}]=\frac{1}{4\pi R^2}\int_{\mathcal{M}_2} d\tilde{\phi}\wedge\star d \tilde{\phi}~.
\end{equation}
Here $\tilde{\phi}\sim \tilde{\phi}+2\pi$ is the dual compact boson with radius $\tilde{R}=1/R$. The dual action also has a $\tilde{U}(1)_{\text{m}}\times\tilde{U}(1)_{\text{w}}$ global symmetry, and  T-duality identifies the  Noether currents for the dual field with those in \eqref{eq_compact boson U1xU1 def}: 
\begin{equation}
\begin{aligned}
    &\tilde{U}(1)_{\text{m}}:~~~ \tilde{J}_{\text{m}}=\frac{i}{2\pi R^2}\star d \tilde{\phi}= J_\text{w}~;\\
    &\tilde{U}(1)_{\text{w}}:~~~\tilde{J}_{\text{w}}=\frac{1}{2\pi} d \tilde{\phi}= J_\text{m}~.
\end{aligned}
\end{equation}
The original winding symmetry operator is now identified as the momentum symmetry operator in the T-dual picture, $\eta_{\text{w}}(\alpha)=\exp(i\alpha \int J_{\text{w}})=\exp(i\alpha \int \tilde{J}_{\text{m}})$, which shifts the dual field as $\tilde{\phi} \to \tilde{\phi}+\alpha$. 

In summary, $U(1)_\text{m}$ shifts the compact boson field $\phi$, while $U(1)_\text{w}$ shifts the dual field $\tilde{\phi}$. The symmetry $U(1)_\text{m}\times U(1)_\text{w}$ is further enhanced at special values of the radius $R$. For example, when $R=1$, it is enhanced to a larger group $(SU(2)_{\text{L}}\times SU(2)_{\text{R}})/\mathbb{Z}_2$. 
T-duality becomes an automorphism at the self-dual radius $R=\tilde{R}=1$, and is associated with an invertible global symmetry \cite{Harvey:2017rko}. More generally, there are also non-invertible symmetries at rational $R^2 \in \mathbb{Q}^+$, which we discuss below.

\subsection{Duality defect action revisited}
\label{sec_duality defect action revisted}

Having discussed some of the invertible global symmetries, we move on to the non-invertible defects associated with T-duality 
\cite{Fuchs:2007tx,Kapustin:2009av,Thorngren:2021yso,Choi:2021kmx,Niro:2022ctq,Pace:2024oys,Arias-Tamargo:2025xdd}. 
We will first address an ambiguity in defining their actions and then provide a precise formulation. 

We start with the free compact boson $\phi$ on a spatial interval $[0,2\pi]$:
\begin{equation}
\label{eq_2d bulk action}
    S_\text{bulk}[\phi]=\frac{R^2}{4\pi}\int_{S^1_\beta\times[0,2\pi ]}d\phi\wedge\star d\phi~,
\end{equation}
where $\tau\sim \tau+ 2\pi \beta$ is the compactified Euclidean time direction, and $0 \leq x\leq 2\pi$ dentoes the spatial coordinate. We impose the Neumann condition $\left. \partial_x \phi\right|_{x=0,2\pi }=0$ on the two boundaries. We denote the boundary values of the compact boson field by:
\begin{equation}
\label{eq_boundary fields def}
    \phi^+(\tau)\equiv \phi(\tau,x=0 )~,\text{ and }
    \phi^-(\tau)\equiv\phi(\tau,x=2\pi)~.
\end{equation}

We would like to couple the two boundaries to create a line defect that is localized in space and extends in the Euclidean time coordinate.  
The naive action for the duality  defect that glues $x=0$ and $x=2\pi $ together is \cite{Kapustin:2009av,Choi:2021kmx,Niro:2022ctq,Kim:2025zdy}
\begin{equation}
\label{eq_naive defect action}
    \text{``}~\mathcal{D}_{N}~\text{''}: ~~~\text{``}-\frac{i N}{2\pi}\int_{S^1_\beta} \phi^-d\phi^+\text{''}~.
\end{equation}
(We put this expression in quotation marks because of a subtlety discussed below.) 
The level $N \in \mathbb{Z}$ is quantized to ensure gauge invariance under $\phi \sim \phi +2\pi$.  
(See also \cite{Kim:2025tvu} for more general codimension-1 defects in the compact boson CFT.)

A topological defect commutes with the stress-energy tensor. This further requires
\begin{equation}\label{R2N}
R^2=|N|~,
\end{equation}
Hence, for every radius satisfying $R^2\in \mathbb{N}^+$, there is a pair of topological defects $\mathcal{D}_{\pm R^2}$. 
In the following, we will assume \eqref{R2N} unless stated otherwise. 
When rotating to Lorenzian signature with $t=-i\tau$, the defect action \eqref{eq_naive defect action} leads to the following modified Neumann boundary condition:
\begin{equation}
\label{eq_compact boson modified Neumann condition}
\begin{dcases}
\left. \partial_x \phi\right|_{x=2\pi}=\left. \partial_t \phi\right|_{x=0}~,~\phantom{-}~\left. \partial_t \phi\right|_{x=2\pi}=\left. \partial_x \phi\right|_{x=0}~,&\text{if}~N>0~,\\
\left. \partial_x \phi\right|_{x=2\pi}=\left. -\partial_t \phi\right|_{x=0}~,~\,~\left. \partial_t \phi\right|_{x=2\pi}=\left. -\partial_x \phi\right|_{x=0}~,&\text{if}~N<0~.
\end{dcases}
\end{equation}
More compactly, \eqref{eq_compact boson modified Neumann condition} can be written as $\left. d\phi\right|_{x=2\pi}=\left.\sgn{N} \star d\phi\right|_{x=0}$. Combined with the bulk equation of motion $(\partial_t^2-\partial_x^2)\phi=0$, we see that for $N>0$, the left-movers are periodic in space, while the right-movers are anti-periodic.
For $N<0$, the boundary conditions for the left- and right-movers are exchanged.  
For simplicity, we will focus on the $N>0$ case below, and comment on the other case later.

Explicitly, the on-shell mode expansion subject to the modified Neumann condition \eqref{eq_compact boson modified Neumann condition} takes the following form:
\begin{equation}
\label{eq_onshell mode expansion}
    \begin{aligned}
\phi(t,x)=\phi_{0}+\frac{Q_\text{L}}{2N}(t+x)+\frac{i}{\sqrt{2N}}\left(\sum_{n\in \mathbb{Z}/\{0\}} \frac{\alpha_n}{n} e^{-in(t+x)}+\sum_{r\in \mathbb{Z}+1/2}\frac{\bar{\alpha}_r}{r}e^{-ir(t-x)}\right)~,
    \end{aligned}
\end{equation}
where $\alpha_n$ and $\bar{\alpha}_r$ denote the left-moving and right-moving oscillators, respectively, and  $\phi_0\sim \phi_0+2\pi$. 
They obey $[\alpha_n,\alpha_{n}']=n\delta_{n+{n}',0}$ and $[\bar{\alpha}_r,\bar{\alpha}_{{r}'}]=r\delta_{r+{r}',0}$.
Here $Q_\text{L}$ is the conjugate momentum of $\phi_0$. 
The quantization of the eigenvalues of $Q_\text{L}$ will be discussed in detail in section \ref{sec:anomaly}.  
While the canonical quantization of the $\alpha_n$ and $\bar{\alpha}_r$ modes is straightforward, the quantization of the zero mode $Q_\text{L}$ is subtle. Crucially, it suffers from an ambiguity in the inaccurate defect action \eqref{eq_naive defect action}, which we now address.

\begin{figure}[thb]
\centering
\includegraphics[width=.95\textwidth]{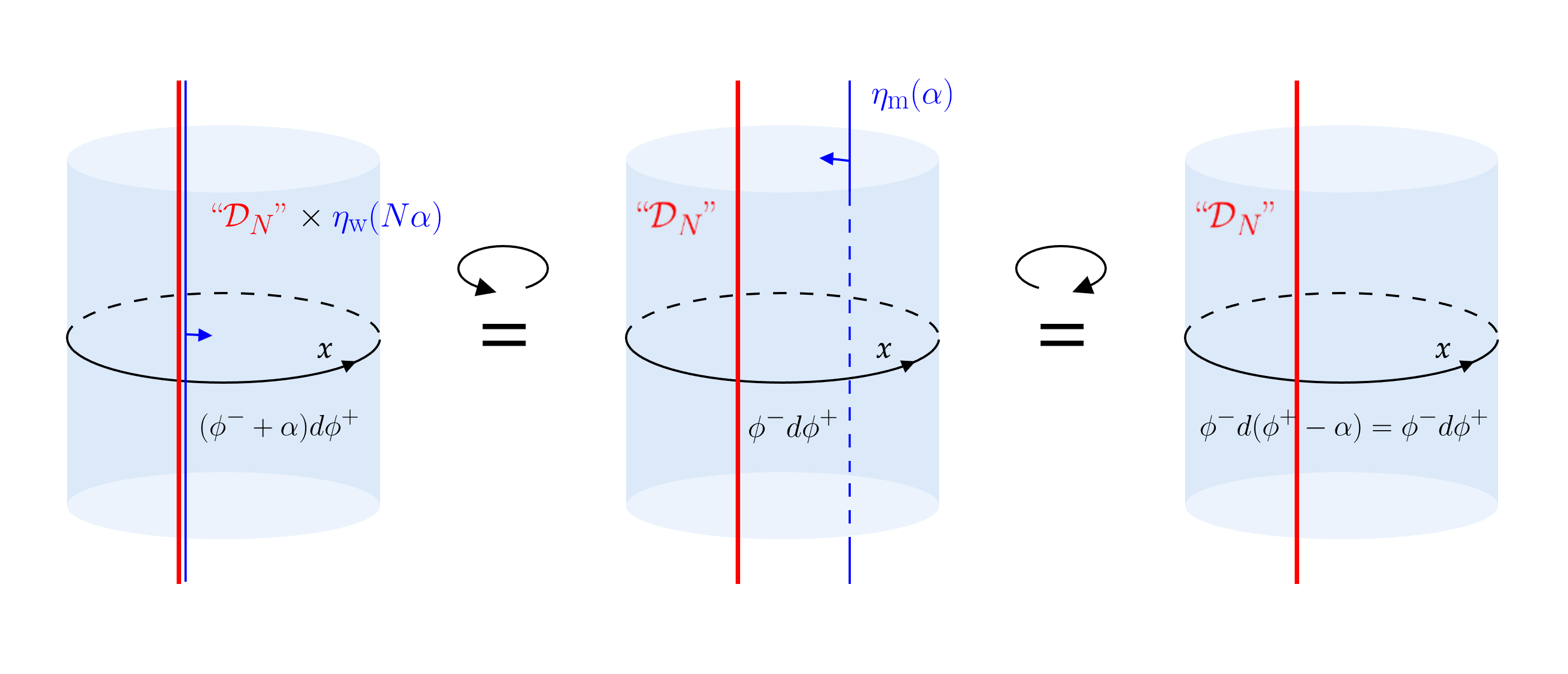}
  \caption{\label{pic_boson puzzle} Inconsistent fusion rule of the naive defect action \eqref{eq_naive defect action}. On the left, we fuse the topological defect $\eta_\text{m}(\alpha)$ to the $x=2\pi$ boundary, and it shifts the boson field by $\phi^-\to \phi^-+\alpha$. On the right, we fuse it to the $x=0$ boundary, and it shifts the field as $\phi^+\to \phi^+-\alpha$. However, they yield different results.} 
\end{figure}

To see the ambiguity, we consider the configuration in figure \ref{pic_boson puzzle} where both ``$\mathcal{D}_{N}$'' and a $U(1)_\text{m}$ symmetry defect $\eta_\text{m}(\alpha)$ are wrapped along the $\tau$-direction. 
However, this leads to the following contradiction. 
On the one hand, merging $\eta_\text{m}(\alpha)$ with the $x=0$ boundary leaves the defect action invariant. On the other hand, merging it with the $x=2\pi$ boundary induces a $U(1)_\text{w}$ defect $\eta_\text{w}(N\alpha)=\exp(\frac{iN\alpha}{2\pi}\int d\phi^+)$. 
See figure \ref{pic_boson puzzle} for details. 
This results in a contradiction: The quantization of the zero mode $Q_\text{L}$ from the latter procedure depends on $\alpha$ (known as the spectral flow), whereas the former does not.  The two procedures lead to different spectra, and therefore, \eqref{eq_naive defect action} can not be correct. 

The inconsistency originates from the fact that $\phi$ is not a gauge-invariant operator, and the naive action \eqref{eq_naive defect action}  depends on a choice of local trivialization, which is unphysical.  
The same issue also arises in Chern-Simons theory, where the naive action is not globally well-defined. 
It is well-known that a precise action can be derived using differential cohomology, which we review in appendix \ref{sec_DB cohomology}. 
For our purpose, since the naive defect action \eqref{eq_naive defect action} is identical to the action for the quantum mechanical system for a particle on a ring \cite{Gaiotto:2017yup}, we can directly apply the precise action derived in \cite{Freed:2006yc,Cordova:2019jnf}.
Let us denote the winding number of $\phi^-$ along the $\tau$-direction as $Q=\frac{1}{2\pi}\int_{S_\beta^{1}} d \phi^- \in \mathbb{Z}$.\footnote{This winding number $Q$ in the Euclidean time direction is not to be confused with $Q_\text{L}$ in \eqref{eq_onshell mode expansion}.} The precise action for the duality defect includes a boundary term:
\begin{equation}
\label{eq_duality defect action 1}
\mathcal{D}_N:~~~S_{\text{defect}}[\phi^-,\phi^+]=-\frac{i N}{2\pi}\int_{0}^{2\pi \beta} d\tau \phi^{-}\partial_\tau \phi^+
+iNQ \phi^+\Big|_{\tau=0}~~~~(\text{Euclidean}).
\end{equation}
Now let us revisit the puzzle in figure \ref{pic_boson puzzle}. 
If we merge $\eta_\text{m}(\alpha)$ to the $x=2\pi$ boundary, this gives a $U(1)_\text{w}$ defect $\eta_\text{w}(N\alpha)=\exp(\frac{iN\alpha}{2\pi}\int d\phi^+)$ as beofre, which equals $\exp(i N Q\alpha)$ because there is no other defect insertion.  
As we merge $\eta_\text{m}(\alpha)$ from the other side to the $x=0$ boundary, it shifts $\phi^+\to \phi^+-\alpha$, and the boundary term in \eqref{eq_duality defect action 1} induces the same defect $\eta_\text{w}(N\alpha)=\exp(i N Q\alpha)$. 
Thus, we conclude that \eqref{eq_duality defect action 1} is the correct action for $\mathcal{D}_N$. 

The preceding discussion leads to the following fusion rule:
\begin{equation}
\label{eq_duality fusion rule 1}
\begin{aligned}
    \eta_{\text{m}}(\alpha)\times \mathcal{D}_{N}=&\mathcal{D}_{N}\times \eta_\text{w}(N\alpha)~,\\
    \mathcal{D}_{N}\times \eta_{\text{m}}(\alpha)=& \eta_\text{w}(N\alpha)\times \mathcal{D}_{N}~.
\end{aligned}
\end{equation}
In particular, it implies that the duality defect $\mathcal{D}_N$ can absorb the $\mathbb{Z}_N\subset U(1)_\text{m}$ momentum symmetry defects:
\begin{equation}
\label{eq_duality fusion corollary}
   \eta_{\text{m}}(\alpha)\times \mathcal{D}_{N}=\mathcal{D}_{N}\times \eta_{\text{m}}(\alpha)=\mathcal{D}_N~,~~~\text{ if  }\alpha\in\frac{2\pi \mathbb{Z}}{N}~.
\end{equation}
Therefore, the duality defect is non-invertible and is not associated with a group-like symmetry when $N>1$.\footnote{Suppose $\mathcal{D}_N$ were invertible, then we multiply $\mathcal{D}_N^{-1}$ on both sides and find $\eta_\text{m}(2\pi/N)=1$, which is a contradiction if $N>1$.} 
In appendix \ref{sec_2d torus partition functions}, we use the precise defect action \eqref{eq_duality defect action 1} to reproduce the fusion rule of two duality defects \cite{Thorngren:2021yso,Choi:2021kmx,Niro:2022ctq}:
\begin{equation}
\label{eq_duality fusion rule 2}
    \mathcal{D}_N\times \mathcal{D}_N=\sum_{n=0}^{N-1}\eta_\text{m}(\frac{2\pi n}{N})~.
\end{equation}
The right-hand side is proportional to the projection operator to the invariant states under $\mathbb{Z}_N\subset U(1)_\text{m}$. 
This $\mathbb{Z}_N$ invertible symmetry together with $\mathcal{D}_N$ forms a fusion category known as the $\mathbb{Z}_N$ Tambara-Yamagami category \cite{Tambara:1998vmj}. 
The $N=2$ case corresponds to the Ising fusion category in the $R=\sqrt{2}$ compact boson CFT, which is obtained by gauging the fermion parity of a free massless Dirac fermion \cite{Fuchs:2007tx,Thorngren:2021yso,Choi:2021kmx,Pace:2024oys}. 

The defect at the opposite level  $\mathcal{D}_{-N}$ is related to $\mathcal{D}_N$ as
\begin{equation}
\label{eq_charge conj defect def}
    \mathcal{D}_{-N}= C\times \mathcal{D}_{N}=\mathcal{D}_{N}\times C~,
\end{equation}
where $C$ is the charge conjugation symmetry that acts as $\phi\to -\phi$.

When $R^2=N=1$, in addition to the $U(1)$ currents $J_{\text{m}}$ and $J_{\text{w}}$, the theory also has the vertex operators $\exp{(\pm i(\phi+\tilde{\phi}))}$ and $\exp{(\pm i(\phi-\tilde{\phi}))}$ as conserved currents, generating the left and right $\mathfrak{su}(2)_1$ Kac-Moody current algebras. 
The global symmetry is $(SU(2)_{\text{L}}\times SU(2)_{\text{R}})/\mathbb{Z}_2$. We use a pair of $SU(2)$ matrices $(g_\text{L},g_\text{R})$, subject to the identification $(g_\text{L}, g_\text{R})\sim (-g_\text{L},-g_\text{R})$, to parametrize a group element. 
In particular, the $U(1)_\text{m}$ and $U(1)_\text{w}$ subgroups are parametrized as:
\begin{equation}
    \eta_{\text{m}}(\alpha)=(e^{\frac{i\alpha}{2}\sigma_z},e^{\frac{i\alpha}{2}\sigma_z})~,~~~\eta_{\text{w}}(\alpha)=(e^{\frac{i\alpha}{2}\sigma_z},e^{-\frac{i\alpha}{2}\sigma_z})~.
\end{equation}
T-duality is associated with two order 4 group elements, $\text{T}_\text{L}=(i\sigma_x,1)$ and $\text{T}_\text{R}=(1,i\sigma_x)$ \cite{Harvey:2017rko}. 
They square to the $\mathbb{Z}_2$ center of the left and right $SU(2)$, which are identified by the quotient, i.e., $\eta_\text{m}(\pi)\eta_\text{w}(\pi)= \text{T}_\text{L}^2 \sim \text{T}_\text{R}^2 $. 
The duality defects $\mathcal{D}_{\pm1}$ are invertible and of order 2. 
They correspond to the following group elements:
\begin{equation}
\label{eq_SU(2) identification}
    \mathcal{D}_1=\eta_\text{w}(\pi)\text{T}_\text{R}=(i\sigma_z,i\sigma_y)~,~~~\mathcal{D}_{-1}=\eta_\text{w}(\pi)\text{T}_\text{L}=(i\sigma_y,-i\sigma_z)~.
\end{equation}
One can verify that the fusion rules of \eqref{eq_SU(2) identification} agree with \eqref{eq_duality fusion rule 1} and \eqref{eq_duality fusion rule 2}, and $\mathcal{D}_{\pm1}$ differ by charge conjugation $C=(i\sigma_x,i\sigma_x)$.

In Lorentzian signature, the defect action becomes 
\begin{equation}
\mathcal{D}_N:~~~S_{\text{defect}}[\phi^-,\phi^+]=\frac{ N}{2\pi}\int dt \phi^{-}\partial_t \phi^+
-NQ \phi^+\Big|_{t=0}~~~~(\text{Lorentzian}),
\end{equation}
which is localized at $x=0$ and extends in the time direction. 
The dual of the defect ${\cal D}_N$ is defined as its image under the canonical CRT operator $\Theta$. 
The $\Theta$ operator acts on the scalar field as
\begin{equation}
\Theta \phi(t,x) \Theta^{-1}  = \phi(-t,-x)\,.
\end{equation}
It follows that the defect $\mathcal{D}_N$ is its own dual, i.e.,\footnote{This is to be contrasted with $\mathcal{D}_{-N}= \bar{\mathcal{D}}_N$ for the duality defect in 4d Maxwell theory as discussed in \eqref{eq_4d orientation reverse}.} 
\begin{align}
\mathcal{D}_N= \bar{\mathcal{D}}_N.
\end{align}

\subsection{An anomaly involving the non-invertible defect}\label{sec:anomaly}

We now demonstrate a mixed 't Hooft anomaly between the non-invertible symmetry $\mathcal{D}_N$ and an invertible $O(2)$ symmetry when $N$ is odd. 
We diagnose this anomaly by identifying a projective $O(2)$ representation in the presence of the $\mathcal{D}_N$ defect. 

This $O(2)$ symmetry can be written as $U(1)_\text{L}\rtimes \mathbb{Z}_2^C$, where $\mathbb{Z}_2^C$ is generated by the charge conjugation operator $C$ and $U(1)_\text{L}$  is a subgroup of $U(1)_\text{m} \times U(1)_\text{w}$, whose current is
\begin{equation}
        U(1)_{\text{L}}:~~~ J_{\text{L}}=J_\text{m}+NJ_{\text{w}}\,.
\end{equation}
In Lorentzian signature, the current is purely left-moving and takes the form   $J_{\text{L}}=\frac{N}{2\pi}(\partial_t\phi+\partial_x\phi)(dt+dx)$. 
The unitary operator for this $U(1)_\text{L}$ symmetry is
\begin{equation}
\eta_{\text{L}}(\alpha)= e^{i \alpha Q_\text{L}}= \eta_{\text{m}}(\alpha)\times \eta_{\text{w}}(N\alpha)\,,
\end{equation}
where we have denoted the charge of this chiral $U(1)_\text{L}$ by $Q_\text{L}=\int_{0}^{2\pi} dx J_{\text{L},x}$, which features in the on-shell mode expansion in   \eqref{eq_onshell mode expansion}. This $O(2)$ symmetry commutes with the non-invertible operator $\mathcal{D}_N$:
\begin{align}
&C\times \mathcal{D}_{N}=\mathcal{D}_{N}\times C \,,\\
& \eta_{\text{L}}(\alpha) \times \mathcal{D}_N= \mathcal{D}_N \times \eta_{\text{L}}(\alpha)\,,
\end{align}
where we have used \eqref{eq_duality fusion rule 1}.\footnote{By contrast, the duality defect $\mathcal{D}_N$ in 4d Maxwell theory only preserves a discrete subgroup of the continuous $U(1)^{(1)}_\text{e}\times U(1)^{(1)}_\text{m}$ one-form global symmetry. See the discussion around \eqref{anyonic}.}

\begin{figure}[thb]
\centering
\includegraphics[width=.95\textwidth]{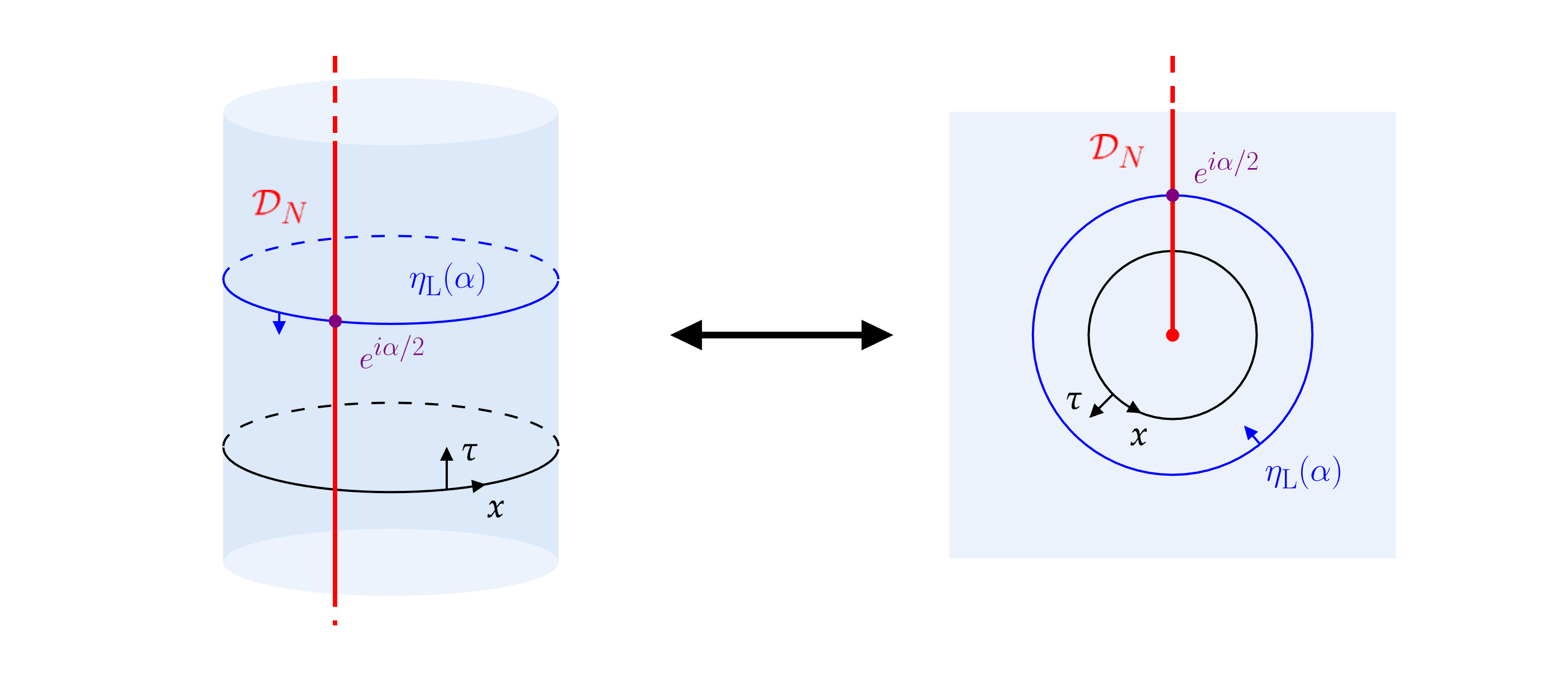}
  \caption{\label{pic_stateoperator and junction}
  Left: The compact boson theory is quantized on a circle with a topological duality defect $\mathcal{D}_N$ insertion. We denote this twisted Hilbert space by ${\cal H}_{{\cal D}_N}$. The charges $Q_\text{L}$ under the chiral $U(1)_\text{L}$ global symmetry in ${\cal H}_{{\cal D}_N}$ are half integers when $N$ is odd. 
  Right: The states in ${\cal H}_{{\cal D}_N}$ on the left are mapped to the twist fields (shown at the origin) attached to the topological defect $\mathcal{D}_N$ under the state/operator correspondence.} 
\end{figure}

Consider the insertion of the topological defect $\mathcal{D}_N$ along the time direction. 
Such an insertion means that the boson field is now subject to the boundary condition in \eqref{eq_compact boson modified Neumann condition}. 
We denote the resulting twisted Hilbert space as ${\cal H}_{{\cal D}_N}$. 
The states in ${\cal H}_{{\cal D}_N}$ are labeled by the charge under the $U(1)_\text{L}$ symmetry and their conformal weights. 
Under the state/operator correspondence, these states in the twisted Hilbert space are mapped to twist fields living at the end of $\mathcal{D}_N$. See figure \ref{pic_stateoperator and junction}.

To understand how the $U(1)_{\text{L}}$ charge is defined in the presence of the defect, we couple the system to a background one-form gauge field $A=A_\tau d\tau+A_x dx$ on the torus. We couple $A$ to the current $J_{\text{L}}=J_\text{m}+NJ_{\text{w}}$ in the bulk action \eqref{eq_2d bulk action} in the standard way:
    \begin{equation}
    \label{eq_bulk gauge coupling}
    S_\text{bulk}[\phi;A]=\frac{R^2}{4\pi}\int_{[0,2\pi\beta]\times[0,2\pi ]}(d\phi-A)\wedge\star (d\phi-A)-\frac{iN}{2\pi }\int_{[0,2\pi\beta]\times[0,2\pi ]}A\wedge(d\phi-A)~,
\end{equation}
where $[0,2\pi\beta]\times [0,2\pi]$ denotes the spacetime torus parametrized by $0\le \tau \le 2\pi\beta$ and $0\le x\le 2\pi$. 

A consistent coupling of the gauge field to the defect action \eqref{eq_duality defect action 1} and the background gauge field $A$ can be determined by 
gauge invariance. The gauge transformation acts on the fields as
\begin{equation}
\label{eq_gaugetransformation}
    \phi\to \phi+\lambda~,~~~A\to A+d\lambda~,~~~Q=\frac{1}{2\pi}\int_{0}^{2\pi\beta} d\tau  \partial_\tau\phi^- \to Q+Q_\lambda~.
\end{equation}
In particular, the boundary fields transform as $\phi^{\pm}(\tau)\to \phi^\pm(\tau)+\lambda(\tau,0)$.
Importantly, the gauge parameter $\lambda(\tau,x)$ is subject to the identification $\lambda\sim \lambda+2\pi$ and need not be a single-valued function. 
It may have a nontrivial winding number in the Euclidean time direction, $Q_\lambda=\frac{1}{2\pi }\int_{0}^{2\pi\beta}d\tau \partial_\tau\lambda \in \mathbb{Z}$. 
Under \eqref{eq_gaugetransformation}, we find the bulk action transforms as follows:
\begin{equation}
\mathtoolsset{multlined-width=0.9\displaywidth}
\begin{multlined}
    S_{\text{bulk}}\to  S_{\text{bulk}}+\frac{iN}{2\pi}\int_{0}^{2\pi \beta}d\tau\lambda(\tau,0)(\partial_\tau \phi^{+}-\partial_\tau \phi^{-})\\
    -iNQ_\lambda(\phi^+-\phi^-)\Big|_{\tau=0}-\frac{iN}{2\pi}\int_{[0,2\pi\beta]\times [0,2\pi]}\lambda dA+2\pi i \mathbb{Z}~,
    \end{multlined}
\end{equation}
where we have used $\lambda(2\pi \beta ,x) -\lambda(0,x) = 2\pi Q_\lambda$. 
For a flat background $dA=0$, the total action $S_\text{bulk}+S_\text{defect}$ remains gauge invariant modulo $2\pi i$ if we introduce a counterterm to the defect action:
\begin{equation}
\label{eq_defect gauge coupling}
\mathtoolsset{multlined-width=0.9\displaywidth}
\begin{multlined}
S_\text{defect}[\phi^-,\phi^+;A]=-\frac{i N}{2\pi}\int_{0}^{2\pi \beta}d\tau \phi^{-}\partial_\tau \phi^++iNQ \phi^+\Big|_{\tau=0}+\frac{i k}{2}\int_{0}^{2\pi \beta}d\tau A_\tau~.
\end{multlined}
\end{equation}
where 
\begin{equation}
k=N\bmod 2.
\end{equation}
Indeed, the gauge transformation of \eqref{eq_bulk gauge coupling} and \eqref{eq_defect gauge coupling} gives
\begin{equation}
    S_\text{bulk}+S_\text{defect}\xrightarrow{dA=0} S_\text{bulk}+S_\text{defect}-i\pi (NQ_\lambda^2-kQ_\lambda)+2\pi i\mathbb{Z}~.
\end{equation}
The second-to-last term is $2\pi i$ times an integer because $NQ_\lambda^2= kQ_\lambda\bmod 2$. 
For a generic background gauge field $A$, the total gauge transformation gives 
\begin{equation}
\label{eq_dual current anomaly}
    S_\text{bulk}+S_\text{defect}\to S_\text{bulk}+S_\text{defect}-\frac{iN}{2\pi}\int _{[0,2\pi\beta]\times[0,2\pi ]}\lambda dA+2\pi i\mathbb{Z}~.
\end{equation}
The second to the last term in \eqref{eq_dual current anomaly} is interpreted as the 't Hooft anomaly of $U(1)_{\text{L}}$, which is canceled by a 3d anomaly inflow action ${iN\over 2\pi}\int AdA$.

Now let us discuss the interesting counterterm ${ik\over2} \int A$ in \eqref{eq_defect gauge coupling}. 
For even $N$, we can set $k=0$ and remove this counterterm. 
Physically, it is interpreted as a heavy charged particle that decouples from the compact boson theory.

For odd $N$, however, this term is nontrivial and cannot be removed by adding any other local, gauge-invariant counterterm. 
It implies that the $U(1)_\text{L}$ charge in the presence of the $\mathcal{D}_N$  defect obeys:\footnote{Strictly speaking, the symmetry operators in the presence and absence of the $\mathcal{D}_N$ defect should be denoted by different symbols. To simplify the notation, we will not make this distinction for $C$, $Q_\text{L}$, and $\eta_\text{L}(\alpha)=\exp(i\alpha Q_\text{L})$ in what follows. See also footnote \ref{fn:modified}.}
\begin{equation}\label{eq_fractional}
Q_\text{L}\in 
\begin{cases}
\mathbb{Z}~~~~~~~~\,~\text{in}~{\cal H}_{{\cal D}_N}~\text{for even $N$} \\
\mathbb{Z}+\frac{1}{2}~~~~\text{in}~{\cal H}_{{\cal D}_N}~\text{for odd $N$}
\end{cases}
\,.
\end{equation}
For odd $N$, one might be tempted to shift the charge $Q_\text{L}$ uniformly by $1/2$ so that the charges are integral. 
However, such a shift is prohibited once we impose charge conjugation symmetry $C$ and the relation $CQ_\text{L}= - Q_\text{L} C$. 
The fractional charge in \eqref{eq_fractional} leads to $\eta_\text{L}(2\pi) = -1$, implying that the $O(2)$ symmetry is realized projectively in the presence of the $\mathcal{D}_N$ defect. 
We interpret this as a mixed anomaly between the non-invertible symmetry ${\cal D}_N$ and the chiral $O(2)$ symmetry. 
Note that these two symmetries are also separately anomalous:  the  $O(2)$ symmetry has an ordinary 't Hooft anomaly of a chiral symmery (as can be seen in \eqref{eq_dual current anomaly}), while the anomaly of the non-invertible symmetry $\mathcal{D}_N$ has been discussed in \cite{Tambara:2000qlz,Chang:2018iay,Thorngren:2019iar,Choi:2023xjw,Zhang:2023wlu}.

In particular, for odd $N$, the operators
$C$ and $\eta_\text{L}(\pi)$ anticommute on ${\cal H}_{{\cal D}_N}$:
\begin{equation}\label{Z2Z2}
C \eta_\text{L}(\pi ) C^{-1} = \eta_\text{L}(- \pi) = -\eta_\text{L}(\pi)~~~\text{in ${\cal H}_{{\cal D}_N}$ for odd $N$}\,,
\end{equation}
where we have used $\eta_\text{L}(2\pi)= -1$. 
This means that the $\mathbb{Z}_2\times \mathbb{Z}_2$ subgroup of $O(2)$ generated by $C$ and $\eta_\text{L}(\pi)$ is also realized projectively. 
The projectivity is labeled by the nontrivial element in $H^2(\mathbb{Z}_2\times \mathbb{Z}_2,U(1))=\mathbb{Z}_2$. 
The algebra \eqref{Z2Z2} does not admit a one-dimensional representation, thus enforcing a two-fold degeneracy in ${\cal H}_{{\cal D}_N}$ at every energy level.

More generally, we can view this anomaly of $O(2)$ (or its $\mathbb{Z}_2\times \mathbb{Z}_2$ subgroup) as a \textit{defect 't Hooft anomaly} \cite{Antinucci:2024izg,Komargodski:2025jbu} in the presence of the defect $\mathcal{D}_N$.  
This perspective allows us to consider general bulk or defect deformations that preserve the $\mathbb{Z}_2\times \mathbb{Z}_2$ symmetry operators in the presence of $\mathcal{D}_N$.
Importantly, such deformations typically render the $\mathcal{D}_N$ defect non-topological. 
Nonetheless, since the projective algebra in 
\eqref{Z2Z2} is robust under continuous deformations, any such deformation cannot lift the 2-fold degeneracy. 
For instance, one can deform the bulk Lagrangian by the following $\mathbb{Z}_2\times \mathbb{Z}_2$-symmetric potential 
\begin{equation}
   V_{\text{int}}(\phi,\tilde{\phi})=\sum_{n\in \mathbb{Z}} a_n\cos{(n(\phi+N\tilde{\phi}))}+\sum_{n\in \mathbb{Z}}b_n\cos{(n(N\phi-\tilde{\phi}))}~.
\end{equation}
This provides a nontrivial application of the anomaly involving $\mathcal{D}_N$. 

\subsection{Twist fields for non-invertible duality symmetries}
\label{sec_twisted sector operators}

We now proceed to compute the spectrum of the Hilbert space ${\cal H}_{{\cal D}_N}$ twisted by the topological duality defect $\mathcal{D}_N$. 
By the state/operator correspondence, these states correspond to the twist fields for $\mathcal{D}_N$. 
Canonical quantization of the on-shell modes in \eqref{eq_onshell mode expansion} gives the following spectrum of states:
\begin{equation}
|\{m_n\},\{m_r\}, Q_\text{L}\rangle=
\prod_{n\in \mathbb{N}^+} \alpha_{-n}^{m_n} \prod_{r\in \mathbb{N}+\frac12} \bar\alpha_{-r}^{m_r} |Q_\text{L}\rangle\,,
\end{equation}
where $|Q_\text{L}\rangle$ is a highest-weight state with eigenvalue $Q_\text{L}$ under the charge operator for $U(1)_\text{L}$, which obeys \eqref{eq_fractional}.
The conformal weights of these states are therefore given by
\begin{equation}
h-{1\over24}= \frac{1}{2}(H+P)=\frac{Q_\text{L}^2}{4N}+\sum_{n\in \mathbb{N}^+}\alpha_{-n}\alpha_{n}-\frac{1}{24}~,
\end{equation}
and
\begin{equation}
\bar h-{1\over24}= \frac{1}{2}(H-P)=\sum_{r\in \mathbb{N}+\frac{1}{2}}\bar{\alpha}_{-r}\bar{\alpha}_{r}+\frac{1}{48}~,
\end{equation}
where $H$ is the Hamiltonian and $P$ is the momentum operator along the $x$-direction. 
This leads to the partition function:
\begin{equation}
    \label{eq_torus partition function 1}\mathcal{Z}_{\mathcal{D}_N}=\text{tr}_{\mathcal{H}_{\mathcal{D}_N}}(e^{\pi i \uptau (H+P)}e^{-\pi i \bar{\uptau}(H-P)})=\begin{dcases}
    \frac{\vartheta_2(\uptau/2N)\bar{\vartheta}_2(\uptau/2)}{2 |\eta(\uptau)|^2}~,& \text{for odd } N\\
    \frac{\vartheta_3(\uptau/2N)\bar{\vartheta}_2(\uptau/2)}{2 |\eta(\uptau)|^2}~,& \text{for even } N        
    \end{dcases}~,
\end{equation}
where $\uptau$ is the modular parameter for the spacetime torus.\footnote{We use $\uptau$ for the modular parameter of the spacetime torus, and $\tau$ for the Euclidean time. We hope this will not cause any confusion.} 
This agrees with the result in \cite{Thorngren:2021yso}, which is obtained using modular covariance. In the special case when $\uptau$ is purely imaginary, this also agrees with the Euclidean path integral calculation in appendix \ref{sec_2d torus partition functions} using the precise defect action \eqref{eq_duality defect action 1}.

Since the topological defect $\mathcal{D}_N$ preserves the holomorphic $\mathfrak{u}(1)$ current algebra corresponding to $U(1)_\text{L}$ and the antiholomorphic Virasoro algebra, we can expand the partition functions on the corresponding torus characters, which are  $e^{2\pi i \uptau (h-\frac{1}{24}) -2\pi i \bar\uptau (\bar h-\frac{1}{24})}/ |\eta(\uptau)|^2$.\footnote{When $N$ is even, $\mathcal{D}_N$ further commutes with the extended $\mathfrak{u}(1)_{2N}$ current algebra  \cite{Thorngren:2021yso}. Here for the left movers, we only expand the partition function with respect to the (unextended) $\mathfrak{u}(1)$ characters, which is $e^{2\pi i \uptau (h-1/24)}/\eta(\uptau)$ (including $h=0$). For the right movers, even though the Virasoro multiplets have extra null states at $\bar h=  n^2/4$ for $n\in \mathbb{Z}$ at $c=1$ (see, e.g., \cite{Ginsparg:1988ui}), this never occurs in our partition functions so the character is always $e^{-2\pi i\bar \uptau (\bar h-1/24)}/\eta(\bar\uptau)$.} We find
\begin{equation}
\mathcal{Z}_{\mathcal{D}_N} = 
{1\over 2|\eta(\uptau)|^2}
\sum_{s \in \mathbb{Z}+\frac12, Q_\text{L}} 
e^{2\pi i \uptau \, {q_\text{L}^2\over 4N}}
e^{-2\pi i \bar\uptau \, { s^2\over4}} \,,
\end{equation}
where the range of $Q_\text{L}$ is over $\mathbb{Z}$ for even $N$ and over $\mathbb{Z}+\frac12$ for odd $N$. 
We conclude that the $\mathfrak{u}(1)\times \overline{\text{Vir}}$ primaries for the twist fields are labeled by a half-integer $s\in \mathbb{Z}+\frac12$ and by $Q_\text{L}$ in the range above, with conformal weights given by 
\begin{equation}
\label{eq_2d twist primary}(h,\bar{h})=\left(\frac{Q_\text{L}^2}{4N},\frac{s^2}{4}\right)~.
\end{equation}

\subsection{Non-invertible duality symmetries at rational $R^2$}
\label{sec_2d general duality defects}

In this section, we discuss generalizations of the topological duality defect \eqref{eq_duality defect action 1} and their twist fields at rational $R^2$. 
Given two coprime positive integers $N_{\text{e}}$ and $N_{\text{m}}$, we define a line defect in spacetime by the following action:
\begin{equation}
\label{eq_duality defect action 2}
\mathtoolsset{multlined-width=0.9\displaywidth}
\begin{multlined}
\mathcal{D}_{N_\text{e}/N_\text{m}}:~S_\text{defect}[\phi^-,\phi^+;\varphi_1,\varphi_2]=-\frac{i N_\text{e}}{2\pi}\int_{0}^{2\pi \beta} d\tau \phi^{-}\partial_\tau \varphi_1\hfill\\
+\frac{iN_\text{m}}{2\pi} \int_{0}^{2\pi \beta} d\tau \varphi_1\partial_\tau \varphi_2-\frac{i}{2\pi}\int_{0}^{2\pi \beta}d\tau \phi^{+}\partial_\tau \varphi_2\\
\hfill+iN_\text{e}Q \varphi_1\Big|_{\tau=0}-iN_\text{m}{Q}'_1 \varphi_1\Big|_{\tau=0}+iQ\varphi_2\Big|_{\tau=0}~,
\end{multlined}
\end{equation}
where $\varphi_i(\tau)\sim \varphi_i(\tau)+2\pi$ with $i=1,2$ are two auxiliary compact scalar fields we introduced along the defect worldline. 
We have added three boundary terms in the last line in addition to the naive defect action presented in  \cite{Niro:2022ctq}. 
Here ${Q}'_{1}=\frac{1}{2\pi }\int_{S^1_\beta}d\varphi_{1}\in \mathbb{Z}$ is the winding number of $\varphi_1$ along the $\tau$-direction.\footnote{The coprime condition $\gcd(N_\text{e}, N_\text{m})=1$ is necessary for the duality symmetry defect \eqref{eq_duality defect action 2} to be simple, i.e., it cannot be written as a direct sum of other defects. See  \eqref{eq_app_general defect action 123} and \eqref{eq_app_semisimple cond}.}  
Naively integrating out $\varphi_1$ and $\varphi_2$ yields an effective defect action \eqref{eq_app_mode expansion}
\begin{equation}
``\mathcal{D}_{N_\text{e}/N_\text{m}}":~~~
S_{\text{defect}}[\phi^-,\phi^+]\sim `` ~
-\frac{iN_\text{e}}{2\pi N_\text{m}}\int_{^1_\beta} \phi^-d\phi^+ ~"~,
\end{equation}
However, this effective action is not gauge invariant and only serves as an intuitive way to understand the more precise, gauge-invariant action in \eqref{eq_duality defect action 2}. 
This duality defect is topological when $R^2=N_\text{e}/N_{\text{m}}\in \mathbb{Q}^+$, which we will assume for the rest of this subsection. For the special case $N_\text{e}=N$ and $N_{\text{m}}=1$, the auxiliary fields $\varphi_1$ and $\varphi_2$ can be globally integrated out and the action reduces to \eqref{eq_duality defect action 1}.

The generalization of the fusion rule \eqref{eq_duality fusion rule 2} is 
 \begin{equation}
 \label{eq_duality fusion rule 3}
    \begin{aligned}
    \eta_{\text{m}}(N_\text{m}\alpha)\times \mathcal{D}_{N_\text{e}/N_\text{m}}=&\mathcal{D}_{N_\text{e}/N_\text{m}}\times \eta_{\text{w}}(N_\text{e}\alpha)~,\\
    \mathcal{D}_{N_\text{e}/N_\text{m}}\times \eta_{\text{m}}(N_\text{m}\alpha)=&\eta_{\text{w}}(N_\text{e}\alpha)\times \mathcal{D}_{N_\text{e}/N_\text{m}}~,
    \end{aligned}
\end{equation}
which we refer to appendix \ref{sec_2d torus partition functions} for further details. The fusion of two duality defects $\mathcal{D}_{N\text{e}/N_{\text{m}}}$ is
\begin{equation}
\label{eq_duality fusion rule 4}
\mathcal{D}_{N_\text{e}/N_\text{m}}\times \mathcal{D}_{N_\text{e}/N_\text{m}}=\mathcal{C}_{N_\text{e},N_\text{m}}~.
\end{equation}
where
\begin{equation}
    \label{eq_2d condensation defect}
    \mathcal{C}_{N_\text{e},N_\text{m}}\equiv\left(\sum_{n=0}^{N_\text{e}-1}\eta_\text{m}(\frac{2\pi n}{N_\text{e}})\right)\times\left(\sum_{m=0}^{N_\text{m}-1}\eta_\text{w}(\frac{2\pi m}{N_\text{m}})\right)~, 
\end{equation}
is a projection operator that satisifes $\mathcal{C}_{N_\text{e},N_\text{m}}\times \mathcal{C}_{N_\text{e},N_\text{m}}=N_\text{e}N_\text{m}\mathcal{C}_{N_\text{e},N_\text{m}}$. 
It can be viewed as the condensation defect for an ordinary symmetry \cite{Roumpedakis:2022aik}. 

Equation \eqref{eq_duality fusion rule 3} implies that $\mathcal{D}_{N_\text{e}/N_\text{m}}$ commutes with the $U(1)_\text{L}$ symmetry whose conserved current is: 
\begin{equation}
\label{eq_U1 subgroup}
J_\text{L}=N_\text{m}J_\text{m}+N_\text{e}J_{\text{w}}~.
\end{equation}
The unitary operator for this $U(1)_\text{L}$ is
\begin{equation}
\eta_{\text{L}}(\alpha)= \eta_{\text{m}}(N_\text{m}\alpha)\times \eta_{\text{w}}(N_\text{e}\alpha)~.
\end{equation}
When $R^2=N_\text{e}/N_\text{m}$, we note that $J_\text{L}$ is a left-moving chiral current and the level of $U(1)_\text{L}$ is $2N_\text{e}N_\text{m}$.

When $R^2=N_\text{e}/N_\text{m}$, the torus partition function with $\mathcal{D}_{N_\text{e}/N_\text{m}}$ inserted along the time direction is found to be
\begin{equation}
\label{eq_torus partition function 2}
\mathcal{Z}_{\mathcal{D}_{N_\text{e}/N_\text{m}}}=\begin{dcases}   \frac{\vartheta_2(\uptau/2N_\text{e}N_\text{m})\bar{\vartheta}_2(\uptau/2)}{2 |\eta(\uptau)|^2}~,& \text{for odd } N_\text{e} N_\text{m}\\
\frac{\vartheta_3(\uptau/2N_\text{e}N_\text{m})\bar{\vartheta}_2(\uptau/2)}{2 |\eta(\uptau)|^2}~,& \text{for even } N_\text{e} N_\text{m}        
    \end{dcases}~.
\end{equation}
The $\mathfrak{u}(1)\times \overline{\text{Vir}}$ primaries of twist fields for $\mathcal{D}_{N_\text{e}/N_\text{m}}$ are therefore labeled by $s\in \mathbb{Z}+\frac12$, and the $U(1)_{2N_\text{e}N_\text{m}}$ charge $Q_\text{L}$, which is an integer if $N_\text{e}N_\text{m}$ is even and a half-integer if $N_\text{e}N_\text{m}$ is odd. Their conformal weights are
\begin{equation}
\label{eq_2d twist primary2}
(h,\bar h ) = 
\left( 
{Q_\text{L}^2\over 4N_\text{e}N_\text{m}}, {s^2\over4}
\right)\,.
\end{equation}
This reduces to the results in section \ref{sec_twisted sector operators} when $N_\text{e}=N$ and $N_\text{m}=1$. 

Fusing \eqref{eq_duality defect action 2} with the charge conjugation symmetry operator $C$ gives another defect:
\begin{equation}
    \mathcal{D}_{-N_\text{e}/N_\text{m}}\equiv C\times \mathcal{D}_{N_\text{e}/N_\text{m}}=\mathcal{D}_{N_\text{e}/N_\text{m}}\times C~.
\end{equation}
The twist fields of $\mathcal{D}_{-N_\text{e}/N_\text{m}}$ can be obtained by exchanging the left- and right-moving sectors in \eqref{eq_torus partition function 2} and \eqref{eq_2d twist primary2}. 

\subsection{Conformal line defects at generic $R^2$}

Finally, let us consider the case where $R^2 \neq N_\text{e}/N_\text{m}$. 
In this case, the defect $\mathcal{D}_{N_\text{e}/N_\text{m}}$ is no longer topological and instead becomes conformal. 
Note that the coupling constants on the defect are fixed integers, $N_\text{e}, N_\text{m}$, and therefore cannot flow as the bulk exactly marginal parameter $R$ is varied. 
This conformal line defect and its fusion were studied in \cite{Bachas:2007td, Becker:2017zai}. 

In appendix \ref{sec_2d torus partition functions} we compute the torus partition function with the insertion of the conformal defect $\mathcal{D}_{N_\text{e}/N_\text{m}}$. Since the stress-energy tensor is not conserved across the defect when $R^2\neq N_\text{e}/N_\text{m}$, the momentum $P$ along the $x$-direction is no longer a conserved quantity that can be used to diagonalize the states in the defect Hilbert space. 
The defect partition function is:
\begin{equation}
\label{eq_torus partition function 3}
\mathcal{Z}=\begin{dcases}
    \frac{\vartheta_2(\uptau/2)}{2 \eta(\uptau)^2}\vartheta_2\left(\frac{\uptau}{(N_\text{m}R)^2+(N_\text{e}/R)^2}\right)~,& \text{for odd } N_\text{e} N_\text{m}\\
    \frac{\vartheta_2(\uptau/2)}{2 \eta(\uptau)^2}\vartheta_3\left(\frac{\uptau}{(N_\text{m}R)^2+(N_\text{e}/R)^2}\right)~,& \text{for even } N_\text{e} N_\text{m}
\end{dcases}~,
\end{equation}
where the torus modulus $\uptau$ is purely imaginary. 
For even $N_\text{e} N_\text{m}$, the ground state in the defect Hilbert space is unique, with Casimir energy $E_{0}=-\frac{1}{48}$. 
For odd $N_\text{e} N_\text{m}$, the ground states are 2-fold degenerate, and the Casimir energy is 
\begin{equation}
    E_0=\frac{1}{8}\left(\frac{1}{(N_\text{m}R)^2+(N_\text{e}/R)^2}-\frac{1}{6}\right)~,~~~\text{for odd}~N_\text{e} N_\text{m}~.
\end{equation}
This degeneracy is enforced by the defect 't Hooft anomaly of the $\mathbb{Z}_2\times \mathbb{Z}_2$ symmetry generated by $\eta_\text{L}(\pi)$ and $C$ in the presence of the conformal defect $\mathcal{D}_{N_\text{e}/N_\text{m}}$, similar to the discussion in section \ref{sec:anomaly}. 

The $g$-function \cite{Affleck:1991tk, Cuomo:2021rkm} for the defect can be obtained from the partition function \eqref{eq_torus partition function 3}, which is
\begin{equation}
g(\mathcal{D}_{N_\text{e}/N_\text{m}})=\sqrt{\frac{(N_\text{m}R)^2+(N_\text{e}/R)^2}{2}}~.
\end{equation}

\section{Discussions and outlook}

We highlight a few future directions and open questions.
\begin{enumerate}
    \item \textbf{Defect conformal anomalies.} It would be interesting to compute the Weyl anomalies and the gravitational anomalies for the twist defect discussed in section \ref{sec_4d Maxwell theory}. In particular, one of the Weyl anomaly terms  \cite{Graham:1999pm, Henningson:1999xi, Schwimmer:2008yh} is proportional to the intrinsic Riemann curvature of the surface defect, whose coefficient defines the $b$-function \cite{Jensen:2015swa}. The $b$-function is an RG monotone that plays an important role in understanding the defects. Likewise, the defect gravitational anomalies can be investigated along the lines of~\cite{Wang:2020xkc}, potentially offering a diagnostic of the chiral nature of our twist defect.

    Another stimulating question is whether defect conformal anomalies can be directly extracted from physical observables, such as the Casimir energy and momentum in the presence of the defect (see discussion in footnote \ref{ft_Casimir} and around equation \eqref{eq_Casimir momentum}). By analogy with the chiral edge modes in 3d Chern-Simons theory, it is natural to conjecture that the Casimir momentum \eqref{eq_Casimir momentum} is proportional to the defect gravitational anomaly. We leave a detailed investigation of this conjecture to future work. For a relevant study, see \cite{Huang:2025eze}.
    
    \item \textbf{Twist defects for the Montonen–Olive duality in supersymmetric gauge theories.} Non-invertible duality symmetries also appear in 4d $\mathcal{N}=4$ super Yang-Mills theories \cite{Kaidi:2021xfk, Kaidi:2022uux, Choi:2022zal,Choi:2022rfe}, and it would be interesting to investigate the corresponding twist surface defects and their holographic duals.

    This future direction also presents an interesting challenge. The Montonen–Olive duality acts on the supercharges by phases \cite{Kapustin:2006pk}. Consequently, the twist defect associated with this duality symmetry breaks all supersymmetries in flat spacetime. To define a BPS twist defect, we need to turn on a nontrivial background connection for the $R$-symmetry, such that the phases are canceled. Since the $R$-symmetry current is in the same supermultiplet as the stress-energy tensor, this indicates that BPS twist defects associated with the Montonen–Olive duality can only be defined in a conical spacetime. 
    
    \item \textbf{Twist defects for electromagnetic duality in higher-form $U(1)$ gauge theories.} Electromagnetic duality can be generalized to $p$-form $U(1)$ gauge theories in $(2p+2)$-dimensional spacetime. In this paper, we have analyzed the non-invertible duality symmetry and its associated twist defect in 2d free compact boson ($p=0$) and 4d Maxwell theory ($p=1$). This framework naturally generalizes to cases with $p\geq 2$. We comment that when the $p$-form $U(1)$ gauge theory is self-dual, the electromagnetic duality as an invertible global symmetry forms a $\mathbb{Z}_2$ group for $p\in 2\mathbb{N}$, while it forms a $\mathbb{Z}_4$ group for $p\in 2\mathbb{N}+1$. We have seen that 2d free compact boson and 4d Maxwell theory are the most elementary examples in these two classes. We leave it to future works to explore anomalies, DCFT spectrum, and other aspects of these twist defects in higher-form gauge theories.

    \item \textbf{Phenomenological implications}.  It is intriguing to contemplate whether our twist defect in the Maxwell theory has any phenomenological relevance to the real world. Since our physical world does not have electromagnetic duality, such a defect can only arise in effective models or as an approximation. 
    On experimentally accessible scales, our twist defect potentially models the chiral interaction between free photons and certain optical devices. In this setting, equation \eqref{eq_defect vector primary} manifests a spin-momentum lock of photons propagating along the defect. It will be interesting to explore applications of this mechanism in quantum optics and condensed matter systems. One may also postulate strings described by our twist defect on cosmological scales. However, given the abundance of electrons in our universe and the absence of observed magnetic monopoles,  such strings are unlikely to pervade the cosmological space.

\end{enumerate}

\acknowledgments
We are especially grateful to Yichul Choi and Zohar Komargodski for many stimulating and helpful discussions on this project. 
We also thank Gabriel Cuomo, Shota Komatsu, Andrew O'Bannon,  Leonardo Rastelli, Avia Raviv-Moshe, Brandon Rayhaun, and Yunqin Zheng for interesting discussions. 
We thank Chris Herzog for thoughtful comments on the draft. 
SHS is supported by the
Simons Collaboration on Ultra-Quantum Matter, which is a grant from the Simons Foundation
(651444, SHS). SHS is also supported in part by NSF grant PHY-2449936. 
 SZ is supported
in part by the Simons Foundation grant 488657 (Simons
Collaboration on the Non-Perturbative Bootstrap), the
BSF grant no.\ 2018204 and NSF award number 2310283. 
Part of this work was completed during the Kavli Institute for Theoretical Physics (KITP) program “Generalized Symmetries in Quantum Field Theory: High Energy
Physics, Condensed Matter, and Quantum Gravity”, which is supported in part by grant NSF
PHY-2309135 to the KITP. 
The authors of this paper are ordered alphabetically. 

\appendix

\section{Differential cohomology}
\label{sec_DB cohomology}

In this appendix, we review differential cohomology \cite{deligne1971theorie,cheeger2006differential,Hopkins:2002rd}, an elegant mathematical framework that combines the local field fluctuations with nontrivial topology. 
It provides a mathematical formalism that underpins the defect actions in \eqref{eq_maxwell precise defect} and \eqref{eq_duality defect action 1}. 
Our exposition below will not be mathematically rigorous; rather we will follow a more physicist-friendly approach as in \cite{Alvarez:1984es, Freed:2006yc,Thuillier:2015vma, Delcamp:2019fdp, Cordova:2019jnf,Chen:2019mjw}. 
(See also \cite{Sulejmanpasic:2019ytl,Chen:2019mjw,Gorantla:2021svj} for a lattice version of this formalism.)
For simplicity, we illustrate the construction using the compact boson $\phi$ (a $0$-form) and the Maxwell gauge field $A$ (a $1$-form), which are the most relevant cases for this work.

We first introduce the notions of an open cover and a polyhedral decomposition of a closed smooth $d$-dimensional manifold $\mathcal{M}_d$. We assign to $\mathcal{M}_d$ a collection of open subsets $\{U_i\}$, such that their union satisfies $\cup_{i}U_i=\mathcal{M}_d$. 
We assume the labels $i$ are ordered.  The collection $\{U_i\}$ thereby defines an ordered open cover of the manifold $\mathcal{M}_d$. We also denote the intersection between $p$ different patches as follows:
\begin{equation}
    U_{i_0i_1...i_p}\equiv U_{i_1}\cap U_{i_2}\cap ...\cap U_{i_p}\text{, where } i_0<i_1<...<i_p~.
\end{equation}
A polyhedral decomposition of  $\mathcal{M}_d$ is defined with respect to a choice of an open cover $\{U_i\}$. The top decomposition component $\omega^{(d)}_{i}$ is a $d$-chain compactly supported on the subset $U_i$. These top components satisfy $\cup_{i}\omega^{(d)}_{i}=\mathcal{M}_d$. Other components are defined by the iteration relation:
\begin{equation}
    \omega^{(d-p)}_{i_0i_1...i_p}\subset U_{i_0i_1...i_p}\text{ and } \partial \omega^{(d-p)}_{i_0i_1...i_p}=\sum_{i_{p+1}}\omega^{(d-p-1)}_{i_0i_1...i_p i_{p+1}}~,
\end{equation}
where $\omega^{(d-p)}_{i_0i_1...i_p}$ is a $(d-p)$-chain endowed with an orientation. 
The indices of the chains are totally antisymmetric, i.e., 
\begin{equation}
    \omega^{(d-p)}_{i_0...i_a...i_b...i_p}=(-1)^{a+b}\omega^{(d-p)}_{i_0...i_b...i_a...i_p}~,
\end{equation}
where the minus sign corresponds to reversing the orientation of the simplex. The collection of these chains $\{\omega^{(d)}_{i},~\omega^{(d-1)}_{i_0i_1}~,...\}$ defines a polyhedral decomposition of the manifold $\mathcal{M}_d$.

\subsection{Particle on a ring}
\begin{figure}[thb]
\centering
\includegraphics[width=.85\textwidth]{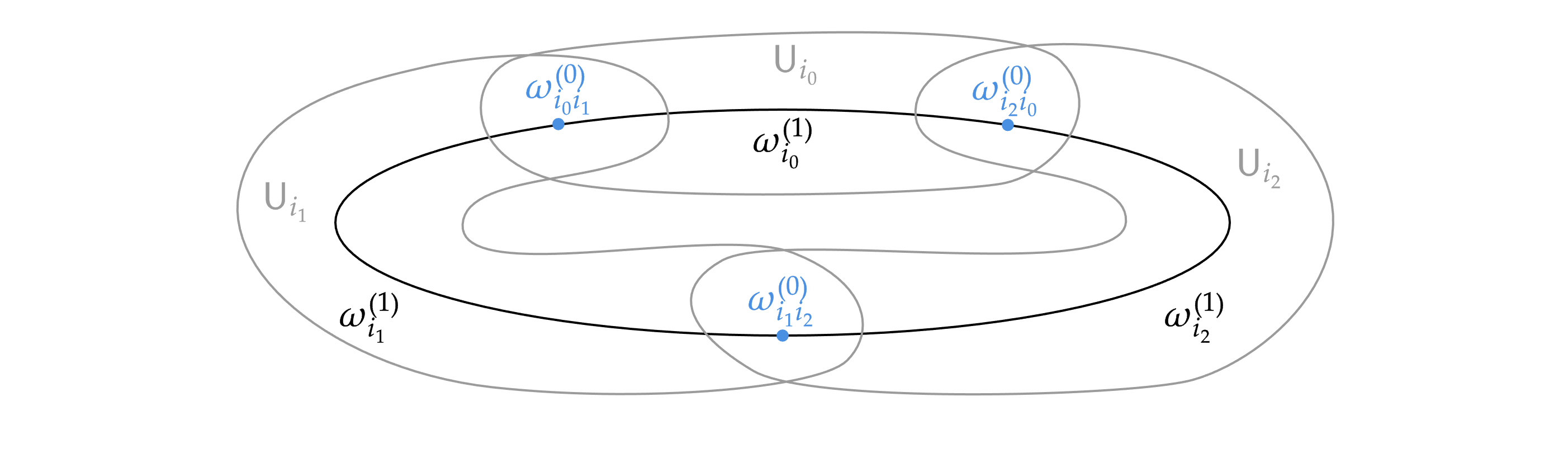}
  \caption{A polyhedral decomposition of $\mathcal{M}_1\cong S^1$.} \label{fig_app}
\end{figure}

Our first example is the quantum mechanics of a particle on a ring. 
Let $\mathcal{M}_1\cong S^1$ be the 1d manifold for the Euclidean time. 
In this case each $\omega^{(1)}_i$ is a line segment in $\mathcal{M}_1$ and each $\omega^{(0)}_{i_0i_1}$ is a point. See figure \ref{fig_app}. 
The coordinate field is a compact scalar field $\phi \in S^1$. 
More precisely, it is characterized by the 2-tuple $(\phi_{i}, n^\phi_{i_0i_1})$.  
In each patch, $\phi_i: U_i\to \mathbb{R}$ is a real-valued local lift of $\phi$ that captures the local fluctuation of the field. 
In each intersection of two patches, $n^\phi_{i_0i_1}: U_{i_0i_1}\to \mathbb{Z}$ is an integer-valued transition function. $(\phi_{i}, n^\phi_{i_0i_1})$ is subject to the cocycle condition
\begin{equation}
\label{eq_appendix_0-form cocycle condition}
\begin{aligned}
    U_{i_0i_1}:&\ \phi_{i_0}-\phi_{i_1}=2\pi n_{i_0i_1}^\phi~,\\
    U_{i_0i_1i_2}:&\ n_{i_0i_1}^\phi+n_{i_1i_2}^\phi+n_{i_2i_0}^\phi=0
    \end{aligned}
\end{equation}
The first condition ensures that $d\phi_{i_0}$ and $d\phi_{i_1}$ agree in $U_{i_0i_1}$. 
We therefore denote this continuous 1-cochain simply as $d\phi$. 
We identify two tuples of $(\phi_{i}, n^\phi_{i_0i_1})$ that differ by a gauge transformation,
\begin{equation}
\label{eq_appendix_0-form gauge transformation}
\begin{aligned}
    U_{i}:&\  \phi_i\to \phi_i+2\pi m^\phi_i~,\\
    U_{i_0i_1}:&\  n^\phi_{i_0i_1}\to n^\phi_{i_0i_1}+m_{i_0}^\phi-m_{i_1}^\phi~,
\end{aligned}    
\end{equation}
with gauge parameter $m^\phi_i :U_i \to \mathbb{Z}$.

The winding number $Q$ in Euclidean time is defined as:
\begin{equation}
Q= {1\over 2\pi}\int_{\mathcal{M}_1} d\phi
= {1\over 2\pi}\sum_i \int_{\omega_i^{(1)} } d\phi_i
=\sum_{i_0<i_1}\int_{\omega^{(0)}_{i_0i_1}}n^\phi_{i_0i_1}
~,
\end{equation}
which is manifestly an integer. Furthermore, it is gauge invariant under \eqref{eq_appendix_0-form gauge transformation}. 

Next, consider two compact scalar fields  $(\phi_i,n_{i_0i_1}^\phi)$ and $(\varphi_i,n_{i_0i_1}^\varphi)$.
The naive action ${iN\over2\pi} \int_{\cal M} \phi d\varphi$, which is used in \eqref{eq_naive defect action}, is more precisely  defined as \cite{Freed:2006yc, Cordova:2019jnf}
\begin{equation}
\label{eq_appendix_0-form DB product}
\frac{iN}{2\pi}\int_{\mathcal{M}_1}\phi d \varphi\equiv\frac{iN}{2\pi}\sum_{i}\int_{\omega^{(1)}_i}\phi_i d\varphi_i-iN\sum_{i_0<i_1}\int_{\omega^{(0)}_{i_0i_1}} n^\phi_{i_0i_{1}} \varphi_{i_1}~.
\end{equation}
Equation \eqref{eq_appendix_0-form DB product} gives the precise defect actions in \eqref{eq_duality defect action 1} and \eqref{eq_duality defect action 2}. It is straightforward to verify that under the gauge transformation \eqref{eq_appendix_0-form gauge transformation}, 
\begin{equation}
    \begin{aligned}
\frac{iN}{2\pi}\int_{\mathcal{M}_1}\phi d\varphi\to \frac{iN}{2\pi}\int_{\mathcal{M}_1}\phi d\varphi+2\pi i N \mathbb{Z}~.
    \end{aligned}
\end{equation}
Gauge invariance of the path integral then requires $N\in \mathbb{Z}$. 

\subsection{Chern-Simons action}\label{app:CS}

Our second example concerns Chern-Simons-type actions in 3d spacetime.

Let $A\in \Omega^1(\mathcal{M}_3,\mathbb{R}/2\pi \mathbb{Z})$ be a $U(1)$ 1-form gauge field on a 3-manifold $\mathcal{M}_3$. 
It is characterized by a tuple $(A_i,h^A_{i_0i_1},n^A_{i_0i_1i_2})$. Locally on each patch, $A_i: U_i\to \Omega^1(U_i,\mathbb{R})$ is a real-valued 1-form gauge field. There are two different sets of transition functions defined at the intersections between patches, namely $h^A_{i_0i_1}: U_{i_0i_1}\to \mathbb{R}$ and $n^A_{i_0i_1i_2}: U_{i_0i_1i_2}\to \mathbb{Z}$. The 3-tuple $(A_i,h^A_{i_0i_1},n^A_{i_0i_1i_2})$ is subject to the cocycle condition
\begin{equation}
\label{eq_app_1-form cocycle condition}
    \begin{aligned}
U_{i_0i_1}:\ & A_{i_0}-A_{i_1}=dh^A_{i_0i_1}~,\\
U_{i_0i_1i_2}:\ & h^A_{i_0i_1}+h^A_{i_1i_2}+h^A_{i_2i_0}=2\pi n^A_{i_0i_1i_2}~,\\
U_{i_0i_1i_2i_3}:\ &n^A_{i_0i_1i_2}-n^A_{i_1i_2i_3}+n^A_{i_2i_3i_0}-n^A_{i_3i_0i_1}=0~.
    \end{aligned}
\end{equation}
As in the compact boson case, these consistency conditions ensure that the field strength $dA$ is continuous across different patches. 
The gauge redundancy of $(A_i,h^A_{i_0i_1},n^A_{i_0i_1i_2})$ also involes two gauge parameters: $\lambda_{i}^A: U_i\to \mathbb{R}$ and $m_{i_0i_1}^A: U_{i_0i_1}\to \mathbb{Z}$, such that the gauge fields and transition functions transform as follows:
\begin{equation}
\label{eq_appendix_1-form gauge transformation}
    \begin{aligned}
U_i:\ & A_i \to  A_i +d\lambda^A_i~,\\
U_{i_0i_1} :\ & h^A_{i_0i_1}\to  h^A_{i_0i_1}+\lambda^A_{i_0}-\lambda^A_{i_1}+2\pi m^A_{i_0i_1}~,\\
U_{i_0i_1i_2}:\ & n^A_{i_0i_1i_2}\to n^A_{i_0i_1i_2}+m^A_{i_0i_1}+m^A_{i_1i_2}+m^A_{i_2i_0}~.
    \end{aligned}
\end{equation}

\begin{figure}[thb]
\centering
\includegraphics[width=.85\textwidth]{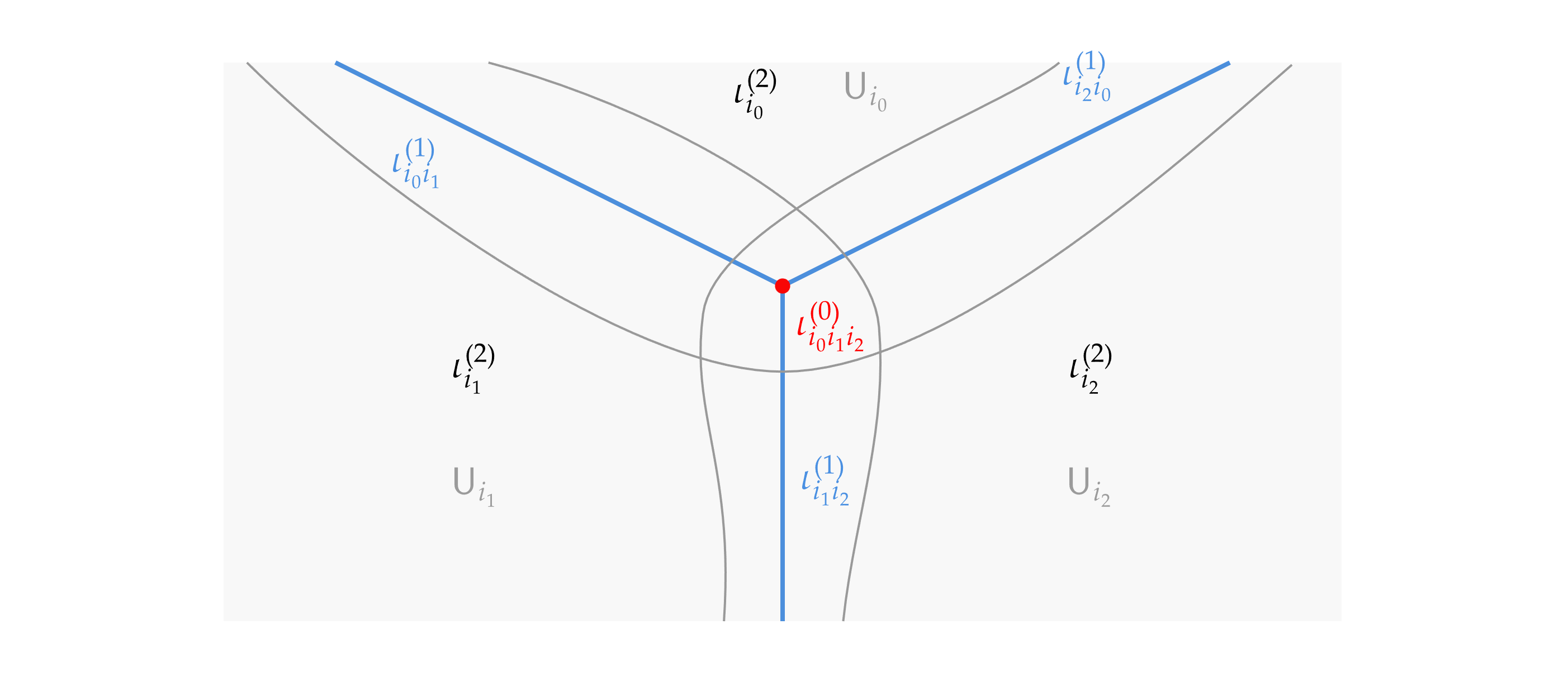}
  \caption{A polyhedral decomposition of $\mathcal{M}_2$.} \label{fig_app 2}
\end{figure}

Dirac quantization follows straightforwardly from the formalism of differential cohomology. 
Consider the integration of the field strength $dA$ on a closed 2d submanifold $\mathcal{M}_2\subset \mathcal{M}_3$. For convenience, we endow $\mathcal{M}_2$ with a polyedral decomposition $\{\iota^{(2)}_{i},\iota^{(1)}_{i_0i_1},\iota^{(0)}_{i_0i_1i_2}\}$, independent of that for $\mathcal{M}_3$. The total magnetic flux through $\mathcal{M}_2$ is 
\begin{equation}
\label{eq_app_magnetic flux=sum}
    Q=\frac{1}{2\pi}\int_{\mathcal{M}_2}dA=\frac{1}{2\pi}\sum_{i}\int_{\iota^{(2)}_i}dA_i=\frac{1}{2\pi}\sum_{i_0<i_1}\int_{\iota^{(1)}_{i_0i_1}}dh^A_{i_0i_1}=\sum_{i_0<i_1<i_2}\int_{\iota^{(0)}_{i_0i_1i_2}}n^A_{i_0i_1i_2}~,
\end{equation}
which is obviously an integer. 
See figure \ref{fig_app 2}. If the two-dimensional surface is contractible, the cocycle condition \eqref{eq_app_1-form cocycle condition} implies $Q=0$, whereas for non-contractible surfaces the magnetic flux can be nonzero.

We now formulate the Chern–Simons action in the framework of differential cohomology. 
Consider two 1-form gauge fields $(A_i,h^A_{i_0i_1},n^A_{i_0i_1i_2})$ and $(B_i,h^{B}_{i_0i_1},n^{B}_{i_0i_1i_2})$. The Chern-Simons coupling between $A$ and $B$ is defined as follows:
\begin{equation}
\label{eq_app_1-form DB product}
\mathtoolsset{multlined-width=0.9\displaywidth}
\begin{multlined}
\frac{iN}{2\pi} \int_{\mathcal{M}_3}A d B
\equiv \frac{iN}{2\pi}\sum_{i}\int_{\omega^{(3)}_i} A_i \wedge d B_i-\frac{iN}{2\pi}\sum_{i_0<i_1}\int_{\omega^{(2)}_{i_0i_1}}h^A_{i_0i_1}dB\\
  +iN \sum_{i_0<i_1<i_2}\int_{\omega_{i_0i_1i_2}^{(1)}} n^A_{i_0i_1i_2} B_{i_2}-i N\sum_{i_0<i_1<i_2<i_3}\int_{\omega_{i_0i_1i_2i_3}^{(0)}}n^A_{i_0i_1i_2}h^{B}_{i_2i_3} ~.
\end{multlined}
\end{equation}
This definition also applies to the case of $A=B$, which is of particular relevance in chiral Chern–Simons theories.

In the rest of this subsection, we show that gauge invariance (modulo $2\pi i$) of the Chern-Simons action \eqref{eq_app_1-form DB product} under  \eqref{eq_appendix_1-form gauge transformation} requires $N\in \mathbb{Z}$. The first term on the RHS of equation \eqref{eq_app_1-form DB product} transforms as follows:
\begin{equation}
\label{eq_app_term by term 1}
\sum_{i}\int_{\omega^{(3)}_i} A_i \wedge dB \to \sum_{i}\int_{\omega^{(3)}_i} A_i \wedge dB+\sum_{ i_0<i_1}\int_{\omega^{(2)}_{i_0i_1}} (\lambda^A_{i_0}-\lambda^A_{i_1}) dB~.
\end{equation}
For the second term in equation \eqref{eq_app_1-form DB product}, we note the following identity:
\begin{equation}
\int_{\omega_{i_0i_1}^{(2)}}dB=\int_{\omega_{i_0i_1}^{(2)}}dB_{i_1}=\sum_{i_{2}}\int_{\omega_{i_0i_1i_2}^{(1)}}B_{i_{2}}+\sum_{i_{2}<i_{3}}\int_{\omega_{i_0i_1i_2i_3}^{(0)}}(2\pi n^B_{i_1i_2i_3}-h^B_{i_{2}i_{3}})~.
\end{equation}
Applying this identity, we obtain
\begin{equation}
\label{eq_app_term by term 2}
\mathtoolsset{multlined-width=0.9\displaywidth}
\begin{multlined}
\sum_{i_0<i_1}\int_{\omega^{(2)}_{i_0i_1}} h^A_{i_0i_1}dB\to \sum_{i_0<i_1}\int_{\omega^{(2)}_{i_0i_1}} h^A_{i_0i_1}dB+\sum_{ i_0<i_1}\int_{\omega^{(2)}_{i_0i_1}} (\lambda^A_{i_0}-\lambda^A_{i_1}) dB\hfill\\
\hfill+2\pi\sum_{i_0<i_1}\sum_{i_2}\int_{\omega_{i_0i_1i_2}^{(1)}} m^A_{i_0i_1}B_{i_2}-2\pi\sum_{i_0<i_1}\sum_{i_2<i_3}\int_{\omega_{i_0i_1i_2i_3}^{(0)}}m^A_{i_0i_1}h^B_{i_2i_3}+4\pi^2\mathbb{Z}~.
\end{multlined}
\end{equation}
Terms involving the gauge parameter $\lambda^A_i$ only appear in \eqref{eq_app_term by term 1} and \eqref{eq_app_term by term 2}, and they add up to zero in \eqref{eq_app_1-form DB product}. 

Next, the third term on the RHS of equation \eqref{eq_app_1-form DB product} transforms as follows:
\begin{equation}
\label{eq_app_term by term 3}
\mathtoolsset{multlined-width=0.9\displaywidth}
\begin{multlined}
\sum_{i_o<i_1<i_2}\int_{\omega_{i_0i_1i_2}^{(1)}} n^A_{i_0i_1i_2} B_{i_2}\to \sum_{i_o<i_1<i_2}\int_{\omega_{i_0i_1i_2}^{(1)}} n^A_{i_0i_1i_2} B_{i_2}\hfill\\
+\sum_{i_0<i_1}\sum_{i_2}\int_{\omega_{i_0i_1i_2}^{(1)}} m_{i_0i_1}^A B_{i_2}+\sum_{i_0<i_1<i_2}\sum_{i_3}\int_{\omega_{i_0i_1i_2i_3}^{(0)}}(m^A_{i_1i_2}h^B_{i_2i_0}+m^A_{i_2i_0}h^B_{i_2i_1})\\
\hfill+\sum_{i_0<i_1<i_2}\sum_{i_3}\int_{\omega_{i_0i_1i_2i_3}^{(0)}}(n_{i_0i_1i_2}^A+m_{i_0i_1}^A+m_{i_1i_2}^A+m_{i_2i_0}^A)\lambda^B_{i_2}~,
\end{multlined}
\end{equation}
 where we have used the cocycle condition \eqref{eq_app_1-form cocycle condition} to rearrange terms in the second line: 
\begin{equation}
\mathtoolsset{multlined-width=0.9\displaywidth}
\begin{multlined}
        \phantom{=}\sum_{i_0<i_1<i_2}\int_{\omega_{i_0i_1i_2}^{(1)}} (m_{i_0i_1}^A+m_{i_1i_2}^A+m_{i_2i_0}^A) B_{i_2}\hfill\\
        =\sum_{i_0<i_1<i_2}\int_{\omega_{i_0i_1i_2}^{(1)}}(m_{i_0i_1}^AB_{i_2}+m_{i_1i_2}^AB_{i_0}+m_{i_2i_0}^AB_{i_1})\hfill\\
        \hfill+\sum_{i_0<i_1<i_2}\int_{\omega_{i_0i_1i_2}^{(1)}}(m_{i_1i_2}^Adh^B_{i_2i_0}+m_{i_2i_0}^Adh^B_{i_2i_1})\\
        =\sum_{i_0<i_1}\sum_{i_2}\int_{\omega_{i_0i_1i_2}^{(1)}} m_{i_0i_1}^A B_{i_2}+\sum_{i_0<i_1<i_2}\sum_{i_3}\int_{\omega_{i_0i_1i_2i_3}^{(0)}}(m^A_{i_1i_2}h^B_{i_2i_0}+m^A_{i_2i_0}h^B_{i_2i_1})~.\hfill\\
\end{multlined}
\end{equation}
We notice that the $m_{i_0i_1}^A B_{i_2}$ term in \eqref{eq_app_term by term 3} cancels with that in \eqref{eq_app_term by term 2}. For the rest of the terms, we need to consider the fourth term on the RHS of equation \eqref{eq_app_1-form DB product}:
\begin{equation}
\label{eq_app_term by term 4}
\mathtoolsset{multlined-width=0.9\displaywidth}
\begin{multlined}
\sum_{i_0<i_1<i_2<i_3}\int_{\omega_{i_0i_1i_2i_3}^{(0)}}n^A_{i_0i_1i_2}h^B_{i_2i_3}\to \sum_{i_0<i_1<i_2<i_3}\int_{\omega_{i_0i_1i_2i_3}^{(0)}}n^A_{i_0i_1i_2}h^B_{i_2i_3}\hfill\\
+\sum_{i_0<i_1<i_2<i_3}\int_{\omega_{i_0i_1i_2i_3}^{(0)}}(m^A_{i_0i_1}+m^A_{i_1i_2}+m^A_{i_2i_0})h^B_{i_2i_3}\\
+\sum_{i_0<i_1<i_2}\sum_{i_3}\int_{\omega_{i_0i_1i_2i_3}^{(0)}}(n_{i_0i_1i_2}^A+m_{i_0i_1}^A+m_{i_1i_2}^A+m_{i_2i_0}^A)\lambda^B_{i_2}+2\pi \mathbb{Z}~.
\end{multlined}
\end{equation}
In this equation, we have also used the cocycle condition \eqref{eq_app_1-form cocycle condition} to rearrange terms in the third line. For example, we have 
\begin{equation}
\begin{aligned}
&\sum_{i_0<i_1<i_2<i_3}\int_{\omega_{i_0i_1i_2i_3}^{(0)}}n^A_{i_0i_1i_2}\left(\lambda^B_{i_2}-\lambda^B_{i_3}\right)\\
=&\sum_{i_0<i_1<i_2<i_3}\int_{\omega_{i_0i_1i_2i_3}^{(0)}}\left(n_{i_0i_1i_2}^A\lambda^B_{i_2}-n_{i_0i_1i_3}^A\lambda^B_{i_3}+n_{i_0i_2i_3}^A\lambda^B_{i_3}-n_{i_0i_1i_3}\lambda^B_{i_3}\right)\\
=&\sum_{i_0<i_1<i_2}\sum_{i_3}\int_{\omega_{i_0i_1i_2i_3}^{(0)}}n_{i_0i_1i_2}^A\lambda^B_{i_2}~,
\end{aligned}
\end{equation}
and a similar identity also holds for the $(m_{i_0i_1}^A+m_{i_1i_2}^A+m_{i_2i_0}^A)\lambda^B_{i_2}$ term. In the Chern-Simons action \eqref{eq_app_1-form DB product}, the third line of \eqref{eq_app_term by term 4} cancels with the third line of \eqref{eq_app_term by term 3} up to $2\pi \mathbb{Z}$. We are therefore left with terms that take the form $m^Ah^B$ in \eqref{eq_app_term by term 2}, \eqref{eq_app_term by term 3}, and \eqref{eq_app_term by term 4}. 
Once again, it follows from the cocycle condition \eqref{eq_app_1-form cocycle condition} that
\begin{equation}
\mathtoolsset{multlined-width=0.9\displaywidth}
\begin{multlined}
\phantom{=}\sum_{i_0<i_1}\sum_{i_2<i_3}\int_{\omega_{i_0i_1i_2i_3}^{(0)}}m^A_{i_0i_1}h^B_{i_2i_3}+\sum_{i_0<i_1<i_2}\sum_{i_3}\int_{\omega_{i_0i_1i_2i_3}^{(0)}}(m^A_{i_1i_2}h^B_{i_2i_0}+m^A_{i_2i_0}h^B_{i_2i_1})\hfill\\
\hfill-\sum_{i_0<i_1<i_2<i_3}\int_{\omega_{i_0i_1i_2i_3}^{(0)}}(m^A_{i_0i_1}+m^A_{i_1i_2}+m^A_{i_2i_0})h^B_{i_2i_3}\\
=\sum_{i_0<i_1<i_2<i_3}\int_{\omega_{i_0i_1i_2i_3}^{(0)}}\left[(m^A_{i_0i_2}+m^A_{i_0i_3})(h^B_{i_1i_2}+h^B_{i_2i_3}+h^B_{i_3i_1})\right.\hfill\\
\hfill\left.+(m^A_{i_1i_2}+m^A_{i_1i_3})(h^B_{i_0i_3}+h^B_{i_3i_2}+h^B_{i_2i_0})+m^A_{i_2i_3}(h^B_{i_0i_1}+h^B_{i_1i_3}+h^B_{i_3i_0})\right]\\
\in 2\pi \mathbb{Z}~.\hfill\\
\end{multlined}
\end{equation}

From equations \eqref{eq_app_term by term 1}, \eqref{eq_app_term by term 2}, \eqref{eq_app_term by term 3}, and \eqref{eq_app_term by term 4}, we find that the total gauge transformation to the Chern-Simons action reads
\begin{equation}
\label{eq_appendix_Chern Simons total gauge transformation}
\frac{iN}{2\pi} \int_{\mathcal{M}_3}AdB\to \frac{iN}{2\pi} \int_{\mathcal{M}_3}AdB+2\pi iN\mathbb{Z}~.
\end{equation}
Similar to the compact boson case,  gauge invariance of \eqref{eq_app_1-form DB product} modulo $2\pi i$ imposes the quantization condition $N\in \mathbb{Z}$. We note that the formalism discussed in this appendix only applies to purely bosonic actions.

\section{Mode expansion and partition functions}\label{app:mode}

In this appendix, we perform an off-shell mode expansion for 4d Maxwell theory (see \eqref{eq_maxwell precise defect}) and the 2d compact boson theory (see \eqref{eq_duality defect action 1}, \eqref{eq_duality defect action 2}) in the presence of a duality defect. 
We then perform a Euclidean path integral to compute the partition functions. 
In 4d we show that the partition function depends only on the combination $e^2N$. 
In 2d, we explicitly compute the partition functions, which depend on $R$, $N_\text{e}$, and $N_\text{m}$.

\subsection{4d Maxwell theory}
\label{sec_4d torus partition functions}

Consider the Maxwell action \eqref{eq_maxwell bulk action} on $T^3 \times [0,2\pi]$. 
We denote the spacetime coordinates as $x_i\sim x_i+2\pi$ for $i=1,2,3$ and $0\leq x_4\leq 2\pi$. 
At the two boundaries $x_4=0$ and $x_4=2\pi$, we impose the Neumann condition:
\begin{equation}
\label{eq_app_maxwell Nuemann}
    \left. F_{i4}\right|_{x_4=0,2\pi}=0~,~~~\text{for}~ i=1,2,3~.
\end{equation}
The magnetic fluxes on the non-contractible cycles are:
\begin{equation}
    Q_{ij}=\frac{1}{2\pi}\int dx_idx_jF_{ij} \in \mathbb{Z}~.
\end{equation}
We have  $Q_{ij}=-Q_{ji}$.

Below we follow the formalism of appendix \ref{sec_DB cohomology} to present the precise defect action of $\mathcal{D}_N$ in \eqref{eq_maxwell duality defect action}. 
Strictly speaking, one should decompose the spacetime manifold into several patches and express the magnetic flux $Q_{ij}$ as a sum of the integer transition functions as in \eqref{eq_app_magnetic flux=sum}. 
However, for simplicity, we will use a single patch on $T^3\times [0,2\pi]$ which intersects with itself, instead of the standard polyhedral decomposition as in Appendix \ref{sec_DB cohomology}. 

Let $h_i (x_1,x_2,x_3,x_4)\in \mathbb{R}$ be the transition function for the gauge field along the $x_i$-direction. These functions satisfy the following consistency conditions 
\begin{equation}
\label{eq_T3 global conditions}
    \begin{gathered}
        \left. A\right|_{x_i=2\pi}-\left. A\right|_{x_i=0}=dh_i~,\\
        (\left.h_i\right|_{x_j=2\pi}-\left.h_i\right|_{x_j=0})-(\left.h_j\right|_{x_i=2\pi }-\left.h_j\right|_{x_i=0})=2\pi Q_{ij}~. 
\end{gathered}
\end{equation}
The third cocycle condition in  \eqref{eq_app_1-form cocycle condition} becomes trivial in this setting of a single patch.

As in \eqref{eq_boundary gauge fields}, we denote the boundary fields $A^+(x_i)=A(x_i,x_4=0)$, $A^-(x_i)=A(x_i,x_4=2\pi)$, $h_i^+(x_i)=h_i(x_i,x_4=0)$, and $h_i^-(x_i)=h_i(x_i,x_4=2\pi)$. The precise defect action is:
\begin{equation}
\label{eq_maxwell precise defect}
\mathtoolsset{multlined-width=0.9\displaywidth}
\begin{multlined}
S_\text{defect}[A^-,A^+]=\frac{iN}{2\pi}\sum_{i,j,k}\frac{\epsilon_{ijk}}{2}\left(\int dx_idx_jdx_k A^-_iF^+_{jk}-\int_{x_i=0}dx_jdx_kh_i^-F_{jk}^+\right)\\
+iN\sum_{i,j,k}\frac{\epsilon_{ijk}}{2}Q_{ij}\left(\int_{x_i=x_j=0}dx_k A^+_{k}-\left.h_k^+\right|_{x_i=x_j=x_k=0}\right)~,
\end{multlined}
\end{equation}
where the boundary field strength is $F^{\pm}=dA^{\pm}$.

Below we demonstrate that the partition function $\mathcal{Z}$ of the Maxwell theory on $T^3 \times [0,2\pi]$ in the presence of this defect depends only on the ratio between the gauge coupling $e^2$ and the defect level $N$, i.e., $\mathcal{Z}=\mathcal{Z}(e^2N)$. 

We first use the gauge redundancy \eqref{eq_appendix_1-form gauge transformation} to fix the transition function $h_i$ to:
\begin{equation}
h_i=\frac{1}{2}\sum_jQ_{ij}x_j~,
\end{equation}
The residual gauge redundancies are parametrized by a single-valued function $\lambda(x_i,x_4)\in \mathbb{R}$ on $T^3\times [0,2\pi]$. We can further choose a gauge to set $A_4=0$. This reduces the gauge redundancy to $\lambda(x_i)\in \mathbb{R}$ being a single-valued function on $T^3$.

Next, we perform a Fourier mode expansion along the $x_i$-directions as follows:
\begin{equation}
A_i=-\frac{1}{4\pi}\sum_j Q_{ij}x_j+\sum_{\vec{n}\in \mathbb{Z}^3}e^{i\sum_{j}n_jx_j}A_{i;\vec{n}}~,\text{ where }A_{i;-\vec{n}}=(A_{i;\vec{n}})^*~.
\end{equation}
The Fourier modes $A_{i;\vec{n}}$ are functions of the $x_4$ coordinate, and we denote $A_{i;\vec n}(0)= A_{i;\vec{n}}^+$, $A_{i;\vec n}(2\pi)= A_{i;\vec{n}}^-$ as their boundary values. In terms of Fourier modes, the defect action \eqref{eq_maxwell precise defect} takes the following form:
\begin{equation}
\label{eq_app_maxwell duality mode expansion}
S_\text{defect}=2\pi i N \sum_{i,j,k}\frac{\epsilon_{ijk}}{2}Q_{ij}(A_{k;\vec{0}}^++A_{k;\vec{0}}^-)-4\pi^2 N \sum_{\vec{n}\in \mathbb{Z}^3} \sum_{i,j,k}\epsilon_{ijk}\vec{n}_i A^+_{j;\vec{n}} A^-_{k;-\vec{n}}~,
\end{equation}
where $\vec{0}=(0,0,0)$. 

We can similarly perform a Fourier expansion along the $x_4$-direction. The Nuemann boundary condition \eqref{eq_app_maxwell Nuemann} at $x_4=0$ and $x_4=2\pi$ in this gauge becomes $\partial_{x_4}A_i^\pm=0$. We therefore find the off-shell mode expansion: 
\begin{equation} A_{i;\vec{n}}(x_4)=A_{i;0;\vec{n}}+\sqrt{2}\sum_{m\in \mathbb{N}^+}(-1)^m \cos \left(mx_4/2\right)A_{i;m;\vec{n}}~.
\end{equation}
Substituting the mode expansion into the bulk action \eqref{eq_maxwell bulk action}, we arrive at 
\begin{equation}
\label{eq_app_maxwell bulk action}
\mathtoolsset{multlined-width=0.9\displaywidth}
\begin{multlined}
S_{\text{bulk}}=\frac{4\pi^2}{e^2}\left(Q_{12}^2+Q_{23}^2+Q_{31}^2\right)+\frac{16\pi^4}{e^2}\sum_{\vec{n}\in\mathbb{Z}^3}\sum_{m\in\mathbb{N}}\\
\hfill \times\left[\sum_i(\vec{n}\cdot\vec{n}+\frac{m^2}{4}-n_i^2)|A_{i;m;\vec{n}}|^2-2\sum_{i\neq j} n_i n_jA_{i;m;\vec{n}}A_{i;m;-\vec{n}}\right]~.
\end{multlined}
\end{equation}

The zero modes $A_{i;0;\vec{0}}$ play an important role in the path integral. $A_{i;0;\vec{0}}$ are invariant under the residual gauge transformation parametrized by $\lambda(x_i)\in \mathbb{R}$, and they do not appear in the bulk action \eqref{eq_app_maxwell bulk action}. They show up in the path integral only via the first term in the defect action \eqref{eq_app_maxwell duality mode expansion}, where they serve as Lagrange multipliers, enforcing the partition function to vanish unless $Q_{12}=Q_{23}=Q_{31}=0$. Therefore, the partition function $\mathcal{Z}$ receives contributions only from off-shell field configurations with zero magnetic flux. Since these field configurations are topologically trivial, we are free to rescale them. 
By rescaling all the remaining gauge fields by $e$, we see that the partition function depends only on $e^2N$. 

In particular, the duality defect becomes topological when $e^2=\frac{2\pi}{N}$. Even though the spacetime manifold $T^4$ has 2-cycles, the partition function in this case, denoted by $\mathcal{Z}_{\mathcal{D}_N}$, is independent of $N$. Since the duality defect $\mathcal{D}_N$ breaks the 1-form symmetry $U(1)_\text{e}^{(1)}\times U(1)_\text{m}^{(1)}$ down to the discrete subgroup generated by \eqref{anyonic}, we note that modes in the path integral do not carry quantized 1-form symmetry charges. This stands in sharp contrast with the 2d compact boson case, as we will discuss below (see \eqref{eq_torus partition function 1}).

\subsection{2d free compact boson}
\label{sec_2d torus partition functions}

In this appendix, we present a detailed analysis of the duality defect action for $\mathcal{D}_{N_\text{e}/N_\text{m}}$ in \eqref{eq_duality defect action 2}. 
We compute the torus partition functions with a defect insertion via the $\zeta$-function regularization. 

We begin with a general 1d Euclidean action involving four compact scalar fields $\phi^\pm \sim \phi^\pm +2\pi$ and $\varphi_{1,2} \sim \varphi_{1,2}+2\pi$:
\begin{equation}
\label{eq_app_general defect action 123}
\mathtoolsset{multlined-width=0.9\displaywidth}
\begin{multlined}
S_\text{defect}[\phi^-,\phi^+;\varphi_1,\varphi_2]=-\frac{i N_1}{2\pi}\int_{0}^{2\pi \beta} d\tau \phi^{-}\partial_\tau \varphi_1 +i N_1 Q^- \varphi_1 \Big|_{\tau=0} \\
\phantom{S_\text{defect}[\phi^-,\phi^+;\varphi_1,\varphi_2]=}-\frac{i N_2}{2\pi}\int_{0}^{2\pi \beta}d\tau \phi^{+}\partial_\tau \varphi_2+iN_2 Q^+\varphi_2 \Big|_{\tau=0}\\
\hfill+\frac{i N_{3}}{2\pi}\int_{0}^{2\pi \beta} d\tau \varphi_1\partial_\tau \varphi_2-iN_{3}Q_1\varphi_2\Big|_{\tau=0}~,
\end{multlined}
\end{equation}
where $Q^{\pm}$, $Q_{1,2}\in \mathbb{Z}$ denotes the winding number of corresponding variables along the $\tau$-direction, and  $N_{1}$, $N_2$, $N_3\in \mathbb{N}^+$. 
We interpret $\phi^{\pm}$ in \eqref{eq_app_general defect action 123} as the bulk fields of the compact boson on the two sides of the defect, and $\varphi_{1,2}$ as the auxiliary degrees of freedom living along the defect. 
When the two Neumann boundaries are connected from the other side of the bulk (as in the main text around \eqref{eq_boundary fields def}), we have $Q^+=Q^-=Q$, but they are generally independent. 

We now integrate out the auxiliary fields $\varphi_{1,2}$ and obtain an effective coupling between the compact boson boundary fields $\phi^\pm$. This can be done using the Fourier mode expansion:
\begin{equation}
    \phi^{\pm}(\tau)=Q^\pm\frac{\tau}{\beta}+\sum_{n\in \mathbb{Z}} e^{in\tau/\beta}\phi^{\pm}_n~,~~~\varphi_{1,2}(\tau)=Q_{1,2}\frac{\tau}{\beta}+\sum_{n\in\mathbb{Z}}e^{in\tau/\beta}\varphi_{1,2;n}~,
\end{equation}
where the reality condition implies $\phi^{\pm}_{-n}=(\phi^{\pm}_{n})^*$ and $\varphi_{1,2;-n}=(\varphi_{1,2;n})^*$. The Fourier modes $\varphi_{1;n}$ and $\varphi_{2;n}$ are Lagrange multipliers of the path integral. In particular, the terms involving the zero modes $\varphi_{1;0}$ and $\varphi_{2;0}$ are:
\begin{equation}
    S_\text{defect}\supset i(N_1Q^-+N_3Q_2)\varphi_{1;0}+i(N_2Q^+-N_3Q_1)\varphi_{2;0}~.
\end{equation}
Integration of $\varphi_{1;0}$ and $\varphi_{2;0}$ from 0 to $2\pi$ imposes the following constraint
\begin{equation}
\label{eq_app_constraint}
    Q^-=-\frac{N_3}{N_1}Q_2~,~~~Q^+=\frac{N_3}{N_2}Q_1~.
\end{equation}
We introduce a pair of integers $N_\text{e}$, $N_\text{m}\in \mathbb{N}^+$ that are coprimes (i.e.,   $\gcd(N_\text{e},N_\text{m})=1$) and satisfiy 
\begin{equation}
{N_\text{e}\over N_\text{m}} ={ N_1N_2\over N_3}.
\end{equation}
We further define $k_1=\gcd(N_1,N_3)$, $k_2=\gcd(N_2,N_3)$. 
In terms of these new numbers, the constraint \eqref{eq_app_constraint} implies that only those $Q^\pm$ obeying 
\begin{equation}
\label{eq_app_sandwhich cond}
    Q^-\in k_1N_\text{m}\mathbb{Z}~,~~~Q^+\in k_2N_\text{m}\mathbb{Z}~,
\end{equation}
will contribute to the partition function. There are also similar constraints on $Q_{1,2}$. 

Integrating out $\varphi_{1,2}$ in \eqref{eq_app_general defect action 123} gives the following action for $\phi^\pm$: 
\begin{equation}
\label{eq_app_mode expansion}
S_\text{defect}=\frac{iN_\text{e}}{N_\text{m}}(\phi^+_0Q^--\phi^-_0Q^+-\pi Q^+Q^-)+\frac{N_\text{e}}{N_\text{m}}\sum_{n \in \mathbb{Z}}n \phi_n^+\phi_{-n}^-~,
\end{equation}
where $Q^+$ and $Q^-$ are subject to the constraint \eqref{eq_app_sandwhich cond}.
The constraint in  \eqref{eq_app_sandwhich cond} implies that $\mathcal{D}_{N_1, N_2, N_3}$ can be decomposed into $k_1k_2$ defects:
\begin{equation}
\label{eq_app_semisimple cond}
\mathcal{D}_{N_1,N_2,N_3}=\left(\sum_{n_1=0}^{k_1-1} \sum_{n_2=0}^{k_2-1}\eta_{\text{w}}(\frac{2\pi n_1}{k_1N_\text{m}})\eta_{\text{m}}(\frac{2\pi n_2}{k_2N_\text{e}})\right)\times \mathcal{D}_{N_\text{e}/N_\text{m}}~,
\end{equation}
where $\mathcal{D}_{N_\text{e}/N_\text{m}}$ is defined by taking $N_1=N_\text{e}$, $N_2=1$, and $N_3=N_\text{m}$ (see also \eqref{eq_duality defect action 2}). Therefore, $\mathcal{D}_{N_1,N_2,N_3}$ is simple if $k_1=k_2=1$.

For the simple defect $\mathcal{D}_{N_\text{e}/N_\text{m}}$, the constraint \eqref{eq_app_sandwhich cond} is symmetric upon exchanging $Q^+$ with $Q^-$. 
This implies that $\mathcal{D}_{N_\text{e}/N_\text{m}}$ is its own dual, 
\begin{equation}
\mathcal{D}_{N_\text{e}/N_\text{m}} 
=\overline{ \mathcal{D}}_{N_\text{e}/N_\text{m}}.
\end{equation}
Indeed, one can check that \eqref{eq_app_mode expansion}, in Lorentzian signature, is its own image under the CRT operator $\Theta$. 
On the other hand, a generic simple defect $\mathcal{D}_{N_1, N_2, N_3}$ \eqref{eq_app_general defect action 123} is not its own dual.

\subsubsection{Fusion rules}

Fusion rules between the defect \eqref{eq_duality defect action 1} and a $U(1)_\text{m}\times U(1)_\text{w}$ symmetry defect can also be derived from the mode expansion \eqref{eq_app_mode expansion}. We first define the momentum charges $Q_{\text{m}}^\pm \in \mathbb{Z}$ by the following coupling of the zero modes $\phi_0^\pm$ in the bulk action:
\begin{equation}
\label{eq_momentum charge def}
    S_{\text{bulk}}\supset -i(Q_{\text{m}}^-\phi_0^-+Q_{\text{m}}^+\phi_0^+)~.
\end{equation}
Indeed, an insertion of $e^{i \alpha Q^\pm_\text{m}}$ in the path integral shifts the fields $\phi^\pm$ by $\alpha$, as expected. 
The winding charge for $\phi^-$ is defined as $Q_{\text{w}}^-=Q^- = {1\over2\pi } \int_0^{2\pi\beta} d\tau \partial_\tau \phi^-$, while that for $\phi^+$ is $Q_{\text{w}}^+=-Q^+ = - {1\over2\pi } \int_0^{2\pi\beta} d\tau \partial_\tau \phi^+$.

Integrating out the zero modes $\phi^+_0$ and $\phi^-_0$ in $S_{\text{bulk}}+S_{\text{defect}}$ imposes the following constraint:
\begin{equation}
\label{eq_app_general fusion rule}
    N_\text{m}Q^-_\text{m}=N_\text{e}Q^+_\text{w}~,~~~N_\text{e}Q^-_\text{w}=N_\text{m}Q^+_\text{m}~.
\end{equation}
\eqref{eq_app_general fusion rule} identifies $U(1)_\text{m}$ and $U(1)_\text{w}$ charges across the defect, from which we conclude the fusion rule \eqref{eq_duality fusion rule 3}.

\begin{figure}[thb]
\centering
\includegraphics[width=.98\textwidth]{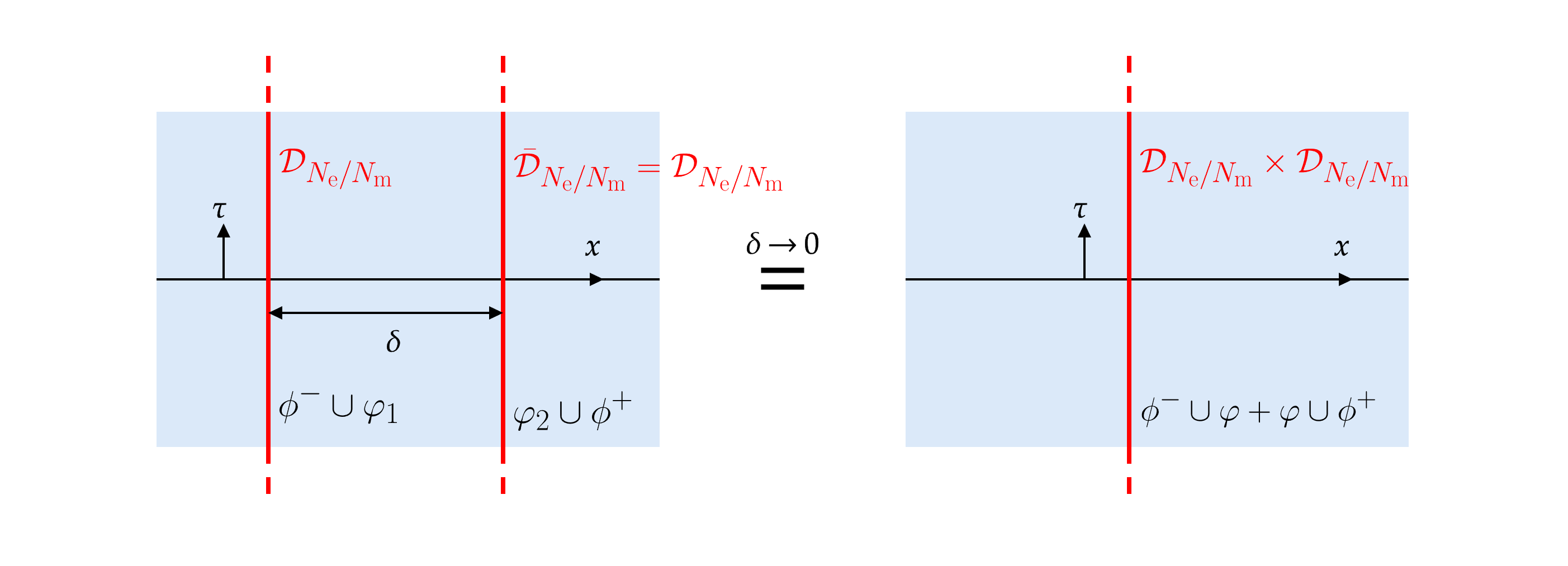}
  \caption{\label{pic_defect fusion}Parallel fusion of two defects. We use $\varphi \sim \varphi+2\pi$ to denote the compact boson field in the intermediate region of width $\delta$.} 
\end{figure}

The fusion of two simple defects $\mathcal{D}_{N_\text{e}/N_\text{m}}$ can be defined as in figure \ref{pic_defect fusion}. We denote $\varphi$ as the compact boson field in the intermediate region between two defects, and $\varphi^\pm$ as the corresponding boundary fields. We adopt two assumptions to derive the fusion rule: First, we assume that the fluctuations of the compact boson field $\varphi(\tau,x)$ along the $x$-direction (transversal to the defect) are suppressed at the limit $\delta\to0$. This allows us to drop the dependence on the $x$-coordinate of the $\varphi$ field as the two defects are brought close to each other. 
Second, we assume the action governing the dynamics of the intermediate field $\varphi(\tau,x)$ scales as $O(\delta)$, and can therefore be neglected in the $\delta \to 0$ limit. Both assumptions are evident if the field $\varphi$ is described by the free compact action, and it extends to a broad class of interacting models.

We perform a Fourier mode expansion of the intermediate field $\varphi$, which under the first assumption reads
\begin{equation}
    \varphi(\tau)={Q}'\frac{\tau}{\beta}+\sum_{n\in\mathbb{Z}}e^{in\tau/\beta}\varphi_{n}~,
\end{equation}
where ${Q}'\in \mathbb{Z}$ and $\varphi_{-n}=(\varphi_n)^*$. By merging two copies of the ${\cal D}_{N_\text{e}/N_\text{m}}$ defects (see \eqref{eq_app_mode expansion}) parallely as in figure \ref{pic_defect fusion}, we obtain
\begin{equation}
\label{eq_app_defect fusion 2}
\mathtoolsset{multlined-width=0.9\displaywidth}
\begin{multlined}
   \mathcal{D}_{N_\text{e}/N_\text{m}}\times \mathcal{D}_{N_\text{e}/N_\text{m}}:~S_\text{defect}=i\frac{N_\text{e}}{N_\text{m}}\left[{Q}'(\phi^+_0-\phi^-_0)\right.\\
   \left. +\varphi_0(Q^--Q^+)-\pi{Q}'(Q^-+Q^+) \right]+\frac{N_\text{e}}{N_\text{m}}\sum_{n\in\mathbb{Z}}n(\phi_n^+-\phi_n^-)\varphi_{-n}~,  
   \end{multlined}
\end{equation}
subject to the constraint \eqref{eq_app_sandwhich cond}, i.e.,  $Q^-$, $Q^+$, $ {Q}'\in N_{\text{m}}\mathbb{Z}$. The integration of the Fourier zero mode $\varphi_0$ yields $Q^-=Q^+$, while the summation over ${Q}'$ enforces $\phi^+_0-\phi^-_0\in \mathbb{Z}/N_\text{e}$. We conclude that \eqref{eq_app_defect fusion 2} is identified as the condensation defect $\mathcal{C}_{N_\text{e},N_\text{m}}$ defined in \eqref{eq_2d condensation defect}, and the fusion rule \eqref{eq_duality fusion rule 4} readily follows. 

\subsubsection{Partition functions}

We now compute the torus partition function with a defect insertion. As in the main text, we employ the coordinate system $\tau\sim \tau+2\pi \beta$ and $0\leq x\leq 2\pi$.  
We consider the free compact boson action \eqref{eq_2d bulk action} in the bulk, with the two boundary fields $\phi^+(\tau)=\phi(\tau,x=0)$ and $\phi^-(\tau)=\phi(\tau,x=2\pi)$ (see also \eqref{eq_boundary fields def}). We couple the $\phi^\pm$ fields via the defect action \eqref{eq_app_general defect action 123} with $N_1=N_\text{e}$, $N_2=1$, and $N_3=N_\text{m}$, such that it represents the simple defect $\mathcal{D}_{N_e/N_\text{m}}$. We assume a generic radius $R$, and the defect $\mathcal{D}_{N_e/N_\text{m}}$ is in general conformal but not topological.

We begin by performing a Fourier mode expansion along the $\tau$-direction that runs parallel to the defect $\mathcal{D}_{N_\text{e}/N_\text{m}}$:
\begin{equation}
\label{eq_app_compact boson Fourier 1}
    \phi(\tau,x)=Q\frac{\tau}{\beta}+\sum_{n\in \mathbb{Z}}e^{in\tau/\beta}\phi_n(x)~,
\end{equation}
where $Q \in \mathbb{Z}$ denotes the compact boson winding number. Similarly, we perform a mode expansion in the $x$-direction:
\begin{equation}
\label{eq_app_compact boson Fourier 2}
\phi_n(x)=\phi_{n,0}+\sqrt{2}\sum_{m\in \mathbb{N}^+}(-1)^m\cos{\left(mx/2\right)}\phi_{n,m}~, \text{where }\phi_{-n,m}=(\phi_{n,m})^*~.
\end{equation}
Substituting the mode expansion into the bulk action \eqref{eq_2d bulk action}, we find 
\begin{equation}
\label{eq_app_compact boson bulk action mode detail}
S_\text{bulk}=\frac{\pi R^2 }{\beta}\left[Q^2+\sum_{n\in \mathbb{Z}}\sum_{m\in\mathbb{N}}\left(n^2+(\beta m/2)^2\right)|\phi_{n,m}|^2\right]~.
\end{equation}
On the other hand, the defect action in mode expansion readily follows from \eqref{eq_app_mode expansion}. 
The Fourier modes of the boundary fields $\phi^\pm_{n}$ can be further expanded in terms of  $\phi_{n,m}$ as in \eqref{eq_app_compact boson Fourier 2}, which leads to
\begin{equation}
\label{eq_app_compact boson defect action mode detail}
\mathtoolsset{multlined-width=0.9\displaywidth}
\begin{multlined}
    S_{\text{defect}}=-i\pi \frac{N_\text{e}}{N_\text{m}}Q^2-i2\sqrt{2}\frac{N_\text{e}}{N_\text{m}}Q\sum_{m\in 2\mathbb{N}+1}\phi_{0,m}\\
    +\frac{2N_\text{e}}{N_\text{m}}\sum_{n\in \mathbb{Z}/\{0\}}\left[n(\sum_{m\in 2\mathbb{N}+1}\phi_{n,m}^*)(\frac{\phi_{n,0}}{\sqrt{2}}+\sum_{m\in 2\mathbb{N}^+}\phi_{n,m})-\text{c.c.}\right]~,
\end{multlined}
\end{equation}
subject to the constrant $Q\in N_\text{m}\mathbb{Z}$.

The defect partition function is defined by the following path integral:
\begin{equation}
\label{eq_compact boson partition function def}
\mathcal{Z}=N_\text{m}\sum_{Q\in N_\text{m}\mathbb{Z}}\int\prod_{n\in \mathbb{Z}}\prod_{m\in \mathbb{N}} d\phi_{n,m} \exp{(-S_\text{bulk}-S_\text{defect})}~,
\end{equation}
 where the overall coefficient $N_\text{m}$ arises from the dimension of the Hilbert space for the auxiliary quantum mechanical degrees of freedom in \eqref{eq_app_general defect action 123}. This is a Gaussian integral of $\phi_{n,m}$, and the partition function boils down to evaluating the inverse determinant of an infinite-dimensional matrix.  
The bulk action \eqref{eq_app_compact boson bulk action mode detail} corresponds to the diagonal terms of this matrix, while the defect action \eqref{eq_app_compact boson defect action mode detail} introduces cross terms. 
The latter couple modes with the same parallel wavenumber $n$ but different transverse wavenumbers $m$ together. 
Importantly, the matrix is block-diagonal in $n$.

To evaluate the partition function \eqref{eq_compact boson partition function def}, the following determinant identity will become handy momentarily:
\begin{equation}
\label{eq_app_matrix det lemma}
\left|\begin{pmatrix}
A & uv^\text{T} \\
-uv^\text{T} & B \\
\end{pmatrix}\right|=|A||B| \left(1+(u^\text{T}A^{-1} u)(v^\text{T}B^{-1} v)\right)~,
\end{equation}
where $A\in \mathbb{C}^{a\times a}$, $B\in \mathbb{C}^{b\times b}$ are non-singular matrices, and $u\in \mathbb{C}^a$, $v \in \mathbb{C}^b$ are vectors.

We first focus on the block matrix corresponding to a fixed nonzero $n$. Without the defect action \eqref{eq_app_compact boson defect action mode detail}, the determinant of this block matrix is simply the infinite products of the diagonal terms:
\begin{equation}
\prod_{m\in \mathbb{N}} \frac{ R^2}{\beta}\left(n^2+(\beta m/2)^2\right)~.
\end{equation}
Now we add the cross terms from the defect action, and denote the determinant of the block matrix as $\Delta_n$. 
Using lemma \eqref{eq_app_matrix det lemma}, we find the following expression for $\Delta_n$:
\begin{equation}
\begin{aligned}
    &\Delta_n\left[\prod_{m\in \mathbb{N}} \frac{ R^2}{\beta}\left(n^2+(\beta m/2)^2\right)\right]^{-1}\\
    =&1+(\frac{n \beta N_\text{e}}{\pi R^2 N_\text{m}})^2(\frac{1}{n^2}+\sum_{m\in \mathbb{N}^+}\frac{2}{n^2+ (\beta m)^2})(\sum_{r\in \mathbb{N}+\frac{1}{2}}\frac{2}{n^2+(\beta r)^2})\\
    =&1+(\frac{N_\text{e}}{N_\text{m}R^2})^2~.
    \end{aligned}
\end{equation}

In the full path integral, we need to further take the infinite product of $\Delta_n$ over $n$. 
Using the $\zeta$-function regularization \cite{quine1993zeta}, we find
\begin{equation}
\label{eq_app_lemma 1}
    \prod_{n \in \mathbb{Z}/\{0\}}\Delta_n=\frac{4\pi \beta^{\frac{3}{2}}N_\text{m}^2R}{(N_\text{m}R)^2+(N_\text{e}/R)^2}\left[\frac{ (\eta(\uptau))^2}{\vartheta_2(\uptau/2)}\right]^2~,
\end{equation}
where $\uptau=i \beta$ is the modular parameter.

Integrating out $\phi_{n\neq 0,m}$, we find the following expression for the partition function \eqref{eq_compact boson partition function def}:
\begin{equation}
\label{eq_app_pathinte result 1}
\begin{aligned}
\mathcal{Z}=&N_\text{m}(\prod_{n \in \mathbb{Z}/\{0\}} \Delta_n^{-\frac{1}{2}})\sum_{Q\in N_\text{m}\mathbb{Z}}\int\prod_{m\in \mathbb{N}} d\phi_{0,m} \exp{(-S_0)}\\
=&\beta^{-\frac{3}{4}}\sqrt{\frac{(N_\text{m}R)^2+(N_\text{e}/R)^2}{4\pi R}}\frac{\vartheta_2(\uptau/2)}{\eta(\uptau)^2}\sum_{Q\in N_\text{m}\mathbb{Z}}\int\prod_{m\in \mathbb{N}} d\phi_{0,m} \exp{(-S_0)}~,
\end{aligned}
\end{equation}
where the action $S_0$ associated with the zero-modes is  
\begin{equation}
S_{0}=\frac{\pi R^2}{\beta} Q^2+\frac{\pi R^2 \beta}{4}\sum_{m\in \mathbb{N}^+}m^2\phi_{0,m}^2-i\pi \frac{N_\text{e}}{N_\text{m}}Q^2-2\sqrt{2}i\frac{N_\text{e}}{N_\text{m}}Q\sum_{m\in 2\mathbb{N}+1}\phi_{0,m}~.
\end{equation}
This part of the path integral can also be evaluated with the $\zeta$-function regularization. We find
\begin{equation}
\label{eq_app_pathinte result 2}
\begin{aligned}
    &\sum_{Q\in N_\text{m}\mathbb{Z}}\int\prod_{m\in \mathbb{N}^+} d\phi_{0,m} \exp{(-S_0)}\\
    =& \left[\prod_{m\in \mathbb{N}^+}\frac{R^2}{4}(\beta m)^2\right]^{-\frac{1}{2}}\sum_{\tilde{Q}\in \mathbb{Z}} \exp{\left[-\frac{\pi }{\beta}((N_\text{m}R)^2+(N_\text{e}/R)^2)\tilde{Q}^2-i\pi N_\text{e} N_\text{m}\tilde{Q}\right]}\\
=&\beta^{\frac{3}{4}}\sqrt{\frac{R/4\pi }{(N_\text{m}R)^2+(N_\text{e}/R)^2}}\sum_{{\tilde{Q}}'\in \mathbb{Z}} \exp{\left[\frac{i\pi\uptau }{(N_\text{m}R)^2+(N_\text{e}/R)^2}\left({\tilde{Q}}'-\frac{N_\text{e} N_\text{m}}{2}\right)^2\right]}~,
\end{aligned}
\end{equation}
where we have used the Poisson
summation in the second equation. Finally, we note the integration $\int d\phi_{0,0}=2\pi $ represents the volume of the compact boson target space. 
Combining \eqref{eq_app_pathinte result 1} and \eqref{eq_app_pathinte result 2}, the partition function \eqref{eq_torus partition function 3} is reproduced.

The defect $\mathcal{D}_{N_\text{e}/N_\text{m}}$ becomes topological when $R^2=N_\text{e}/N_\text{m}$. 
In this case, the spatial momentum operator commutes with the defect, and the partition function can be generalized to have a generic complex modular parameter $\uptau$ (which is not necessarily purely imaginary). 
This corresponds to inserting $e^{i\pi (\uptau -\bar\uptau)P}$ in the trace over the Hilbert space. 
The special case with $N_\text{e}=N$, $N_\text{m}=1$ was discussed in the main text, with the final partition function given in \eqref{eq_torus partition function 1}. 
The more general case with generic coprime $N_\text{e},N_\text{m}$ was presented in \eqref{eq_torus partition function 2}. 
Importantly, the ground state of the defect Hilbert space is two-fold degenerate whenever $N_\text{e}N_\text{m}\in 2\mathbb{N}+1$. As we discussed in section \ref{sec:anomaly}, this degeneracy is a consequence of the defect 't Hooft anomaly.

\section{ODEs for twist defects}
\label{sec_hypergeometric solutions}

In this appendix, we study the solutions to the differential equations \eqref{eq_maxwell ODE}  and \eqref{eq_three point closure condition}. 

\subsection{Electromagnetic waves with a twist defect}

We start with the ODE \eqref{eq_maxwell ODE} for the electromagnetic fields $(f_{t\psi},f_{t\theta})$ in the presence of a 2d conformal twist defect.  

When $s+n\neq \pm \omega$,   \eqref{eq_maxwell ODE} admits two linearly independent hypergeometric solutions:
\begin{equation}
\label{appendix_eq_maxwell ODE solutions}
\mathtoolsset{multlined-width=0.9\displaywidth}
\begin{multlined}
  \left.  \begin{pmatrix}
f_{t\psi}\\
f_{t\theta}
\end{pmatrix}\right|_{s\in \mathbb{Z}\pm \frac{1}{4}}=\left[\begin{pmatrix}
 n(n+s)-\omega^2(1-\rho)\\
s(n+s)-\omega^2\rho
\end{pmatrix}+\begin{pmatrix}
 \pm 2\omega \rho (1-\rho) \\
\mp  2\omega \rho (1-\rho)
\end{pmatrix}\partial_\rho\right]\hfill\\
\times (1-\rho)^{\frac{n}{2}}\left[C_1\rho^{\pm \frac{s}{2}}\, _2F_1\left(\frac{n\pm s\mp \omega}{2},1+\frac{n\pm s\pm \omega}{2} ;1\pm s;\rho \right)\right.\\
\hfill\left.+C_2\rho^{\mp \frac{s}{2}}\, _2F_1\left(\frac{n\mp s\mp \omega }{2} ,1+\frac{n\mp s\pm \omega
   }{2};1\mp s;\rho \right)\right]~.
\end{multlined}
\end{equation}
where $C_1$, $C_{2}$ are   constants. Close to the twist defect $\rho \to 0$, the solution \eqref{appendix_eq_maxwell ODE solutions} has the following asymptotic behavior:
\begin{equation}
\mathtoolsset{multlined-width=0.9\displaywidth}
\begin{multlined}
  \left.  \begin{pmatrix}
f_{t\psi}\\
f_{t\theta}
\end{pmatrix}\right|_{s\in \mathbb{Z}\pm \frac{1}{4}}=\frac{C_1\rho^{\pm \frac{s}{2}}}{n+s+\omega}\left[\begin{pmatrix}
 n+\omega\\
 s
\end{pmatrix}+O(\rho)\right]+\frac{C_2\rho^{\mp\frac{s}{2}}}{n+s-\omega}\left[\begin{pmatrix}
 n-\omega\\
 s
\end{pmatrix}+O(\rho)\right]~.
\end{multlined}
\end{equation}
As explained in Sections \ref{sec_defect Hilbert space} and \ref{sec_Generalized free field sector}, the divergent solution in \eqref{appendix_eq_maxwell ODE solutions} leads to non-normalizable wavefunctions and is excluded by unitarity. 
We therefore set $C_2=0$ if $|s|\in \mathbb{N}+\frac{1}{4}$, and $C_1=0$ if $|s|\in \mathbb{N}+\frac{3}{4}$.

On the other hand, in the $\rho\to 1$ limit, the solution \eqref{appendix_eq_maxwell ODE solutions} becomes:
\begin{equation}
\label{appendix_eq_maxwell ODE solutions expand 2}
\begin{aligned}
  \left.  \begin{pmatrix}
 f_{t\psi}\\
f_{t\theta}
\end{pmatrix}\right|_{s\in \mathbb{Z}\pm \frac{1}{4}}=&\frac{C_1(1-\rho)^{\frac{n}{2}}\Gamma (-n) \Gamma (1\pm s)}{(n+s+\omega ) \Gamma \left(\frac{\pm s-n-\omega}{2}
\right) \Gamma \left(1+\frac{\omega\pm s-n}{2}\right)}\left[\begin{pmatrix}
 n\\
 s \pm \omega
\end{pmatrix}+O(1-\rho)\right]\\
&+\frac{C_1 (1-\rho)^{-\frac{n}{2}}\Gamma (n) \Gamma (1\pm s)}{(n+s+\omega ) \Gamma \left(\frac{n\pm s+\omega}{2}\right) \Gamma \left(1+\frac{n\pm s-\omega}{2}\right)}\left[\begin{pmatrix}
 n\\
 s\mp \omega
\end{pmatrix}+O(1-\rho)\right]\\
&+\frac{C_2 (1-\rho)^{\frac{n}{2}}\Gamma (-n) \Gamma (1\mp s)}{(n+s-\omega)\Gamma \left(\frac{\omega\mp s+n}{2}\right) \Gamma \left(1-\frac{\omega\pm s+n}{2}\right)}\left[\begin{pmatrix}
 n\\
 s\pm \omega
\end{pmatrix}+O(1-\rho)\right]\\
&+\frac{C_2 (1-\rho)^{-\frac{n}{2}}\Gamma (n) \Gamma (1\mp s)}{(n+s-\omega ) \Gamma
   \left(\frac{n\mp s-\omega }{2}\right) \Gamma \left(1+\frac{n\mp s+\omega }{2}\right)}\left[\begin{pmatrix}
 n\\
 s\mp \omega
\end{pmatrix}+O(1-\rho)\right]~.
\end{aligned}
\end{equation}
Physically, the limit $\rho \to 1$ is the regime far from the conformal twist defect. 
See figure \ref{pic_S3 coordinate}. 
We therefore require the wavefunction to be finite at $\rho=1$. 
This can only be the case if the corresponding gamma functions in the denominator of \eqref{appendix_eq_maxwell ODE solutions expand 2} have poles, leading to the quantized eigenfrequencies in \eqref{eq_eigenfrequencies}.

The solutions to the ODE \eqref{eq_maxwell ODE} become singular when $s+n=\pm \omega$. For $\omega=s+n$, we find the following two solutions (multipled by $\tilde C_1$ and $\tilde C_2$, respectively below)
\begin{equation}
\label{appendix_eq_maxwell ODE special sol}
\mathtoolsset{multlined-width=0.9\displaywidth}
\begin{multlined}
    \left.  \begin{pmatrix}
 f_{t\psi}\\
 f_{t\theta}
\end{pmatrix}\right|_{s\in \mathbb{Z}\pm  \frac{1}{4}}=\tilde{C}_2(1-\rho )^{\mp n/2} \rho ^{\mp s/2}\begin{pmatrix}
1\\
-1
\end{pmatrix}\\
+\tilde{C}_1(1-\rho )^{\pm n/2} \rho ^{\pm s/2} \left[\begin{pmatrix}
 1\\
1
\end{pmatrix}-\frac{n}{s+n} \, _2F_1(1,\pm (n+s);1\pm s;\rho )\begin{pmatrix}
1\\
-1
\end{pmatrix}\right]~.
\end{multlined}
\end{equation}
Clearly, these two solutions  scale as $\sim \rho^{\pm\frac{s}{2}}$ and $\sim \rho^{\mp\frac{s}{2}}$ near the defect, respectively. 
Following the same reasoning as in the general case, we find $\tilde{C}_2=0$ for $|s|\in \mathbb{N}+\frac{1}{4}$ and $\tilde{C}_1=0$ for $|s|\in \mathbb{N}+\frac{3}{4}$. Far away from the defect where $\rho\to 1$, the singular solution \eqref{appendix_eq_maxwell ODE special sol} yields the following expansion:
\begin{equation}
\label{appendix_eq_maxwell ODE special expand}
\mathtoolsset{multlined-width=0.9\displaywidth}
\begin{multlined}
  \left.  \begin{pmatrix}
f_{t\psi}\\
 f_{t\theta}
\end{pmatrix}\right|_{s\in \mathbb{Z}\pm \frac{1}{4}}=\tilde{C}_1\frac{(1-\rho )^{\pm n/2}}{s+n}\left[\begin{pmatrix}
2s+n\\
n
\end{pmatrix}+O(1-\rho)\right]\hfill\\
\hfill+(1-\rho )^{\mp n/2}\left[\tilde{C}_2-\tilde{C}_1\frac{\Gamma (1\pm n) \Gamma (1\pm s)}{\Gamma (1\pm (s+n   ))}\right] \left[\begin{pmatrix}
1\\
-1
\end{pmatrix}+O(1-\rho)\right]~.
\end{multlined}
\end{equation}
Solutions that are continuous at $\rho=1$ obey $|s|\in \mathbb{N}+\frac{3}{4}$ and $ns>0$. 
The $\omega>0$ branch is a special case in the second line of the dispersion relation \eqref{eq_eigenfrequencies}.

Finally, for $\omega=-s-n$ we find 
\begin{equation}
\label{appendix_eq_maxwell ODE special sol 2}
\mathtoolsset{multlined-width=0.9\displaywidth}
\begin{multlined}
    \left.  \begin{pmatrix}
f_{t\psi}\\
f_{t\theta}
\end{pmatrix}\right|_{s\in \mathbb{Z}\pm  \frac{1}{4}}=C_2(1-\rho )^{\pm n/2} \rho ^{\pm s/2}\begin{pmatrix}
1\\
-1
\end{pmatrix}\\
+C_1(1-\rho )^{\mp n/2} \rho ^{\mp s/2} \left[\begin{pmatrix}
 1\\
1
\end{pmatrix}-\frac{n}{s+n} \, _2F_1(1,\mp (n+s);1\mp s;\rho )\begin{pmatrix}
1\\
-1
\end{pmatrix}\right]~.
\end{multlined}
\end{equation}
\eqref{appendix_eq_maxwell ODE special sol 2} can be obtained by exchanging the two solutions in \eqref{appendix_eq_maxwell ODE special sol} with $s\in \mathbb{Z}+\frac{1}{4}$ and $s\in \mathbb{Z}-\frac{1}{4}$. 
The spectrum from the singular solution \eqref{appendix_eq_maxwell ODE special sol 2} is then completely parallel to that from \eqref{appendix_eq_maxwell ODE special sol} and \eqref{appendix_eq_maxwell ODE special expand}.

\subsection{DCFT three-point functions}

Next, we discuss the differential equation \eqref{eq_three point closure condition} for the bulk-defect-defect three-point functions. In general, \eqref{eq_three point closure condition} admits two linearly independent hypergeometric solutions. 
To avoid cluttering, we only explicitly present the $f_3^\pm$ component, while the corresponding $f_1^\pm$ and $f_2^\pm$ are determined by \eqref{eq_three point linear} and \eqref{eq_three point closure condition}. We find that
\begin{equation}
\label{appendix_eq_three point sol}
\mathtoolsset{multlined-width=0.9\displaywidth}
\begin{multlined}
    f_3^\pm= (\xi +1)^{\Delta_2}\xi^{1-\bar{h}_2-\bar{h}_3}\left[C_3\frac{(s/2)^2-(h_2-h_3)^2}{1\pm s} \right.\hfill\\
    \times \xi
   ^{1\pm \frac{s}{2}} \,
   _2F_1\left(h_2-h_3+1\pm\frac{s}{2},\bar{h}_2-\bar{h}_3+1\pm \frac{s}{2},;2\pm s;-\xi \right)\\
   \left. \mp sC_4\xi ^{\mp \frac{s}{2}} \,
   _2F_1\left(h_2-h_3\mp \frac{s}{2},\bar{h}_2-\bar{h}_3\mp \frac{s}{2},;\mp s;-\xi \right)\right]~,
\end{multlined}
\end{equation}
where $C_3$, $C_4$ are  constants. In the limit $\xi\to 0$, these solutions exhibit the following asymptotic scaling behavior:
\begin{equation}
\mathtoolsset{multlined-width=0.9\displaywidth}
\begin{multlined}
(f^\pm_1,f^\pm_2,f^\pm_3)=C_3\xi ^{1\pm\frac{s}{2}-\bar{h}_2-\bar{h}_3} \left[(2s,-\frac{s}{2}\pm(h_3-h_2),0)+O\left(\xi\right)\right]\\
+C_4\xi ^{1\mp \frac{s}{2}-\bar{h}_2-\bar{h}_3}\left[(2s,-\frac{s}{2}\pm (\bar{h}_3-\bar{h}_2),\mp s)+O\left(\xi\right)\right]~.
\end{multlined}
\end{equation}
We note that, with $s\in \mathbb{Z}\mp \frac{1}{4}$, the $C_3$ solution in \eqref{appendix_eq_three point sol} corresponds to a defect primary operator with scaling dimension $\Delta=1\pm s$ in the bulk-to-defect OPE of $F_{\mu\nu}$, while the $C_4$ solution corresponds to  one with $\Delta=1\mp s$. This is in agreement with the analysis in \eqref{eq_defect vector primary}. 
These defect primaries have (parallel) spin $\ell = h-\bar h = \pm1$, and the  unitarity bound reads $\Delta \ge1$, which enforces $C_3=0$ for $|s|\in \mathbb{N}+\frac{1}{4}$ and $C_4=0$ for $|s|\in \mathbb{N}+\frac{3}{4}$.

Next, we consider the opposite limit $\xi \to +\infty$ of \eqref{appendix_eq_three point sol} by taking $z_1=0$, $z_2=|w|\frac{\sqrt{\xi}-1}{\sqrt{\xi}+1}$, and $z_3=-|w|$. No two operators become coincident in this limit, so the three-point function $\langle F_{\mu \nu}\mathcal{O}_2 \mathcal{O}_{3}\rangle$ is expected to remain finite. It suffices to consider the $\langle F_{z\bar{z}}\mathcal{O}_2\mathcal{O}_3\rangle$ component, which yields the following double expansion:
\begin{equation}
\label{eq_three point large argument} 
\mathtoolsset{multlined-width=0.9\displaywidth}
\begin{multlined}
    \left. \langle F_{z\bar{z}}\mathcal{O}_2\mathcal{O}_3\rangle\right|_{s\in \mathbb{Z}\pm \frac{1}{4}}=C_3\frac{(w/|w|)^s \Gamma(1\mp s)}{2|w|^{\Delta_2+\Delta_3+2}}\hfill\\
    \times \left\{\mp\xi^{\frac{\ell_3-\ell_2+1}{2}}\left[\frac{\Gamma (\ell_3-\ell_2+1)}{\Gamma
   (h_3-h_2\mp\frac{s}{2}) \Gamma (\bar{h}_2-\bar{h}_3+1\mp \frac{s}{2})}+O(\xi^{-1/2})\right]\right.\\
  \hfill \left.\pm  \xi^{\frac{\ell_2-\ell_3+1}{2}}\left[\frac{ \Gamma (\ell_2-\ell_3+1)}{\Gamma
   (h_2-h_3\mp \frac{s}{2}) \Gamma (\bar{h}_3-\bar{h}_2+1\mp \frac{s}{2})}+O(\xi^{-1/2})\right]\right\}\\
   \phantom{\left. \langle F_{z\bar{z}}\mathcal{O}_2\mathcal{O}_3\rangle\right|_{s\in \mathbb{Z}\pm \frac{1}{4}}}+C_4\frac{(w/|w|)^s \Gamma(1\pm s)}{2|w|^{\Delta_2+\Delta_3+2}}\hfill\\
    \times \left\{\pm \xi^{\frac{\ell_3-\ell_2+1}{2}}\left[\frac{ \Gamma (\ell_3-\ell_2+1)}{\Gamma
   (\bar{h}_2-\bar{h}_3\pm \frac{s}{2}) \Gamma (h_3-h_2+1\pm \frac{s}{2})}+O(\xi^{-1/2})\right]\right.\\
  \hfill \left.\mp  \xi^{\frac{\ell_2-\ell_3+1}{2}}\left[\frac{\Gamma (\ell_2-\ell_3+1)}{\Gamma
   (\bar{h}_3-\bar{h}_2\pm \frac{s}{2}) \Gamma (h_2-h_3+1\pm \frac{s}{2})}+O(\xi^{-1/2})\right]\right\}~.
\end{multlined}
\end{equation}
For simplicity, we will focus on cases where $\ell_3 \geq \ell_2$ and $|s|\in \mathbb{N}+\frac{3}{4}$, while other cases follow analogously. When $\ell_3>\ell_2$, the finiteness of the first term in \eqref{eq_three point large argument} requires the gamma functions in the denominator to have poles. Therefore, at least one of $h_3-h_2+\frac{|s|}{2}$ and $\bar{h}_2-\bar{h}_3+1+\frac{|s|}{2}$ is a non-positive integer. 
Assume $h_2\in h_3+\frac{|s|}{2}+\mathbb{N}$, then it follows from $\ell_2,\ell_3\in \mathbb{Z}$ and $\ell_3>\ell_2$ that we also have $\bar{h}_2\in h_3+1+\frac{|s|}{2}+\mathbb{N}$. On the other hand, when $\ell_3=\ell_2$, a similar finiteness condition requires $|h_2-h_3|=|\bar{h}_2-\bar{h}_3|\in 1+\frac{|s|}{2}+\mathbb{N}$. 
In summary, we have found the following condition:
\begin{equation}
    (h_2,\bar{h}_2)\in (h_3,\bar{h}_3)+\left(\frac{|s|}{2},1+\frac{|s|}{2}\right)+(\mathbb{N},\mathbb{N})~.
\end{equation}
The pole analysis for other cases is similar, and altogether we obtain the double twist condition in \eqref{eq_double twist condition}.

\section{Tensor structures in DCFT correlation functions}
\label{sec_dcft tensors}

In this appendix, we review the covariant tensor structures in DCFT correlation functions discussed in section \ref{sec_Generalized free field sector}. 
The possible tensor structures can be enumerated by polynomial combinations of the building blocks. 
These building blocks can be tabulated using the embedding formalism \cite{Billo:2016cpy} or  from taking derivatives of the cross ratios \cite{Herzog:2020bqw,Herzog:2022jqv}, among other equivalent methods. 

In section \ref{sec_Generalized free field sector}, we have used the four building blocks $X_\mu$, $Y_{\nu}$, $I_{\mu z}$, and $I_{\mu \bar{z}}$ for the bulk-defect two-point function. We note that both parity-even and parity-odd structures can be constructed using these blocks \cite{Herzog:2022jqv}. 
Indeed, our DCFT is chiral. 

We now turn to the bulk–defect–defect three-point function. As in the main text, we denote the bulk point by $(z_1,\bar{z}_1,w_1,\bar{w}_1)$ and two points on the twist defect by $(z_2,\bar{z}_2,0,0)$, $(z_3,\bar{z}_3,0,0)$. 
We also define $z_{ij}=z_i-z_j$ with $i,j=1,2,3$. 
The covariant tensor structures in the correlation function are spanned by polynomial combinations of four vector building blocks: $X_{12,\mu}$, $X_{13,\mu}$, $Y_{\mu}$, and $Z_{\mu}$. The vectors $X_{12,\mu}$ and $X_{13,\mu}$ are defined by
\begin{equation}
    X_{ij,\mu}=\frac{|w|}{|z_{ij}|^2+|w|^2}\left(\bar{z}_{ij},z_{ij},\frac{\bar{w}}{2}(1-|z_{ij}/w|^2),\frac{w}{2}(1-|z_{ij}/w|^2)\right)~.
\end{equation}
The transverse pseudo-vector $Y_\mu$ follows the same definition as in \eqref{eq_tensor building blocks}. Finally, the vector $Z_\mu$ is given by
\begin{equation}
    Z_\mu=\frac{z_{12}\bar{z}_{13}+|w|^2}{2|w|z_{23}(|z_{13}|^2+|w|^2)}\left(-|w|^2,z_{13}^2,\bar{w}z_{13},wz_{13}\right)~.
\end{equation}
Index contractions between these vectors yield zero except for the following cases:
\begin{equation}
\label{eq_app_3pt vector contraction}
    \begin{aligned}    &X_{12}^{\phantom{12,}\mu}X_{12,\mu}=X_{13}^{\phantom{13,}\mu}X_{13,\mu}=-Y^\mu Y_{\mu}=\frac{1}{2}~;\\
        &X_{12}^{\phantom{12,}\mu}X_{13,\mu}=\frac{1-\xi}{2(1+\xi)}~,~~X_{12}^{\phantom{12,}\mu}Z_\mu=\frac{1}{2(1+\xi)}~,
    \end{aligned}
\end{equation}
where $\xi$ is the cross ratio defined in \eqref{eq_crossration}. For generic $\xi$, we note that $X_{12,\mu}$, $X_{13,\mu}$, $Y_{\mu}$, and $Z_{\mu}$ form a complete basis for the four-dimensional vector space.

For three-point functions where the field strength $F_{\mu \nu}$ is involved, we consider the following combinations of our building blocks
\begin{equation}
\label{eq_app_rank2 antisymmetric base}
    X_{12,[\mu}X_{13,\nu]}~,~~X_{12,[\mu}Y_{\nu]}~,~~X_{12,[\mu}Z_{\nu]}~,~~X_{13,[\mu}Y_{\nu]}~,~~X_{13,[\mu}Z_{\nu]}~,~~Y_{[\mu}Z_{\nu]}~.
\end{equation}
These six tensors are linearly independent, and they provide a complete basis for rank-2 anti-symmetric tensors. We can identify linear combinations of \eqref{eq_app_rank2 antisymmetric base} that are self-dual $(+)$ and anti-self-dual $(-)$:
\begin{equation}
\label{eq_app_rank2 antisymmetric tensor def}
    \begin{aligned}
        \mathcal{X}_{1,\mu\nu}^\pm=& X_{13,[\mu}Z_{\nu]}\mp Y_{[\mu} Z_{\nu ]}~,\hfill\\
        \mathcal{X}_{2,\mu\nu}^\pm=&X_{12,[\mu}X_{13,\nu]}-2X_{12,[\mu}Z_{\nu]}+2X_{13,[\mu}Z_{\nu]}\pm (X_{12,[\mu}Y_{\nu ]}+ X_{13 [\mu }Y_{\nu ]})~,\hfill\\
        \mathcal{X}_{3,\mu\nu}^\pm=& \frac{1}{\xi}(X_{12,[\mu}Y_{\nu ]}-X_{13 [\mu }Y_{\nu ]})\pm\left (2X_{12,[\mu}Z_{\nu]}+2X_{13,[\mu}Z_{\nu]}+\frac{1}{\xi}X_{12,[\mu}X_{13,\nu]}\right)~,
    \end{aligned}
\end{equation}
such that they satisfy
\begin{equation}
\frac{1}{2}\epsilon_{\mu \nu \rho \sigma}\mathcal{X}_{a}^{\pm,\rho \sigma}=\pm \mathcal{X}_{a,\mu\nu}^{\pm}~, \text{ for }a=1,2,3~.
\end{equation}

\bibliography{ref}
\bibliographystyle{JHEP.bst}
    
\end{document}